\documentclass[%
reprint,
amsmath,amssymb,
aps,
pra,
]{revtex4-2}

\usepackage{graphicx}
\usepackage{dcolumn}
\usepackage{bm}
\usepackage[colorlinks=true,citecolor=blue,linkcolor=red,urlcolor=blue]{hyperref}
\def\bra#1{\mathinner{\langle{#1}|}}
\def\ket#1{\mathinner{|{#1}\rangle}}
\def\braket#1{\mathinner{\langle{#1}\rangle}}

\def\proj#1{\ket{#1}\bra{#1}}

\usepackage[]{changes}
\definechangesauthor[name={Benedikt Tissot}, color=purple]{BT}
\definechangesauthor[name={Ferdinand Omlor}, color=blue]{FO}

\begin{document}



\title{Entanglement generation using single-photon pulse reflection in realistic networks}

\author{Ferdinand Omlor}
\thanks{These authors contributed equally to this work}
\thanks{Present address: Division of Solid State Physics and NanoLund, Lund University, S-22100 Lund, Sweden}
\affiliation{Department of Physics, University of Konstanz, 78457 Konstanz, Germany}

\author{Benedikt Tissot}
\thanks{These authors contributed equally to this work}
\thanks{Present address: Center for Hybrid Quantum Networks, Niels Bohr Institute, University of Copenhagen, Blegdamsvej 17, 2100 Copenhagen Ø, Denmark}
\affiliation{Department of Physics, University of Konstanz, 78457 Konstanz, Germany}

\author{Guido Burkard}
\affiliation{Department of Physics, University of Konstanz, 78457 Konstanz, Germany}


\begin{abstract}
A general entanglement generation protocol between remote stationary qubits using single-photon reflection in a photonic network is explored theoretically.
The nodes of the network consist of single qubits that are typically represented by the spin of a color center, each localized in a separate optical cavity and linked to other nodes via photonic links such as optical fibers.
We derive a model applicable to a wide range of parameters and scenarios to describe the nodes and the local spin-photon interaction accounting for the pulsed (finite bandwidth) nature of flying single photons while optimizing the rate and fidelity.
We investigate entanglement generation between remote qubits and tailor protocols to a variety of physical implementations with different properties.
Of particular interest is the regime of weak coupling and low cooperativity between spin and cavity which is relevant in the cases of the nitrogen and silicon vacancy centers in diamond.
We also take into account the variability of the properties between realistic (stationary) nodes.
\end{abstract}

\maketitle


\section{\label{sec:intro}Introduction}

The existence of quantum entanglement is one of the fascinating aspects of quantum physics. Entanglement is not only interesting from a fundamental physics perspective, e.g., where it plays a key role in understanding the violation of local realism \cite{bell64,clauser69},
but also a cornerstone of multiple promising technological applications that aim to use entangled quantum systems as a resource.
Most notable applications include provably secure communication via quantum key distribution \cite{bennett_quantum_2014,ekert_quantum_1991,bennett92,mayers98,acin07,gisin07,arnon-friedman19,pirandola20} and quantum sensing \cite{bongs23} where entanglement can be used to enhance the sensitivity \cite{xia-2023-entan_enhan_optom_sensin}.
Some of these technologies are even envisioned to become accessible to a broader user base within a so-called quantum internet \cite{Kimble2008,Wehner2018}.
Additionally, reliable generation of remote entanglement can be used as a stepping stone towards modular quantum computing \cite{nemoto_photonic_2014,monroe_large-scale_2014}.
Measurement-based quantum computation and quantum gate teleportation provide methods to implement general quantum computation using entangled states as a resource  \cite{gottesman_demonstrating_1999, raussendorf_one-way_2001, eisert_optimal_2000, jiang_distributed_2007}.

Motivated by this wide range of applications we study approaches to generate remote entanglement in this paper.
Following Refs.~\cite{nemoto_photonic_2014, monroe_large-scale_2014}, we assume that the stationary qubits are implemented by a (central electron) spin, connected via (impinging and reflected) flying qubits implemented by photons, where the interaction between flying and stationary qubits is mediated via cavities.
In this article, we use the nitrogen-vacancy center in diamond (NV) \cite{doherty_nitrogen-vacancy_2013} and silicon-vacancy center in diamond (SiV) \cite{hepp_electronic_2014} as paradigmatic examples.
Potentially, the entanglement generated between the spins can also be saved in a quantum memory, e.g., a nuclear spin coupled to the central spin \cite{taminiau_detection_2012, bradley_ten-qubit_2019}.
This helps when trying to repeat the above process between multiple nodes to generate cluster states or long-range entanglement \cite{childress_fault-tolerant_2005, nemoto_photonic_2014, monroe_large-scale_2014}.

\begin{figure}
    \centering
    \includegraphics[width=8.6cm]{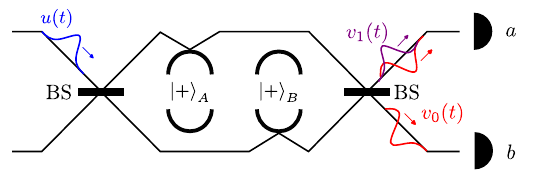}
    \caption{Entanglement generation setup. A single photon in the temporal mode $u(t)$ is sent into a Mach-Zehnder interferometer setup.
    Each interferometer arm is coupled to a separate qubit-cavity node ($A$ or $B$) which is initialized in the state $\ket{+}_{A,B} = (\ket{0}_{A,B} + \ket{1}_{A,B})/\sqrt{2}$. 
    The conditional reflection of the photon at the nodes imprints the qubits states on the photon.
    Here, we indicate the relevant output modes as $v_0(t),v_1(t)$.
    After interference of the photon at the second beamsplitter (BS), the ``which path'' information is erased and the heralding photon detection at channel $a$ or $b$ projects the qubits into an entangled Bell state.
    A detailed description of the entanglement generation protocol is given in Sec.~\ref{sec:Ent_Gen}.}
    \label{fig:setup_one_direction}
\end{figure}
We consider a photonic setup consisting of a Mach-Zehnder interferometer in which a single optical photon is reflected at two qubit-cavity modules, see Fig.~\ref{fig:setup_one_direction}. Their qubit-state dependent reflection amplitudes enable probabilistic entanglement generation which is heralded by the detection of the photon \cite{santori_single_2010, nemoto_photonic_2014}. In case of high cooperativities and one-sided cavities, the protocol can be used to generate high-fidelity Bell states with close to unit probability \cite{santori_single_2010}.
While previous work has shown that a reduction in cooperativity decreases only the success probability and not the generated state fidelity for perfectly identical nodes \cite{nemoto_photonic_2014}, we extend the protocol and analysis to account for non-identical nodes.
This variability can affect the fidelity where the susceptibility to differences between nodes in some quantities is enhanced for low cooperativities.
Additionally, we model the photonic interaction of the defect-cavity module taking the pulse nature of the photon into account and finally investigate possible implementations of the protocol using NV and SiV centers.
Furthermore, we assume that nodes are located sufficiently close to each other such that we can neglect dark counts. 

\begin{figure}[t]
    \centering
    \includegraphics[width=5cm]{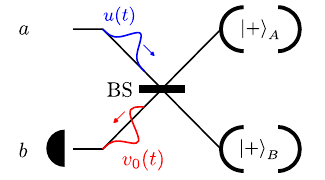}
    \caption{Entanglement generation setup using a single BS.
    The setup and notation are equivalent to the two-BS setup depicted in Fig.~\ref{fig:setup_one_direction}. 
    Note that in this setup the pulses propagate from left to right until they impinge on the qubit-cavity nodes and then propagate in the reverse direction from right to left to the detector.
    }
    \label{fig:setup_reflection}
\end{figure}


The remainder of this article is structured as follows.
We begin our analysis by introducing the model of the nodes and photon pulses in Sec.~\ref{sec:CQED_model}, where we also provide an analytic solution to the dynamics according to the model.
In Sec.~\ref{sec:Ent_Gen} we introduce the setup and entanglement generation protocol.
In the remaining article, we then go through the results based on the previous analysis.
We discuss different flying qubit encodings (Sec.~\ref{sec:flying_enc}), and relevant mechanisms that affect the fidelity (Sec.~\ref{sec:fidelity}).
Then apply the model and protocol to the exemplary implementations in Sec.~\ref{sec:node_implementations}.
Here, we focus on the NV center that can be modeled by a three-level system (Sec.~\ref{sec:three_level}) and the SiV center where we need to include four levels (Sec.~\ref{sec:SiV}).
Finally, we provide a brief comparison to emission-based entanglement generation protocols using cavities in Sec.~\ref{sec:proto_comparison} and conclude in Sec.~\ref{sec:conclusion}.

\section{\label{sec:CQED_model} Node and Photon Pulse Model}
\begin{figure}
\includegraphics[width=8cm]{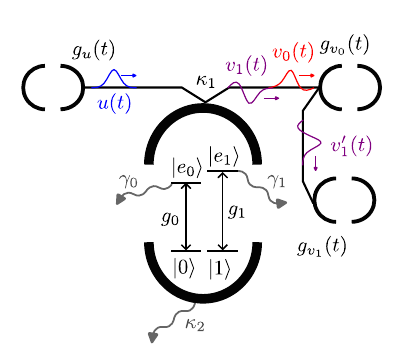}
\caption{Schematic of the model to study the input-output relation of a four-level system in the Schr\"odinger picture. The interaction with quantum pulses is described with a single channel within the Hermitian part of the dynamics.
We assume a virtual cavity that emits the input pulse $u(t)$ via a time-dependent coupling $g_u(t)$ into the channel and two sequentially coupled virtual cavities with coupling constants $g_{v_k}(t)$ ($k=0,1$)  that absorb the relevant outputs after the interaction with the cavity hosting the system. The system in the cavity interacts with the channel at strength $\kappa_1$ and can be modeled by four levels $\ket{k},\ket{e_k}$ with pairwise cyclic interaction with the cavity at strength $g_k$. The gray arrows indicate loss channels at rates $\gamma_k,\kappa_2$.}
\label{fig:CQED_4lvl}
\end{figure}

Before we discuss the entanglement generation, we introduce a cavity quantum electrodynamics (CQED) model to describe the local nodes and their interaction with single-photon pulses.
This interaction is central to the model, understanding, and optimization of the state-dependent reflection of the pulse.
This so-called conditional reflection plays a key role in the entanglement generation protocol discussed in the next section.

We consider an atom(like) system that possesses two pairs of ground and excited states embedded in a cavity. The ground states $\ket{0}_s,\ket{1}_s$ implement the qubit (e.g., for storage) while cyclic transitions to the excited states $\ket{e_0}_s,\ket{e_1}_s$ give rise to the optical interaction. 
The system Hamiltonian describing the interaction between the cavity and the atomic system is
\begin{align}
\label{eq:system-Hamiltonian}
  H_s / \hbar = \sum_{k=0,1} \delta_k \ket{e_k}_s \bra{e_k}_s + g_k \left( \ket{e_k}_s \bra{k}_s \hat{c} + \mathrm{H.c.} \right) ,
\end{align}
with the cavity annihilation operator $\hat{c}$
where we chose an appropriate rotating frame such that $\delta_k$ is the detuning between the atomic transition $\ket{k}_s \leftrightarrow \ket{e_k}_s$ and the cavity resonance frequency.

To describe the coupling of the CQED system to the environment, we introduce four processes. 
Two decay channels of the cavity with rates $\kappa_1$ and $\kappa_2$ 
describing the outcoupling into the channel used for entanglement generation ($\kappa_1$) and additional cavity losses ($\kappa_2$).
These decays can be envisioned to be induced by the mirrors of a two-sided Fabry-P\'{e}rot cavity.
The other two decay channels with rates $\gamma_k$ describe the coupling of the optical transitions to optical modes perpendicular to the cavity, i.e., spontaneous decays of the excited states $\ket{e_k}$.
Combined, these channels are described by the jump operators
\begin{align} \label{eq:L_sys}
    \hat{\vec{L}}_s = \left(\begin{matrix} \sqrt{\kappa_1} \hat{c}\\ \sqrt{\kappa_2} \hat{c} \\ \sqrt{\gamma_0} \ket{0}_s\! \bra{e_0}_s \\ \sqrt{\gamma_1} \ket{1}_s\! \bra{e_1}_s
   \end{matrix} \right)\,.
\end{align}

To model the input and output (IO) of a single photon pulse to the system cavity, we use the recently developed input-output theory for quantum pulses \cite{kiilerich_input-output_2019, kiilerich_quantum_2020}. 
The idea central to this approach is to add virtual cavities with time-dependent out-couplings to the model which can absorb or emit the propagating optical modes of the environment, see Fig.~\ref{fig:CQED_4lvl}.
As shown in \cite{gough_generating_2015,kiilerich_input-output_2019, kiilerich_quantum_2020}, the time-dependent coupling $g_u(t)$ of a virtual cavity, which emits the state inside the virtual cavity into a photon pulse with temporal mode $u(t)$ (the input for the real system), is given by,
\begin{align} \label{eq:g_in}
    g_u(t) = \frac{u^*(t)}{\sqrt{1 - \int_0^t  |u(t')|^2\mathrm{d}t' }}\,,
\end{align}
where we use  square modulus normalization $\int_{0}^\infty  |u(t)|^2\mathrm{d}t = 1$.
Note that here, we assumed a pulse of finite duration but when pulses of infinite duration are considered the reference time $0$ should be replaced with $-\infty$.
This requires that the (real) out-coupling of the nodes (cavities) is constant over the spectrum of $u(t)$ and the validity of the Born-Markov-approximation which is satisfied if $g_u(t)$ changes slowly within an oscillation period of carrier frequency of the photon pulse \cite{kiilerich_quantum_2020}, i.e., if the photon pulse contains multiple optical cycles.

Furthermore, we can capture output channels of the system by cascading them into virtual output cavities which ensure the absorption of an incoming photon in the normalized temporal mode $o(t)$ using the time-dependent coupling strength \cite{nurdin_perfect_2016,kiilerich_input-output_2019, kiilerich_quantum_2020},
\begin{align}
    g_o(t) = -\frac{o^*(t)}{\sqrt{\int_0^t  |o(t')|^2\mathrm{d}t' }}\,.
    \label{eq:g_out} 
\end{align}
As displayed in Fig.~\ref{fig:CQED_4lvl} we are going to capture two orthogonal modes for the decay channel $\propto \kappa_1$ using two virtual cavities in series.
Thus we need to account for the scattering of the second mode $v_1$ on the virtual cavity absorbing the first mode $v_0$,
see also Refs.~\cite{kiilerich_input-output_2019,kiilerich_quantum_2020}.
For brevity we will denote $g_{v_1}(t) = g_{v_1'}(t)$, where $g_{v_1'}(t)$ follows Eq.~\eqref{eq:g_out}
and $v_1'$ denotes the mode $v_1$ scattered on the first virtual cavity.

The single-photon state within the virtual cavities for mode $u(t)$  is linked to the state in terms of the single-mode creation operators $b_p^\dag$ for channel $p$ applied to the vacuum state $\ket{0}_p$ \cite{kiilerich_input-output_2019, kiilerich_quantum_2020},
\begin{align}
\label{eq:pulse_state}
    \ket{u}_p = \int_{-\infty}^{\infty} \mathrm{d}\omega\, \tilde u(\omega) \hat b_p^\dag(\omega) \ket{0}_p\, 
    = \int_{-\infty}^{\infty} \mathrm{d}t\, u(t) \hat b_p^\dag(t) \ket{0}_p\, .
\end{align}
Here, we use $\tilde{f}$ to denote the Fourier transform of $f$, i.e., $\tilde f(\omega) = \sqrt{1/2\pi} \int_{-\infty}^\infty  e^{i\omega t} f(t) \, \mathrm{d}t$.
Note that the Fourier transform leaves the orthogonality and normalization of the temporal modes intact.

Application of the cascade and concatenation rule of the SLH-framework \cite{combes_slh_2017} enables us to describe the combined system including the virtual cavities using a single combined Hamiltonian $\hat{H}(t)$ and jump operators $\hat{L}(t)$.
In agreement with \cite{kiilerich_input-output_2019,kiilerich_quantum_2020} we find that the coupling to the virtual cavities leads to the combined Hamiltonian
\begin{align}
  \label{eq:kiilerlich_Hamiltonian}
  H / \hbar = 
& H_s / \hbar +  \frac{i}{2} \Big\{ g_u(t) \hat{a}_u^{\dag} [\sqrt{\kappa_1} \hat{c} + \sum_{l=0,1} g_{v_l}^{*}(t) \hat{a}_{v_l}] \notag \\
& + [\sqrt{\kappa_1} \hat{c}^{\dag} + g_{v_0}(t) \hat{a}_{v_0}^{\dag}] g_{v_1}^{*}(t) \hat{a}_{v_1} \notag \\
& + \sqrt{\kappa_1} \hat{c}^{\dag} g_{v_0}^{*}(t) \hat{a}_{v_0} - \mathrm{H.c.} \Big\} ,
\end{align}
and the first component of the jump operator changes to
\begin{align}
  \label{eq:jump-4lvl}
  \hat{L}_1(t) = g_u^{*}(t) \hat{a}_u(t) + \sqrt{\kappa_1} \hat{c} + \sum_{l=0,1} g_{v_l}^{*}(t) \hat{a}_{v_l} ,
\end{align}
while the remaining components remain unchanged.

We are interested in two cases that will be relevant to the protocol introduced in the next section.
In both the atomic system is prepared in a superposition of the qubit states $\ket{k}$ ($k=0,1$) and we either have no initial excitations or a single photon pulse as the initial excitation.
In the first case, there is no excitation in the combined system leading to a trivial time evolution (in the chosen frame of reference).

For the second case, we consider a single excitation in the system which is initially within the input pulse $u(t)$.
Considering that in the protocol the state will be heralded conditioned on a click of a detector that corresponds to the existence of a photon in the channel interacting with the system via $\kappa_1$, we treat the remaining decay channels as losses within the framework of a non-Hermitian Hamiltonian \cite{dalibard92,moelmer93,carmichael93,daley14}.
We use an ansatz wavefunction for the single excitation subspace
\begin{align}
  \ket{\Psi(t)} =  \sum_{k=0,1} &\Big(\alpha_u^k(t) \ket{k,u} + \alpha_c^k(t) \ket{k,c} + \alpha_{e_k}(t) \ket{e_k} \notag \\
  \label{eq:ansatz_single_excitation}
  &+ \sum_{l=0,1}\alpha_{v_l}^k(t) \ket{k,v_l} \Big),
\end{align}
where the states refer to the states with a single excitation in the atom, \(\ket{e_k}\) (\(k=0,1\)), the cavity \(\ket{k,c}\), or the virtual cavities describing the photon pulses for input \(\ket{k,u}\), and the used output \(\ket{k,v_l}\) (\(l=1,2\)). 
For the latter three, we used the first entry of the ket \(k\) to indicate the qubit state (atomic ground state) \(\ket{k}_s\) because the excitation can coexist with both qubit states.

We use temporal mode matching \(\hat{L}_1(t) \ket{\Psi(t)} = 0\) \cite{tissot_efficient_2024}
combined with a non-Hermitian Schrödinger equation
\(i \frac{\partial }{\partial t} \ket{\Psi(t)} = \left[ H/\hbar - \frac{i}{2} (\kappa_2 c^{\dag} c + \sum_{l=0,1} \gamma_l \proj{e_l}) \right] \ket{\Psi(t)}\)
to derive the solution of the system dynamics in Appendix~\ref{app:solving_node_dynamics}.

\subsection{Solution for the final amplitudes and output modes}
After the complete reflection of the pulse on the node at time $T$ we find the (non-normalized) state,
\begin{align}
  \label{eq:final_state}
  \ket{\Psi(T)} =  \sum_{k,l=0,1} \alpha_{v_l}^k(T) \ket{k,v_l} ,
\end{align}
where the amplitudes and pulse shapes can be related to the initial state and input pulse shape in the frequency domain
\begin{align}
    \label{eq:MM-FT-final}
  & \alpha_{v_1}^k(T) \tilde{v}_1(\omega) + \alpha_{v_0}^k(T) \tilde{v}_0(\omega) = r_k(\omega) \alpha_u^k(0) \tilde{u}(\omega) ,
\end{align}
here $\omega$ refers to the detuning between the actual photon pulse frequencies and the cavity frequency due to the choice of the rotating frame.
Again, we assumed pulses of finite duration, if pulses of infinite duration are considered (e.g., a Gaussian pulse) the appropriate initial and final times need to be changed to $\pm \infty$ (from $0$ and $T$).
Further, we defined the transfer function (for the reflection/scattering process)
\begin{align}
    \label{eq:refl_function}
    r_k(\omega) =\,& 1 - \frac{\kappa_1}{\frac{\kappa}{2} - i \omega + \frac{g_k^2 [ i (\omega - \delta_k) + \gamma_k/2 ]}{(\omega - \delta_k)^2 + \gamma_k^2/4}} \notag \\
    =\,& 1 - \frac{2\frac{\kappa_1}{\kappa}(1-2i\frac{\omega - \delta_k}{\gamma_k})}{(1-2i\frac{\omega - \delta_k}{\gamma_k})(1-2i\frac{\omega}{\kappa}) + C_k}\,, 
\end{align}
with the state dependent cooperativity $C_k = \frac{4 g_k^2}{\kappa \gamma_k}$.
The transfer function satisfies $|r_k|^2 < 1$ \footnote{To proof this we use that we can write $r_k = 1 - \frac{\kappa_1}{a+ib}$, where $a,b\in \mathbb{R}$ and $2a > \kappa \ge \kappa_1$, such that $|r_k|^2 = 1 - \frac{\kappa_1 (\kappa_1 - 2a)}{a^2 + b^2} \le 1$ which follows because the second term is positive and $\frac{\kappa_1}{a} (2 - \frac{\kappa_1}{a}) \le 1$ for all $0 \le \frac{\kappa_1}{a} \le 2$.}
and agrees with the transfer function of the empty cavity for $g_k = C_k = 0$.
Additionally, we stress that the reflection transfer function for a single cyclic transition is compatible with established results assuming a two-level atom in the cavity (Jaynes-Cummings model) \cite{hu_giant_2008, hu-2009-propos_entan_beam_split_using,santori_single_2010,nemoto_photonic_2014}. 
However, our solution accounts for the full 4-level system without relying on approximations like adiabatic elimination and discerns the involved orthogonal temporal modes.
Then the amplitudes and temporal modes can be determined by accounting for the orthonormality of $v_0,v_1$ which is conveyed to the frequency domain $\tilde{v}_0, \tilde{v}_1$.
In other words, the frequency dependence of Eq.~\eqref{eq:MM-FT-final} determines the shapes of $\tilde{v}_k$ (where there is some room to choose one $v_k$ to simplify the calculations) and then the $\alpha_{v_l}^k(T)$ are fixed by the equality combined with the normalization of $\tilde{v}_l(\omega)$.

\section{\label{sec:Ent_Gen}Entanglement Generation Protocol}

Having discussed the model to describe the interaction between stationary nodes and photon pulses, we now turn to the description of a general protocol using a Mach-Zehnder interferometer setup to generate entanglement between two qubits at (spatially separated) nodes $A$ and $B$.
A single (input) flying photon pulse is split by a beamsplitter (BS) into two channels, each connected to one of the cavities containing the qubits.
After interacting with the qubits, the outputs undergo interference at a BS once again.
This can be understood as removing the ``which path information'', such that a measurement of the outputs of the BS can be used for heralded entanglement generation between the qubits.
Two implementations of this protocol are sketched in Fig.~\ref{fig:setup_one_direction} and Fig.~\ref{fig:setup_reflection}. Both are mathematically equivalent, but the latter only requires a single BS.

To explain the protocol, we consider two identical nodes $A$ and $B$.
Therefore, the systems are described by the same parameters and follow the same dynamics according to the results of Sec.~\ref{sec:CQED_model}.
With this, we can account for the multimode nature of the photon,
which takes us beyond previous considerations in the single-mode domain \cite{nemoto_photonic_2014, santori_single_2010}.
The case of two different nodes is analyzed in detail in Appendix~\ref{app:diff_modules}.
We describe the protocol to achieve the heralded entanglement generation of a Bell state between qubits $A$ and $B$ within this setup using the following five steps.

\subsubsection{Initialization\label{sec:step1}}
First, each qubit ($A$ and $B$) is initialized to 
\begin{align}
    \ket{+}_{s} = \frac{1}{\sqrt{2}}\big(\ket{0}_{s} + \ket{1}_{s} \big)\,,
\end{align}
where $s=A,B$.

\subsubsection{Photon injection\label{sec:step2}}
Second, a single-photon pulse is emitted into the upper channel of the interferometer setup (see Figs.~\ref{fig:setup_one_direction} and \ref{fig:setup_reflection}) and is then split at a symmetric BS. 
The BS transforms a single photon pulse $u$ according to
\begin{align}
    \begin{cases}
        \ket{u}_a\ket{0}_b \\
        \ket{0}_a \ket{u}_b
    \end{cases} \rightarrow ~\frac{1}{\sqrt{2}} (\ket{u}_a \ket{0}_b \pm \ket{0}_a\ket{u}_b)\,.
\end{align}
Thus, the resulting state can be written as
\begin{align}
    \frac{1}{\sqrt{2}}\ket{+}_A\ket{+}_B \big(\ket{u}_a \ket{0}_b + \ket{0}_a \ket{u}_b \big)\,,
\end{align}
where $\ket{\cdot}_a$ and $\ket{\cdot}_b$ describe the upper ($a$) and lower ($b$) photonic channels which respectively couple to system $A$ and $B$ after the BS.

\subsubsection{Photon reflection\label{sec:step3}}
Third, the photon interacts with the qubit-cavity modules $A$ and $B$. We can cast the state into the form Eq.~\eqref{eq:final_state},
\begin{align} 
 \label{eq:setup_state_pre_BS}
    \frac{1}{2} \hspace{-3mm} \sum_{k,k',l = 0,1} \hspace{-3mm} \ket{k}_A  \ket{k'}_B  \big[ \alpha_{v_l}^{k}(T) \ket{v_l}_a \ket{0}_b + \alpha_{v_l}^{k'}(T) \ket{0}_a \ket{v_l}_b \big] \,,
\end{align}
where Eq.~\eqref{eq:MM-FT-final} determines the amplitudes for the initial condition $\alpha_u^k = 1/\sqrt{2}$.
We remind the reader that this wavefunction is not normalized, because it represents the state conditional on the absence of losses. Thus, the norm of Eq.~\eqref{eq:setup_state_pre_BS} quantifies the probability that the photon was reflected back into the photonic channel.
The amplitudes and photonic wavefunctions follow the notation introduced in Sec.~\ref{sec:CQED_model} where the indices distinguish the different systems and channels.

\subsubsection{Photon interference\label{sec:step4}}
Next, the reflected photon components are interfered again using a symmetric BS.
As depicted in Fig.~\ref{fig:setup_one_direction} and Fig.~\ref{fig:setup_reflection}, two physical setups are equivalent within the theoretical description: Either, the photon is reflected back to the initial BS or sent further to a second symmetric BS.
In both cases, the resulting state can be written as (see Appendix~\ref{app:diff_modules}),
\begin{align}
    \ket{f} = \frac{1}{\sqrt{2}} \Big[ ( \ket{\Phi^+} + \ket{\Psi^+} )   ( & \alpha_{v_0}^+ \ket{v_0}_a + \alpha_{v_1}^+ \ket{v_1}_a) \ket{0}_b \nonumber \\
    + \ket{\Phi^-} \qquad \qquad & \alpha_{v_0}^- \ket{v_0}_a \ket{0}_b \notag \\
   \label{eq:setup_state_final}
    + \ket{\Psi^-} \qquad \qquad & \alpha_{v_0}^- \ket{0}_a \ket{v_0}_b \Big] ,
\end{align}
with the Bell states
\begin{align}
    \ket{\Phi^\pm} = (\ket{0}_A \ket{0}_B \pm \ket{1}_A\ket{1}_B)/\sqrt{2} \,, \label{eq:bell_phi} \\
    \ket{\Psi^\pm} = (\ket{0}_A\ket{1}_B \pm \ket{1}_A\ket{0}_B)/\sqrt{2}\,.
    \label{eq:bell_psi}
\end{align}
Here we define $\sqrt{2} \alpha_{v_l}^{\pm} = \alpha_{v_l}^0(T) \pm \alpha_{v_l}^1(T)$ and
without loss of generality choose $\alpha_{v_1}^- =0$ which then uniquely determines the pulse shapes $v_0,v_1$,
i.e., they satisfy the relations
\begin{align}
    \alpha_{v_0}^{-} \tilde{v}_0(\omega) & = r_-(\omega) \tilde u(\omega)\,, \label{eq:complete_transfer}\\
    \alpha_{v_1}^{+} \tilde{v}_1^+(\omega) & = r_+(\omega) \tilde u(\omega) - \alpha_{v_0}^{+} \tilde{v}_0(\omega)\,, \label{eq:complete_transfer_plus}\\
    r_\pm(\omega) & = \frac{ r_0(\omega) \pm r_1(\omega)}{2}\,,
\end{align}
with $r_k(\omega)$ ($k=0,1$) from Eq.~\eqref{eq:refl_function}.
Note that from $|r_k(\omega)| \le 1$ it also follows that $|r_\pm(\omega)| \le 1$ due to the triangle inequality.
The product of the transfer functions $r_{\pm}(\omega)$ and the Fourier transform of the temporal mode $\tilde{u}(\omega)$ is a convolution between the temporal mode and the Fourier transform of $r_\pm$ in the time domain.
Combined with the orthonormality of the modes ($v_0,v_1$), Eq.~\eqref{eq:complete_transfer} yields $v_0$ and $\alpha_{v_0}^-$ and then Eq.~\eqref{eq:complete_transfer_plus} gives $v_1$ as well as $\alpha_{v_l}^+$ ($l=0,1$).
We stress that from Eq.~\eqref{eq:setup_state_final} it is visible that the states $\propto \alpha_{v_l}^+$ cannot be discriminated and are therefore not suited for entanglement generation.
At the same time, both states $\propto \alpha_{v_0}^-$ can be used for entanglement generation as we will discuss below.
Note that also for two differing systems only the states $\propto \alpha^-_{v_l}$ (where the amplitudes and modes can also differ between the modes) can be used for entanglement generation, see Appendix \ref{app:diff_modules}.

\subsubsection{Photon detection\label{sec:step5}}
Finally, the detection of the photon in one of the channels can be used to herald the success of the protocol and project the qubits onto an entangled state with the appropriate choice of channel and transfer functions.
Assuming a photon detector that cannot discriminate between the pulse shapes $v^{\pm}$ implies that the state after a single-photon detection event on port $p=a,b$ is a statistical superposition of pure quantum states $\hat\rho_{\mathrm{det}}^p$.
In the next section (Sec.~\ref{sec:flying_enc}) we show two approaches to tailor the setup such that the resulting states are maximally entangled pure states upon the successful detection of a photon. In Appendix~\ref{app:photon_detection}, we provide further details on the detection model.
On the other hand, if no photon heralding the entanglement was detected, the entanglement protocol must be restarted from the beginning with the initialization of the qubits, Sec.~\ref{sec:step1}.

\subsection{Flying Qubit Encoding\label{sec:flying_enc}}

In the previous section, we discussed the general protocol. We now detail approaches to implement a photonic encoding and measurement to herald a maximally entangled state.
As is discernible from the previous section, in particular  Eq.~\eqref{eq:setup_state_final}, 
the wavefunctions $\propto \alpha_{v_0}^-$ can be discriminated by their photonic component, such that one can view $|\alpha_{v_0}^-|^2 > 0$ as a necessary requirement for the protocol at hand.
Note that the success probability of the protocol depends on the specific difference in the reflection transfer functions.
%
As is visible in Eq.~\eqref{eq:setup_state_final}, a photon detection at the lower port ($b$) projects onto the maximally entangled Bell state $\ket{\Psi^-}$.
Therefore, the success probability of the protocol generating $\ket{\Psi^-}$ is given by the photon detection probability
\begin{align}
    P_b = \eta \frac{|\alpha_{v_0}^-|^2}{2} &=  \frac{\eta}{2}\int \mathrm{d}\omega |r_-(\omega)|^2 |\tilde u(\omega)|^2 \le \frac{\eta}{2}\,, \label{eq:Pdet}
\end{align}
where we introduced the photon transmittivity of the setup $\eta<1$
and obtained $\alpha_{v_0}^-$ by integrating the modulus squared of Eq.~\eqref{eq:complete_transfer} and using the pulse shape normalization.
Here, we assumed that the exact time of the photon detection is lost and the detector integrates over the complete photon pulse duration (and cannot discriminate different pulse shapes).
As will be further pointed out in the following, the upper bound $\eta/2$ is reached if $A$ and $B$ act as perfect mirrors with a state-dependent phase, i.e., $r_0(\omega) = - r_1(\omega) $ and $|r_{0,1}(\omega) |^2 = 1$ where  $\tilde u(\omega)\neq 0$.

We now investigate two useful special cases of the conditional reflection which imprints the qubit state in the transfer function (and can entangle qubit and photon).
We detail these ``intensity'' and ``phase'' encoding below.

\subsubsection{Intensity Encoding}

Intensity encoding refers to the case where the qubit-cavity modules are tuned such that the photon is only reflected for one of the qubit states, e.g., $\ket{0}$, such that the qubit state is imprinted in the reflected photon intensity.
Starting with one of the qubits $s=A,B$ in a general state $\alpha_u^0(0) \ket{0}_s + \alpha_u^1(0) \ket{1}_s$, the photon interaction is ideally described by
\begin{align}
    \left[\alpha_u^0(0) \ket{0}_s + \alpha_u^1(0) \ket{1}_s \right]\ket{u}_p \rightarrow \alpha_{v_0}^0(T) \ket{0}_s \ket{v_0}_p ,
\end{align}
where $\alpha^0_{v_0}(T)$ and $v_0$ are given by the qubit-cavity transfer function~\eqref{eq:refl_function} and $p=a,b$ is the photon channel connected to $q$.
Note that to satisfy this condition, we need $r_{1}(\omega) = 0$ which implies $r_{\pm}(\omega) = r_0(\omega)/2$, such that we can solve Eqs.~\eqref{eq:complete_transfer} and \eqref{eq:complete_transfer_plus} with a single mode $v_0$, i.e., we can choose $\alpha_{v_1}^{\pm} = 0$ and $\alpha_{v_0}^{\pm} = \alpha_{v_0}$.
While in general $r_1(\omega) = 0$ cannot be satisfied for all $\omega$, our model can be used to tune the system such that $r_1(\omega) \approx 0$ 
by finding the root of the second derivative, analogous to the approach we use for $r_-$ in Eq.~\eqref{eq:opt_kappa} to optimize the rate.
The final pre-detection state (i.e., the state after step 4, see Sec.~\ref{sec:step4}) in the ideal case is given by
\begin{align}
    \ket{f} = \frac{\alpha_{v_0}}{\sqrt{2}} \big[ \left( \ket{\Phi^+} + \ket{\Phi^-} + \ket{\Psi^+} \right) & \ket{v_0}_a \ket{0}_b \nonumber \\
    + \ket{\Psi^-} \hspace{13mm} & \ket{0}_a \ket{v_0}_b \big] \,,
    \label{eq:setup_state_final_intensity}
\end{align}
and the protocol success probability is $\eta |\alpha_{v_0}|^2/2 \le \eta/8$ which is below the upper bound of the general protocol $\eta/2$. 

On the other hand, the encoding of the qubit state in the intensity of the reflected photon allows for easy optical qubit-state readout: Repeatedly sending a photon onto the qubit-cavity system and counting the reflected photons enables a quantum non-demolition measurement of the qubit \cite{nemoto_photonic_2014}.
The entanglement generation protocol using intensity encoding, while disregarding the pulse nature of the photon, was investigated in Refs.~\cite{nemoto_photonic_2014, koshino_entangling_2012}.

\subsubsection{Phase Encoding}

In the phase encoding case, the qubit-cavity module acts as a phase switch that applies a $\pi$ phase to the reflected photon dependent on the qubit state i.e. the interaction of qubit $s=A,B$ with an incoming photon (in the respective channel $p=a,b$) is ideally described by
\begin{align}
    &\big[ \alpha_u^0(0) \ket{0}_s + \alpha_u^1(0) \ket{1}_s \big]\ket{u}_p  \nonumber\\
    &\qquad\rightarrow \frac{\alpha_{v_0}^0(T)}{\alpha_u^0(0)}\big[ \alpha_u^0(0) \ket{0}_s - \alpha_u^1(0) \ket{1}_s  \big] \ket{v_0}_p\,,
\end{align}
where $\alpha_{v_0}^0(T)/\alpha_u^0(0) = - \alpha_{v_0}^1(T)/\alpha_u^1(0)$.
If the perfect phase difference is achieved this corresponds to $\alpha_{v_l}^+ = 0$ ($l=0,1$), from which we see that this case can (ideally) also be described with a single relevant temporal mode.
The final pre-detection state is given by
\begin{align}
    \ket{f} &= \frac{\alpha_{v_0}^-}{\sqrt{2}}\bigg( \ket{\Phi^-} \ket{v_0}_a \ket{0}_b + \ket{\Psi^-} \ket{0}_a \ket{v_0}_b \bigg) ,
    \label{eq:setup_state_final_phase}
\end{align}
where it is immediately visible that the clicks of both detectors now enable the heralding of entanglement.

This makes phase encoding especially useful: not only can the maximum photon detection probability at the lower port $\eta/2$ can be reached, but also detection of the photon at the upper port projects the qubits into a Bell state, namely $\ket{\Phi^-}$. 
Because both possible photon detection events result in a known Bell state, both can be used, leading to a combined protocol success probability of $P_a + P_b = \eta |\alpha_{v_0}^-|^2$ \cite{santori_single_2010} which makes this encoding favorable for entanglement generation for ideal devices.
We will, however, show below that only using port $b$ for the heralding measurement leads to less susceptibility to the pulse nature of the photons.
Note that while here we rely on $\alpha_{v_l}^+ = 0$ for $l=0,1$ [which according to Eq.~\eqref{eq:complete_transfer_plus} implies $r_+(\omega) = 0$ where $\tilde u(\omega) \neq 0$], it is also possible to use both ports when the detector can discriminate the $v_l$ modes and $\alpha_{v_0}^+ = 0$.
However, demanding $\alpha_{v_l}^+ = 0$ leads to fewer requirements on the detectors and simultaneously maximizes the detection probability such that we consider this more optimal phase encoding in the following. 

When considering different applications, further steps including single qubit gates and readout of the qubit should be accounted for.
Therefore we note that in the case of the intensity encoding the CQED system can be used for (non-demolition) qubit readout without an interferometer \cite{volz_measurement_2011, nemoto_photonic_2014} while an interferometer setup is necessary to use the CQED system for readout in the phase encoding case, see Appendix \ref{app:readout} for more details on optical state detection.
In both cases we envision single qubit gates to be implemented using microwaves.

\subsection{Sources of Infidelity and Rate Reduction of the Entanglement Generation\label{sec:fidelity}}

Apart from the access of the entanglement for further processing and utilization, we have to take deviations of the real devices from the idealized theory into account when implementing the entanglement generation protocol.
These deviations can lead to a reduced rate of the entanglement generation as well as decreased fidelity $F_a = \braket{\Psi^{-}|\hat\rho^a_{\mathrm{det}}|\Psi^{-}}$ ($F_b = \braket{\Phi^{-}|\hat\rho^b_{\mathrm{det}}|\Phi^{-}}$) of the generated entangled state upon detection in channel $a$ ($b$).
Here, we expressed the fidelity in terms of the density operator of the two qubits $\hat\rho_{\mathrm{det}}^p$ after the heralding photon detection at channel $p=a,b$ (see Appendix~\ref{app:photon_detection}).
In the following, we will briefly discuss the main imperfections that need to be accounted for in real-world applications.

\subsubsection{Photon Loss}

The successful entanglement generation is heralded by a photon detection event. Therefore, if the photon is lost with the probability $1-\eta$ or is not reflected by the qubit-cavity modules, only the detection probability $P_{p}$ is lowered while the fidelity $F_p$ of the generated entangled state stays unaffected.

\subsubsection{Differences between modules and photon pulse effects} \label{sec:different_modules_general}

The qubit-cavity modules $A$ and $B$ may have slightly different properties which might also change over time, e.g., due to spectral diffusion.  This leads to differences in their transfer functions, i.e., $r_{k}^A(\omega) \ne r_{k}^B(\omega)$ for $k=0,1,\pm$.
Additionally, a photon pulse has a bandwidth, i.e., not only a single relevant frequency $\omega$ for which one can tune the transfer function.
Both can hinder the interference at the second BS and therefore lead to an admixture of additional (unwanted) states after the heralding detection of the photon.
These result in a decreased fidelity $F$ by reducing the indistinguishability of the reflected photon components.

For concreteness, we assume a Gaussian photon pulse as the input,
$\tilde u(\omega) = (\pi \sigma_u^2)^{-1/4} \exp \left(-\frac{(\omega - \Delta)^2}{2 \sigma_u^2} \right)$,
with the detuning $\Delta$ between the pulse center frequency and the cavity frequency,
and two nearly identical nodes $|r_{\pm}^-(\omega)| \ll |r_{-}^+(\omega)|$, where $2 r_{\sigma}^{\pm} = r_{\sigma}^A \pm r_{\sigma}^B$ denotes the average and difference of the transfer function.
For sufficiently narrow pulses the transfer function only changes slightly over the bandwidth $\sigma_u$ around $\Delta$, 
thus we find the approximate fidelity (see Appendix \ref{app:diff_modules_fidelity}),
\begin{widetext}
\begin{align}
    \label{eq:fidelity_nearly_identical_narrow_pulse}
    \left.\begin{aligned}
    F_a \\
    F_b
    \end{aligned}\right\}
    \approx 1 
    - \frac{2 |r_+^{\pm}|^2 + |r_-^-|^2}{ |r_-^+|^2}
    - \frac{\sigma_{u}^{2}}{4} \frac{\partial^2 \left(2 |r_+^{\pm}|^2 + |r_-^-|^2\right)}{\partial \omega^2} \frac{1}{ |r_-^+|^2}
    + \frac{\sigma_u^2}{4} \frac{\partial^2 |r_-^+|^2}{\partial \omega^2} \frac{2 |r_+^{\pm}|^2 + |r_-^-|^2}{|r_-^+|^4} .
\end{align}
\end{widetext}
In this expression and until the end of this Section, the transfer functions and their derivatives are evaluated at the detuning $\Delta$.
We note that also the second derivatives satisfy $\left|\frac{\partial^2|r_{\pm}^-|^2}{\partial \omega^2}\right| \ll |r_-^+|^2$ for nearly identical modules. However, in general, the phase encoding does not ensure this condition for $r_+^+$, i.e., in general $\left|\frac{\partial^2|r_+^+|^2}{\partial \omega^2}\right| \not\ll |r_-^+|^2$.
Therefore, the main insight these equations provide is that the fidelity for detection on port $b$ for identical systems is one and is not affected by the finite bandwidth of the photon pulse.
However, the fidelity is decreased when the systems differ and additionally, differing systems also become susceptible to the finite bandwidth (as a combined effect).
On the other hand, the fidelity $F_a$ is reduced by the finite bandwidth even for identical systems and further reduced by differing systems.

If an intensity or phase encoding is achieved for module $A$ and $B$ individually, the differences in their reflection transfer functions can be counteracted because each node has only one relevant output mode.
This would trade off a lower entanglement generation rate for a higher fidelity.
For example, if a phase encoding is achieved, then $r_+^s=0$ (for $s=A,B$) and $r_+^{\pm} = 0$.
Linear optics components in one of the interferometer arms (such as a BS and a phase plate) can achieve $r_-^- = 0$ such that the second term in Eq.~\eqref{eq:fidelity_nearly_identical_narrow_pulse} vanishes and the fidelity approaches unity for sufficiently narrow band pulses.
Analogously, this applies to the intensity encoding for $F_b$, where $r_+^s = r_-^s$ (for $s=A,B$), and thus using linear optics to achieve $r_-^- = 0$ also achieves $r_+^- = 0$.

Additionally, Eq.~\eqref{eq:fidelity_narrow_pulse_ab} shows a path towards reducing the effect of a finite bandwidth.
Here, the nodes are optimized such that their CQED parameters minimize the second derivative of the modulus square of the transfer function with respect to $\omega$.
We will further elaborate this in Sec.~\ref{sec:three_level} and Figs.~\ref{fig:r_minus} and \ref{fig:Pdet_Fa_NV}.

\subsubsection{Phase Stability}

The protocol analyzed here requires phase stability of the interferometer which has been demonstrated for similar optical setups \cite{chou_measurement-induced_2005, stockill_phase-tuned_2017, humphreys_deterministic_2018}. 
A phase difference between the two interferometer arms $\phi$ causes a mixing of the photon components which results in the final state,
\begin{align}
    \ket{f(\phi)} = \cos (\phi/2) \ket{f} + i \sin(\phi/2) \ket{f}_{a \leftrightarrow b},
\end{align}
where $\ket{f}_{a \leftrightarrow b}$ is obtained by interchanging the optical channels $a$ and $b$ in the state $\ket{f}$ in  Eq.~\eqref{eq:setup_state_final}.
After the detection of the photon in the lower port $b$, the generated state has the fidelity,
\begin{align}
    F_b = \frac{\cos^2(\phi/2) |\alpha_{v_0}^-(T)|^2}{|\alpha_{v_0}^-(T)|^2 + 2\sin^2(\phi/2)[|\alpha_{v_0}^+(T)|^2 + |\alpha_{v_1}^+(T)|^2]}\,,
\end{align}
where we assumed that the nodes are identical and that the phase difference is unknown and not counteracted.
From this expression, we deduce that the fidelity can be increased by minimizing $|\alpha_{v_l}^+(T)|^2$.
Therefore, the phase encoding is most robust against small phase fluctuations, and the fidelity on both ports is given by $F_{a,b} = \cos^2(\phi/2)$.


\subsubsection{Dark Counts}

For long distances between $A$ and $B$, 
the probability of detecting the heralding photon can be very low and approach the probability of a detector dark count \cite{rozpedek_near-term_2019}. 
However, we focus on data-center scale applications of the entanglement protocol and, therefore, assume that the relevant success probability $P_a$ or $P_b$ is significantly larger than the dark count probability.
Note that our analysis can be extended to account for dark counts, e.g., by introducing a depolarizing channel before the detection \cite{rozpedek_near-term_2019}.

\section{Stationary Node Implementations}\label{sec:node_implementations}

In the following, we apply the multimode CQED model, established in Sec.~\ref{sec:CQED_model}, to the general entanglement protocol description introduced in the preceding Sec.~\ref{sec:Ent_Gen}.
First, we investigate a three-level atomic system using the nitrogen vacancy (NV) defect center in diamond as a physical example.
Then, we briefly discuss a significant additional limitation to the rate and fidelity if the second transition is not decoupled from the cavity and therefore the full four-level model needs to be accounted for.
This is necessary for systems such as the silicon vacancy (SiV) defect center in diamond.

We choose these particular systems as examples, because
defects in wide-bandgap semiconductors are promising platforms for multiple applications in quantum technology, such as quantum sensing, computing, and communication \cite{ruf_quantum_2021, awschalom_quantum_2018}.
While several different defects are under investigation, the nitrogen-vacancy (NV) center \cite{doherty_nitrogen-vacancy_2013} and group IV defects in diamond, first and foremost the silicon vacancy (SiV) center \cite{hepp_electronic_2014}, are among the most prominent.
For both the NV and the SiV center, the coupling of the electronic spin states to optical photons enables spin-photon \cite{togan_quantum_2010, nguyen_integrated_2019} and spin-spin entanglement generation \cite{humphreys_deterministic_2018, pompili_realization_2021}. 
In its single negatively charged configuration, the NV ground states possess long coherence times \cite{abobeih_one-second_2018} and can be manipulated using microwave pulses. 
Additionally, for both defects, the electronic spin couples to nearby long-lived nuclear spins through the hyperfine interaction \cite{bradley_ten-qubit_2019, nguyen_integrated_2019} which enables long-lived quantum memory \cite{dutt2007,fuchs2011}, entanglement swapping \cite{childress_fault-tolerant_2005} and distillation \cite{kalb_entanglement_2017}.
This makes such defects good candidates for quantum network nodes and repeaters.

\begin{figure}
    \centering
    \includegraphics[width=8cm]{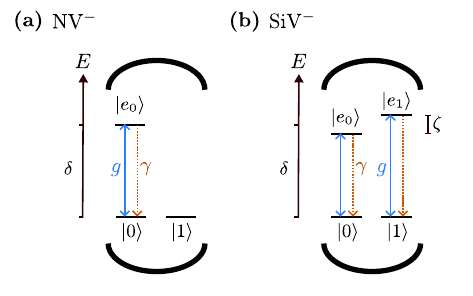}
    \caption{Effective level schemes for the NV$^-$ and SiV$^-$ centers in diamond.
    (a) The NV$^-$ defect center can be modeled as a three-level L-system \cite{nemoto_photonic_2014} where the transition $\ket{0}\leftrightarrow\ket{e_0}$ couples to the cavity mode with detuning $\delta$ and coupling strength $g$.
    The spontaneous decay rate from $\ket{e_0}\rightarrow\ket{0}$ is denoted $\gamma$, and the state $\ket{1}$ is decoupled from the cavity.
    (b) The optical interaction of a SiV$^-$ defect can effectively be described by a four-level system \cite{nguyen_quantum_2019}.
    Both ground (qubit) states can be excited with transition frequencies split by $\zeta$.
    }
    \label{fig:level_schemes}
\end{figure}

Another reason for the choice of these particular examples is that they are modeled by two distinct level structures, see Fig.~\ref{fig:level_schemes}.
While the NV can be described by three levels, where only two couple to the cavity, for the SiV two pairs of states that couple to the cavity are necessary within the model.

\subsection{Three Level System - NV Center} \label{sec:three_level}

\begin{figure*}
    \centering
    \includegraphics[width=17.8cm]{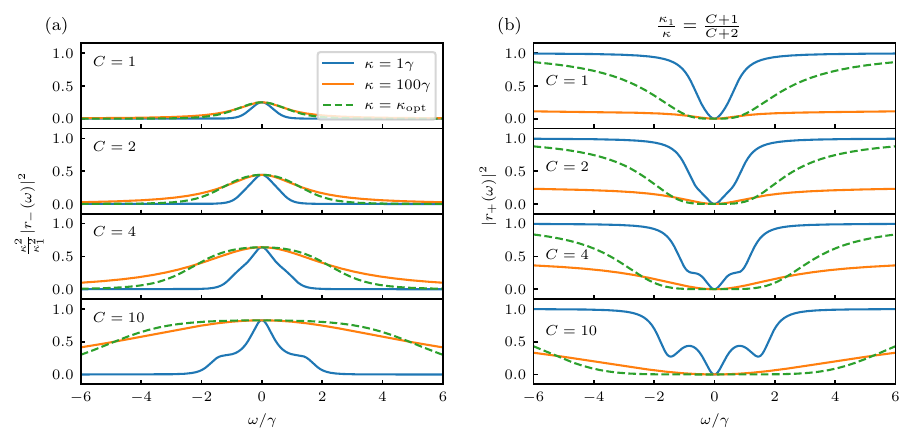}
    \caption{
    The modulus squared of the complete setup transfer functions $|r_\pm(\omega)|^2$ for parameters inspired by the NV center in diamond which links the pulse shape to the rate and fidelity, see Eqs.~\eqref{eq:Pdet} and \eqref{eq:fidelity_nearly_identical_narrow_pulse}.
    The rows display different cooperativities $C$ and the line styles stand for different cavity decay rates $\kappa$ (see legend).
    For the optimal cavity decay rate $\kappa_{\mathrm{opt}}$ which satisfies Eq.~\eqref{eq:opt_kappa} the maximum (minimum) of $|r_\pm(\omega)|^2$ becomes flatter which is beneficial for rate and fidelity (see also Fig.~\ref{fig:Pdet_Fa_NV}).
    (a) $|r_-(\omega)|^2 \kappa^2/\kappa_1^2$ 
    which is independent of $\kappa_1$.
    (b) symmetric transfer function $|r_+(\omega)|^2$ for the coupling ratio $\kappa_1/\kappa$ tuned to the phase encoding condition Eq.~\eqref{eq:phase_enc_cond}. This ensures that $|r_+(\omega)|^2$ has a root at $\omega = 0$.
    To use the upper port for heralding we want $r_+(\omega)$ to (approximately) vanish over the bandwidth of the photon pulse.}
    \label{fig:r_minus}
\end{figure*}

We begin with the model of the optical properties of an NV center, where we assume a three-level L-system consisting of two ground states $\ket{0}$ and $\ket{1}$ and an excited state $\ket{e_0}$ (see Fig.~\ref{fig:level_schemes}).
The transition between $\ket{0}$ and $\ket{e_0}$ can be excited using a cavity photon, whereas $\ket{1}$ is decoupled from the cavity.
The decoupling can be due to large detuning from all allowed transitions or polarization selection rules. 
Therefore, we obtain the reflection transfer functions for the initial states $\ket{0}$ and $\ket{1}$ from Eq.~\eqref{eq:refl_function} where for the decoupled state $\ket{1}$ the cooperativity $C_1 = 0$ (and $\delta_1,\gamma_1$ become irrelevant).
The complete setup transfer functions are given by,
\begin{align}
    r_-(\omega) =&\frac{\kappa_1}{\kappa}\frac{C}{(1-2i\frac{\omega - \delta}{\gamma})(1-2i\frac{\omega}{\kappa})^2 + C\left(1-2i\frac{\omega}{\kappa}\right)}\,, \label{eq:hat_r_minus} \\
    r_+(\omega) =& 1 - \frac{\kappa_1}{\kappa} \frac{2\left(1-2i\frac{\omega - \delta}{\gamma}\right)\left(1-2i\frac{\omega}{\kappa}\right) + C}{(1-2i\frac{\omega - \delta}{\gamma})(1-2i\frac{\omega}{\kappa})^2 + C\left(1-2i\frac{\omega}{\kappa}\right)}~ , \label{eq:hat_r_plus}
\end{align}
where we used $C_0 = C$, $\delta_0=\delta$, and $\gamma_0=\gamma$ for simplicity.
Maximizing $|r_{-}(\omega)|^2$ in terms of the detunings $\delta, \omega$, yields $\delta = \omega = 0$.
This is in agreement with the Purcell effect being strongest on resonance.
Thus we will focus on $\delta = 0$ in the following.
The maximum of the modulus squared of the transfer function found at $\delta=\omega=0$ is,
\begin{align}
    |r_-(0)|^2 = \frac{\kappa_1^2}{\kappa^2} \frac{C^2}{(C+1)^2}\,.
    \label{eq:ref_resonance}
\end{align}
This result is the square of the single photon emitter out-coupling efficiency found in \cite{gorshkov_photon_2007, santori_single_2010, dilley_single-photon_2012, mucke_generation_2013, morin_deterministic_2019, tissot_efficient_2024}
and only depends on the cooperativity $C$ and the cavity coupling ratio $\kappa_1/\kappa$. 
Thus, using over-coupled cavities with ideally $\kappa_1 = \kappa$ directly increases the entanglement generation success probability.
An increase in $C$ not only increases the reflection maximum $|r_-(0)|^2$, but also its width through the Purcell effect [see Fig.~\ref{fig:r_minus}(a)]. 
In particular, for $\kappa \gg (C+1)\gamma$ the width of the maximum is approximately given by the Purcell-enhanced decay rate $(C+1)\gamma$.

Alternatively, to maximize the entanglement generation rate [proportional to the detection probability in Eq.~\eqref{eq:Pdet}], we want $r_-(\omega)$ to be broad around the resonance frequency.
Therefore, we can choose an optimized $\kappa$ by minimizing $\left| \frac{\partial^2 |r_-|^2}{\partial \omega^2}(\omega = \delta = 0) \right|^2$
for fixed $\kappa_1/\kappa$.
We find that the optimal $\kappa$ satisfies the relation,
\begin{align}
    \frac{\kappa}{\gamma} = \frac{C^2 +2C +2}{C}\,. \label{eq:opt_kappa}
\end{align}
We show in Fig.~\ref{fig:r_minus}(a) that this choice indeed achieves broad modulus squared transfer functions in the frequency domain.
This in turn helps to maximize the overlap for a pulse (which has non-zero bandwidth) and thus helps to increase the detection rate, see Eq.~\eqref{eq:Pdet}.
The broadening enables the use of shorter photon pulses (which have a broader spectrum) and could make the protocol more robust against spectral diffusion of the defects. 

Phase encoding can be reached (for sufficiently narrow banded pulses) by tuning the cavity coupling ratio and cooperativity to fulfill the condition,
\begin{align}
    \frac{\kappa_1}{\kappa} = \frac{C + 1}{C + 2}\,. \label{eq:phase_enc_cond}
\end{align}
In this case, the symmetric reflection transfer function Eq.~\eqref{eq:hat_r_plus} has a root at $\omega = 0$ as shown in Fig.~\ref{fig:r_minus}(b).
Simultaneously satisfying Eq.~\eqref{eq:opt_kappa} and Eq.~\eqref{eq:phase_enc_cond} also makes $|r_+(0)|^2 = 0$ a flat minimum where $\left| \frac{\partial |r_+|^2}{\partial \omega^2} \right|^2(\omega=\delta=0) = 0$
which makes the fidelity $F_a$ upon a photon detection in channel $a$ less susceptible to the bandwidth of the photon, see Eqs.~\eqref{eq:fidelity_nearly_identical_narrow_pulse} and \eqref{eq:Fa_NV}.
If $C > \sqrt{2}$ can be reached, the additional detection at the upper channel enables higher success probabilities compared to single-sided cavities $\kappa_1 = \kappa$ used with detection only at the lower port $b$.
Therefore, if systems with $C \ge \kappa/\kappa_1 > \sqrt{2}$ can be produced, lowering the cooperativity by adjusting the placement of the defect or polarization of the incident photon to use phase encoding enhances the overall entanglement generation probability. 

However, so far NV centers integrated into cavities reached cooperativities $C \approx 1$ \cite{barclay_chip-based_2009, faraon_coupling_2012, li_coherent_2015, riedel_deterministic_2017},
corresponding to a parameter range where intensity encoding can be beneficial.
Hence, by using single-sided cavities and the single detector setup, a maximum success probability of  about $\eta/8$ could be reached.
Approaching this limit is possible by tuning $\kappa$ to satisfy Eq.~\eqref{eq:opt_kappa} and by employing active spectrum stabilization of the cavity coupled NV center to reduce the inhomogeneous broadening (usually $\gamma_{\mathrm{inhom}} \gg \gamma$) \cite{bassett_electrical_2011, bernien_heralded_2013} and its impact on entanglement generation. 
Additionally, reducing the laser power incident on the NV centers reduces the spectral diffusion rate drastically \cite{orphal-kobin_optically_2023} from which the entanglement protocol presented in this paper may benefit by relying on single photon reflections.

\subsubsection{Fidelity for finite bandwidth and differing nodes}\label{sec:fidelity_NV}

Differences between the two cavity-defect modules under investigation and the finite bandwidth of pulses result in a reduced fidelity of the generated entangled state as described in Sec.~\ref{sec:different_modules_general}.
Here we delve into more detail using the three-level system established in this section.
Without loss of generality, we assume that system $B$ deviates from the reference parameters according to
\begin{align}
    C^B &= C + \epsilon_C, \kappa^B = \kappa + \epsilon_\kappa,  \\
    \kappa_1^B &= \kappa_1 + \epsilon_{\kappa_1}, \gamma^B = \gamma + \epsilon_\gamma~.
\end{align}
Additionally, we consider small detunings $\delta_A, \delta_B$ of the optical transition from the cavity resonance for both systems independently where spectral diffusion of the transition frequency of both individual defects is a potential source for this deviation.
Using Eq.~\eqref{eq:fidelity_nearly_identical_narrow_pulse} and accounting for the phase encoding condition for $F_a$ [see Eq.~\eqref{eq:phase_enc_cond}], we calculate the fidelities to herald the correct entangled state upon detection on one of the ports $a$ and $b$,
\begin{widetext}
\begin{align}
     F_a = \,&
     1
- \frac{4 \sigma_{u}^{2}}{\left(C + 1\right)^{2}} \left(\frac{1}{\gamma} - \frac{C^{2} + 2 C + 2}{C \kappa}\right)^{2}
- \frac{3 (\delta_{A} - \delta_{B})^2 + 8 \delta_{A} \delta_{B}}{\gamma^{2} \left(C + 1\right)^{2}} 
- \frac{3 \epsilon_{C}^{2}}{4 C^{2} \left(C + 1\right)^{2}} \nonumber \\
& - \frac{\epsilon_{C} (C + 4)}{2 C^{2} \left(C + 1\right)} \left[\frac{\epsilon_{\kappa}}{\kappa} - \frac{\epsilon_{\kappa_1} \left(C + 2\right)}{\kappa \left(C + 1\right)}\right]
- \frac{3 C^{2} + 8 C + 8}{4 C^{2}} \left[\frac{\epsilon_{\kappa}}{\kappa} - \frac{\epsilon_{\kappa_1} \left(C + 2\right)}{\kappa \left(C + 1\right)}\right]^{2} ,
\label{eq:Fa_NV}\\
     F_b = \,&
     1 - \frac{3 \left(\delta_{A} - \delta_{B}\right)^{2}}{\gamma^{2} \left(C + 1\right)^{2}} - \frac{3 \epsilon_{C}^{2}}{4 C^{2} \left(C + 1\right)^{2}} - \frac{\epsilon_{C} \left(C + 4\right)}{2 C^{2} \left(C + 1\right)} \left(\frac{\epsilon_{\kappa}}{\kappa} - \frac{\epsilon_{\kappa_1}}{\kappa_1}\right) - \frac{3 C^{2} + 8 C + 8}{4 C^{2}} \left(\frac{\epsilon_{\kappa}}{\kappa} - \frac{\epsilon_{\kappa_1}}{\kappa_1}\right)^2, 
     \label{eq:Fb_NV}
\end{align}
\end{widetext}
where we neglected terms of higher than the combined second order in the bandwidth and the deviations between the systems from the ideal configuration.

For high cooperativities $C \gg 1$, all terms but the last are suppressed for the fidelity $F_b$ upon detection on port $b$.
Thus we conclude that differences in the mirror ratios of the cavities can negatively impact the fidelity even for high cooperativities $C$.
Additionally, despite low cooperativities not directly decreasing the fidelity of the generated state, systems with low cooperativities are more susceptible to deviations negatively affecting the fidelity of the generated state.
Especially the term $\propto (\delta_A - \delta_B)^2$ which contains the spectral diffusion of defects $A$ and $B$, is expected to have a big negative impact on the fidelity if spectral diffusion reaches the order of the Purcell enhanced linewidth $(C+1)\gamma$.

\begin{figure}
    \centering
    \includegraphics[width=8.3cm]{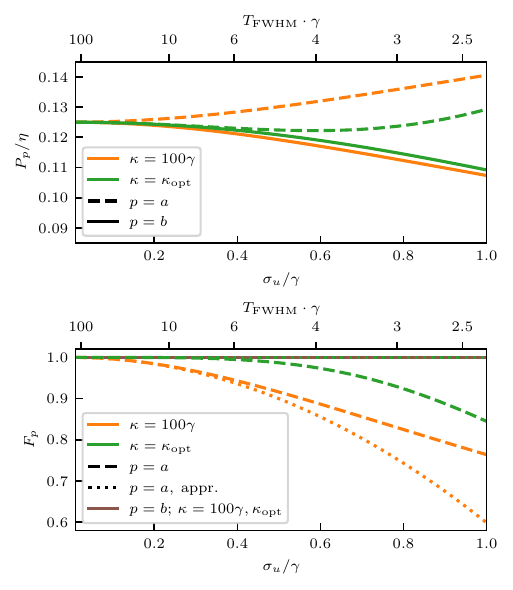}
    \caption{
    Detection probability $P_p$ divided by the setup transmittivity $\eta$ (a) and fidelity $F_p$ (b) of the generated entangled state as a function of the linewidth $\sigma_u$ (full-width at half-maximum duration $T_{\mathrm{FWHM}}$) of a Gaussian input photon for both channels $p=a,b$ (dashed and solid lines).
    We assume two identical nodes with cooperativity of $C=2$ and the phase encoding configuration [Eq.~\eqref{eq:phase_enc_cond}].
    The linestyle shows the choice of $\kappa$ (see legend), where $\kappa_{\mathrm{opt}}$ satisfies Eq.~\eqref{eq:opt_kappa}.
    The fidelity $F_b$ remains constant at $F_b=1$ because both nodes are identical while $P_b/\eta$ decreases for broader pulses.
    In contrast, $P_a/\eta$ increases for broader pulses while the fidelity $F_a$ decreases.
    By choosing $\kappa = \kappa_\mathrm{opt}$, the decrease in fidelity can be suppressed for small $\sigma_u$.
    The dotted lines show the approximation of the fidelity given in Eq.~\eqref{eq:Fa_NV} which stays at unity for $\kappa = \kappa_{\mathrm{opt}}$.}
    \label{fig:Pdet_Fa_NV}
\end{figure}

For detection at the other port the fidelity $F_a$ [assuming phase encoding condition in Eq.~\eqref{eq:phase_enc_cond}]
also most terms can be suppressed by large cooperativities.
In addition to the last term that is not suppressed even for large cooperativities,
this fidelity is in the combined second-order affected by the finite bandwidth $\sigma_u$ of the photon pulse.
We attribute the differences mainly to the additional requirement of 
$|r_+(\omega)|^2 = 0$ only being (identically) satisfied at the resonance $\omega = 0$, as displayed in Fig.~\ref{fig:r_minus}(b).
Therefore, the fidelity can also be decreased for identical $A$ and $B$ if the photon width $\sigma_u$ exceeds the width of the minimum of $|r_+(\omega)|^2$. 
We stress, however, that the term $\propto \sigma_u^2$ in the mixed second order vanishes for the optimal $\kappa/\gamma$ according to Eq.~\eqref{eq:opt_kappa}.
Therefore, we conclude that this optimal choice combined with the phase-encoding condition [Eq.~\eqref{eq:phase_enc_cond}] not only increases the entanglement-generation rate but also suppresses detrimental effects (up to the second mixed order) of a finite bandwidth of the photon pulse.
We display the resulting tradeoff between fidelity and rate in Fig.~\ref{fig:Pdet_Fa_NV}.

\subsection{Four Level System - SiV Center\label{sec:SiV}}

We now turn to the SiV center, which is effectively a four-level system (Fig.~\ref{fig:level_schemes}), and thus not well described by the three-level model discussed above. The ground state doublet $\ket{0}, \ket{1}$ couples  optically to the excited state doublet $\ket{e_0}, \ket{e_1}$.
By applying strain, differing Land\'{e} factors for the ground and excited states can be achieved which allows for a lifting of the degeneracy of the spin-conserving transitions by $\zeta$ using a magnetic field \cite{nguyen_integrated_2019}.

Here, we focus on the negatively charged SiV$^-$ center, which has been studied extensively during the last years \cite{meesala_strain_2018,nguyen_integrated_2019,metsch_initialization_2019,bhaskar_experimental_2020,bersin_telecom_2023}.
Compared to the NV center, the SiV center has lower ground-state coherence times but possesses favorable optical properties.
Its inversion symmetry leads to significantly lower spectral diffusion and higher ZPL emission probability, allowing for high cooperativities when coupled to a cavity \cite{nguyen_integrated_2019, bhaskar_experimental_2020}.
The system can be described by the model Hamiltonian \eqref{eq:system-Hamiltonian} introduced in Sec.~\ref{sec:CQED_model}.
Here we assume matching coupling strengths $g_k = g$ and excited-state decay rates $\gamma_k=\gamma$ ($k=0,1$) between the two pairs of states $\ket{k}, \ket{e_k}$ such that they share the same cooperativity $C=C_0=C_1$.
This is a good approximation because the mixing of orbital and spin wavefunction contributions due to strain is usually small.
Additionally, substituting $\delta_k = \delta - (-1)^k \zeta/2$,
the reflection transfer functions are given by Eq.~\eqref{eq:refl_function}.

\begin{figure*}
    \centering
    \includegraphics[width=17.8cm]{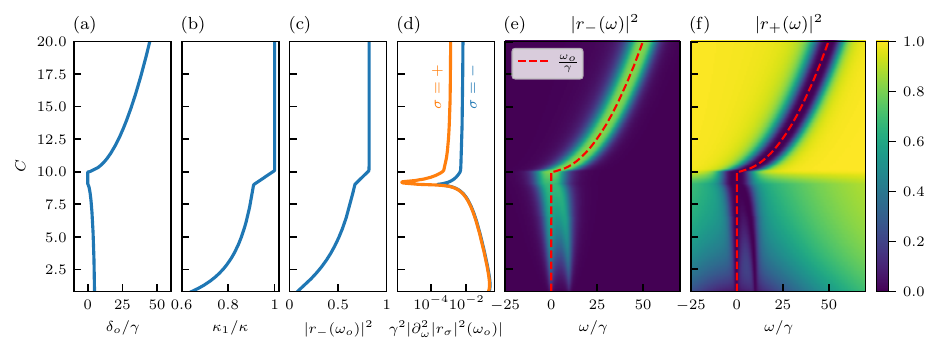}
    \caption{ \label{fig:four_level_C_zeta}
    Optimized detunings, out-coupling ratios, and resulting modulus-square transfer functions for the four-level model as functions of the cooperativity $C$.
    We plot the optimized detunings $\delta_o$ in panel (a) and $\omega_o$ as the red line in (e),(f).
    Furthermore, $\kappa_1$ is optimized to ensure the phase encoding condition, here via $|r_+(\omega_o)|^2 < 10^{-6}$ (panel (b)).
    In panel (c), we show the corresponding maximum modulus square of $|r_-(\omega_o)|^2$, and in (d) we plot the second derivative of $|r_\pm(\omega)|^2$ at $\omega_o$ to quantify the flatness.
    In (e) and (f), we display $|r_\pm(\omega)|^2$ as functions of the cooperativity and the pulse mode detuning $\omega$ for the optimized $\delta_o$. 
    For all plots, $\kappa = 100 \gamma$ was used. For $C>\sqrt{\zeta^2/\gamma^2 + 1} \approx 10$, the Purcell enhanced linewidths of both qubit states overlap at resonance which reduces $|r_-(\omega)|^2$. Therefore, it is beneficial to increase $\delta$, thus decreasing the Purcell broadening. This also has the effect of limiting $|r_-(\omega)|^2$: the entanglement generation success probability can not be increased any further by increasing $C$ beyond $\sqrt{\zeta^2/\gamma^2 + 1}$.}
\end{figure*}

In contrast to the three-level system described in the previous section, now for high cooperativities $C$ the two Purcell-broadened transitions overlap which limits $|r_-(\omega)|^2$. 
To counteract this effect, we numerically optimize $\delta$ to maximize the peak of $|r_-(\omega)|^2$ which is displayed in Fig.~\ref{fig:four_level_C_zeta}. 
The maximum reflection probability on resonance is already reached for a cooperativity $C = \sqrt{\zeta^2/\gamma^2 + 1}$, and for higher cooperativities it is not possible to increase the reflection probability any further.
Instead, for $C >  \sqrt{\zeta^2/\gamma^2 + 1}$ it is favorable to increase the detuning $\delta$ to reduce the overlap of the Purcell-broadened reflection peaks of the two different qubit states. This leaves the antisymmetric reflection probability and linewidth constant but introduces a Lamb shift of the defect transition frequency.

In summary, the entanglement protocol is also applicable to the silicon-vacancy defect in diamond, which is best described by a four-level model.
However, the performance of the protocol is now additionally limited by the transition frequency difference $\zeta$ of the two considered transitions. For a magnetic field of $B \approx 1\,$T, one can expect $\zeta \approx 1\,$GHz \cite{nguyen_integrated_2019} which translates to $\zeta/\gamma \approx 10$.
Therefore, the entanglement generation success probability is limited by $|r_-(\omega)|^2 \approx 0.82 \frac{\kappa_1^2}{\kappa^2}$ which is reached for $C \gtrsim 10$, see also Fig.~\ref{fig:four_level_C_zeta}.

\section{Comparison to Photon Emission Protocols}\label{sec:proto_comparison}

The gained insights about the general reflection-based single-photon entanglement generation protocol can now be used to compare the protocol to approaches using (conditional) photon emission of the nodes.
As we were able to derive analytic expressions for the three-level model (see Sec.~\ref{sec:three_level}) we focus on the insights that can be gained from those in this comparison.
Furthermore, we focus on the comparison of the rates, as we have shown in Sec.~\ref{sec:fidelity_NV} that the fidelity approaches unity for sufficiently broad pulses and identical nodes, while for differing nodes, we can trade a lower rate for a higher fidelity.

The protocol presented by Barrett and Kok \cite{barrett_efficient_2005} uses a similar setup as displayed in Fig.~\ref{fig:setup_reflection}. First, both defects $s=A,B$ are also initialized in $\ket{+}_{s}$. Then, an optical $\pi$ pulse is applied to both defects which coherently excites the transition $\ket{0}_s \rightarrow \ket{e_0}_s$. After a sufficiently long time, the atomic excitation is converted to a cavity photon which outcouples into the optical setup with the probability \cite{santori_single_2010, childress_fault-tolerant_2005},
\begin{align}
    P_{\rm em} = \frac{\kappa_1}{\kappa + \gamma} \frac{C}{C+1}\,,
\end{align}
where we assumed the resonance condition $\delta = 0$. The emitted photons can now be detected at the upper ($a$) or lower ($b$) detector.  Next, a $X$ operation is applied to both defects which coherently flips the spins. Subsequently, the optical excitation and detection steps are repeated. If one photon is detected both times, the defects are in an entangled Bell state: $\ket{\Psi^+}$ if both are detected at the same detector and $\ket{\Psi^-}$ if they are detected at different detectors. An advantage of this approach is that the phase of the interferometer setup only needs to remain stable on the timescale of one entanglement attempt which generally leads to increased fidelity and can enable this protocol on setups without active phase stabilization.
Due to the required two subsequent photon emissions, the success probability of this protocol is given by $\eta P_{\rm em}^2/2$ where $\eta$ is the photon loss probability when traveling two times through the setup (independent of the direction). For the case of $\gamma \ll \kappa$ and photon linewidths $\sigma_u \ll (C+1)\gamma$, this matches the success probability of the single detector reflection-based protocol presented in this paper. If the phase encoding and a second detector are used, the success probability is doubled $\eta P_{\rm em}^2$ and the success probability of the double-detection protocol is exceeded. 

Another protocol using conditional emission that only needs a single detection relies on the preparation of the defect qubits in,
\begin{align}
    \sin(\alpha) \ket{0}_s + \cos(\alpha) \ket{1}_s,
\end{align}
with $\alpha \ll 1$ instead of $\ket{+}_s$, and has a success probability $P = 2\sqrt{\eta}\sin(\alpha)^2 P_{\rm em}$ \cite{childress_fault-tolerant_2005, hermans_qubit_2022}.
If the reduction of the success probability is dominated by $P_{\rm em} \ll 1$ or $\eta \ll 1$, this protocol may be advantageous \cite{rozpedek_near-term_2019}.
However, the possible entanglement heralding for the defect state $\ket{0}_A \ket{0}_B$ introduces a trade-off between fidelity and success probability.
Therefore, to achieve high-fidelity entanglement, $\alpha$ has to be low which counteracts the initial success probability advantage.
The reflection-based protocol does not suffer from this effect, and we have shown approaches to optimize the cavity for a higher rate and robust fidelity.

\section{\label{sec:conclusion}Conclusions}

We developed a cavity quantum electrodynamics model to study spin-photon entanglement.
We first calculated an analytic solution in the single-excitation subspace by incorporating various decay channels into a (time-dependent) non-Hermitian Schr\"odinger equation.
Using the solution of the dynamics we accounted for the pulse nature of the propagating photons.
We then applied this solution to study remote entanglement generation protocols using single photon reflection at the nodes.

We showed that small cooperativities lead to larger susceptibility of the fidelity of the heralded entangled state to deviations between the remote nodes.
Additionally, our analysis shows that ``phase encoding'' can ideally lead to higher entanglement generation rates and herald entangled states upon photon detection in any of the two used photon channels.
However, the fidelity of this protocol is reduced when using short photon-pulse durations (broadband pulses).
Therefore, we proposed one approach where the nodes are tailored to suppress the susceptibility to the bandwidth of the pulse in the leading order.
Furthermore, we note that only using one port to herald the entanglement generation reduces the rate but ensures that the fidelity is unaffected when using short pulses. This even holds for slightly differing nodes up to mixed fourth order in the difference and bandwidth.
We discussed all of the above using the NV and SiV centers in diamond as examples.

All in all, the outlined model and approach can be used to tailor remote entanglement protocols to given implementations, accounting for the different CQED parameters and their variation between nodes.
Future work applying the insights in various applications would provide valuable validation, e.g., using the phase-encoding condition and minimizing the modulus squared of the second derivatives of the transfer functions as constraints when designing integrated and optimized cavities.
Finally, it would be interesting to try to extend our model to account for additional effects, e.g., cross-talk, shelving levels, and pure dephasing.

\begin{acknowledgments}
    FO and BT contributed equally to this work. BT and GB acknowledge funding from the German Federal Ministry of Education and Research (BMBF) under Grant Agreement No.~13N16212 (SPINNING).
\end{acknowledgments}

\appendix

\section{Solving the node and pulse mode dynamics}\label{app:solving_node_dynamics}
To find the dynamics of the ansatz wavefunction in Eq.~\eqref{eq:ansatz_single_excitation} according to the Hamiltonian \eqref{eq:kiilerlich_Hamiltonian} including the virtual cavities and the various decay channels [see Eq.~\eqref{eq:L_sys} with the first component adapted to include the virtual cavities in Eq.~\eqref{eq:jump-4lvl}],
we first use the temporal mode matching
\(\hat{L}_1(t) \ket{\Psi(t)} = 0\) to incorporate the cavity out-couplings into the relevant channel into the Hermitian dynamics by tuning the virtual cavities and thereby determining the relevant pulse shapes \cite{tissot_efficient_2024}.
The mode matching applied to the ansatz [Eq.~\eqref{eq:ansatz_single_excitation}] results in
\begin{align}
  \label{eq:mode-matching-v}
  & g_u^{*}(t) \alpha_u^k(t) + \sqrt{\kappa_1} \alpha_c^k(t) + \sum_{l=0,1} g_{v_l}^{*}(t) \alpha_{v_l}^k(t) = 0 , 
\end{align}
with \(k=0,1\) and which can be satisfied by choosing the orthogonal modes $v_k$.
With the mode-matching the non-Hermitian
time-dependent Schrödinger equation becomes \(i \frac{\partial }{\partial t} \ket{\Psi(t)} = H_{\mathrm{NH}}/\hbar \ket{\Psi(t)}\)
with the corresponding non-Hermitian Hamiltonian
$H_{\mathrm{NH}}/\hbar = H/\hbar - \frac{i}{2} (\kappa_2 c^{\dag} c + \sum_{l=0,1} \gamma_l \proj{e_l})$.
Combining this non-Hermitian Schr\"odinger equation with the mode-matching conditions Eq.~\eqref{eq:mode-matching-v} to exploit the cascaded nature of the virtual cavity model leads to the dynamics of the amplitudes of the state
\begin{align}
  \label{eq:SGMM-uk}
  & \dot{\alpha}_u^k(t) = - \frac{1}{2} | g_u(t) |^2 \alpha_u^k(t) , \\
  \label{eq:SGMM-ck}
  & \dot{\alpha}_c^k(t) = -i g_k \alpha_{e_k}(t) - \sqrt{\kappa_1} g_u^{*}(t) \alpha_u^k(t) - \frac{\kappa}{2} \alpha_c^k(t) , \\
  \label{eq:SGMM-ek}
  & \dot{\alpha}_{e_k}(t) = -i \delta_k \alpha_{e_k}(t) - i g_k \alpha_c^k(t) - \frac{\gamma_k}{2} \alpha_{e_k}(t) , \\
  \label{eq:SGMM-v0}
  & \dot{\alpha}_{v_0}^k(t) = g_{v_0}(t) g_{v_1}^{*}(t) \alpha_{v_1}^k(t) + \frac{1}{2} |g_{v_0}(t)|^2 \alpha_{v_0}^k(t) , \\
  \label{eq:SGMM-v1}
  & \dot{\alpha}_{v_1}^k(t) = \frac{1}{2} |g_{v_1}(t)|^2 \alpha_{v_1}^k(t) , 
\end{align}
where $\kappa = \kappa_1 + \kappa_2$.

Note that we are interested in the initial state where the excitation is completely in the input pulse and the ground state is prepared in a superposition state, i.e., $\ket{\Psi(0)} = \sum_{k=0,1} \alpha_u^k(0) \ket{k}_s$.
We note that if one considers a pulse of infinite duration such as a Gaussian, the appropriate (formal) initial time is $-\infty$ rather than $0$. However, this does not influence the following argument apart from the change of reference time and necessary accommodation of integration bounds.

\subsection{Solving the input mode dynamics}\label{app:input_mode_dynamics}

To solve the dynamics for this initial condition, we start by formally integrating Eq.~\eqref{eq:SGMM-uk}
\begin{align}
\label{eq:auk-formal}
    \alpha_u^k(t) =\,& \alpha_u^k(0) \exp \left[ - \frac{1}{2} \int_0^{t} |g_u(t')|^2 dt' \right]
    = \alpha_u^k(0) \frac{u(t)}{g_u^{*}(t)} ,
\end{align}
where in the last equality we used the relation between the time-dependent coupling strength and the temporal shape of the input mode \cite{kiilerich_input-output_2019,kiilerich_quantum_2020}.

\subsection{Solving the node system dynamics}\label{app:system_dynamics}

Inserting Eq.~\eqref{eq:auk-formal} into Eqs.~\eqref{eq:SGMM-ck} and \eqref{eq:SGMM-ek} results in a differential equation that can be solved by a Fourier transform.
The transformed equations are
\begin{align}
  \label{eq:FT-ck}
  - i \omega \tilde{\alpha}_c^k(\omega) =\,& - \frac{\kappa}{2} \tilde{\alpha}_c^k(\omega) - i g_k \tilde{\alpha}_{e_k}(\omega) 
  - \sqrt{\kappa_1} \alpha_u^k(0) \tilde{u}(\omega) , \\
  \label{eq:FT-ek}
  - i \omega \tilde{\alpha}_{e_k}(\omega) =\,& - (\frac{\gamma_k}{2} + i \delta_k) \tilde{\alpha}_{e_k}(\omega) - i g_k \tilde{\alpha}_c^k(\omega) ,
\end{align}
which can be algebraically solved yielding
\begin{align}
    \label{eq:sol-ek}
    \tilde{\alpha}_{e_k}(\omega) =\,& \frac{g_k}{(\omega - \delta_k) + i \gamma_k/2} \tilde{\alpha}_c^k(\omega) , \\
    \label{eq:sol-ck}
    \tilde{\alpha}_c^k(\omega) =\,& \frac{\sqrt{\kappa_1}}{i \omega - \frac{\kappa}{2} - \frac{i g_k^2}{(\omega - \delta_k) + i \gamma_k/2}} \alpha_u^k(0) \tilde{u}(\omega) .
\end{align}

\subsection{Solving the output mode dynamics}\label{app:output_mode_dynamics}

It remains to determine the (orthogonal) output modes $v_0,v_1$ and their occupation.
To determine those we use the relation between the time-dependent coupling constants and the modes, see Eq.~\eqref{eq:g_out}.
We further know that the coupling constants that perfectly absorb the normalized modes $v$ satisfy the relation \cite{kiilerich_input-output_2019,kiilerich_quantum_2020}
\begin{align}
    \label{eq:cavity_coupling_relation}
    v(t) = -g_v^*(t) \exp[ - \frac{1}{2} \int_{t}^{\infty} |g_v(t')|^2 dt' ],
\end{align}
 which we use for $(v,g_v)=(v_0,g_{v_0}), (v_1', g_{v_1})$.
%
To apply these relations we first (formally) solve the scattering problem to account for
the reflection of mode \(v_1(t)\) on the cavity with time-dependent coupling \(g_{v_0}(t)\) that perfectly absorbs the mode \(v_0(t)\) [perpendicular to \(v_1(t)\)].
The scattering leads to a classical amplitude within the virtual cavity of the mode \(v_0(t)\) described by the differential equation \cite{kiilerich_input-output_2019,kiilerich_quantum_2020}
\begin{align*}
  \dot{\alpha}_{v_1}' = - g_{v_0}(t) v_1(t) - \frac{|g_{v_0}|^2}{2} \alpha_{v_1}',
\end{align*}
with initial value \(\alpha_{v_1}'(0) = 0\).
Variation of constants leads to \(\alpha_{v_1}'(t) = d(t) \exp \left[ \frac{1}{2} \int_t^{\infty} |g_{v_0}(t')|^2 dt' \right]\)
with
\(d(t) = - \int_0^t dt'' g_{v_0}(t'') \exp \left[ -\frac{1}{2} \int_{t''}^{\infty} |g_{v_0}(t')|^2 dt' \right] v_1(t'') \).
We can rewrite this to
\(d(t) = \int_0^t dt' v_0^{*}(t')v_1(t') \)
and thus
\begin{align}
    \label{eq:scatt_formal_ampl}
\alpha_{v_1}'(t) = \exp \left[ \frac{1}{2} \int_t^{\infty} |g_{v_0}(t')|^2 dt' \right] \int_0^t v_0^{*}(t')v_1(t') dt' .
\end{align}
Inserting Eq.~\eqref{eq:scatt_formal_ampl} into the expression for the reflected mode \cite{kiilerich_input-output_2019,kiilerich_quantum_2020} leads to
\begin{align}
  v_1'(t)
  =\,& v_1(t) + g_{v_0}^{*}(t) \alpha_{v_1}'(t) \notag \\
  \label{eq:v1'}
=\,& v_1(t) - \frac{v_0(t) \int_0^t v_0^{*}(t')v_1(t') dt'}{\int_0^t |v_0(t')|^2 dt'} .
\end{align}

We combine Eqs.~\eqref{eq:cavity_coupling_relation} and \eqref{eq:v1'} with the formal solution of Eq.~\eqref{eq:SGMM-v1} and find
\begin{align}
    \label{eq:av1k-formal-app}
\alpha_{v_1}^k(t) =\,& \alpha_{v_1}^k(T) \exp[ - \frac{1}{2} \int_{t}^{\infty} |g_{v_1}(t')|^2 dt' ] \notag \\
=\,& - \alpha_{v_1}^k(T) \frac{v_1'(t)}{g_{v_1}^{*}(t)}
. 
\end{align}
Note that this equation precisely relates a single mode $v_1'(t)$ with a time-independent proportionality constant $\alpha_{v_1}^k(T)$ to $g_{v_1}^{*}(t) \alpha_{v_1}^k(t)$ which is a term relevant in the mode-matching condition \eqref{eq:mode-matching-v}.
Here we write $T$ for the time when the input pulse is completely reflected from the node,
which can be $T\rightarrow \infty$, e.g., for a Gaussian pulse.

Finally, we solve Eq.~\eqref{eq:SGMM-v0} by using variation of constants.
To this end we rewrite the remaining equation using $\alpha_{v_0}^k(t) = [c(t) + \alpha_{v_0}^k(T)] \exp [ - \frac{1}{2} \int_t^{\infty} |g_{v_0}(t')|^2 dt' ]$, 
such that the time-dependent constant is determined by
\begin{align}
  c(t) = & \int_0^t g_{v_0}(t'') g_{v_1}^{*}(t'') \alpha_{v_1}^k(t'') \notag \\
           & \times \exp \left[ \frac{1}{2} \int_{t''}^{\infty} |g_{v_0}(t')|^2 dt' \right] dt''  \label{eq:separation-ansatz} \\
 =& \: \alpha_{v_1}^k(T) \frac{v_1(t) - v_1'(t)}{v_0(t)} , \notag
\end{align}
where we again used Eqs.~\eqref{eq:cavity_coupling_relation} and \eqref{eq:g_out} and in the final step applied partial integration and Eq.~\eqref{eq:v1'}.
Inserting this back into the ansatz yields the relation
\begin{align}
    \alpha_{v_0}^k(t) 
    = [\alpha_{v_1}^k(T) \frac{v_1(t) - v_1'(t)}{v_0(t)} + \alpha_{v_0}^k(T)] \frac{-v_0(t)}{g_{v_0}^{*}(t)} .
    \label{eq:av0k-formal-app}
\end{align}

Summarizing, we can express the amplitudes in terms of the mode shape and time-dependent coupling strength by the relations
\begin{align}
    \label{eq:av1k-formal}
    \alpha_{v_1}^k(t) =\,& \alpha_{v_1}^k(T) \frac{v_1'(t)}{g_{v_1}^{*}(t)} , \\
    \label{eq:av0k-formal}
    \alpha_{v_0}^k(t) =\,& \alpha_{v_1}^k(T) \frac{v_1'(t) - v_1(t)}{g_{v_0}^{*}(t)} - \alpha^k_{v_0}(T) \frac{v_0(t)}{g_{v_0}^{*}(t)} .
\end{align}
Substituting Eqs.~\eqref{eq:av1k-formal} and \eqref{eq:av0k-formal} into the mode-matching condition Eq.~\eqref{eq:mode-matching-v} followed by a Fourier transform results in Eq.~\eqref{eq:MM-FT-final} of the main text.

Note that an analogous approach can be followed to capture the output of the second mirror $\kappa_2$ (or the excited state decays) to incorporate them in the cascaded Hermitian dynamics and calculate their amplitude and temporal mode, these are, however, not of interest for this work, and therefore not accounted for apart from the loss they induce.

\section{Interference for two differing nodes}\label{app:interfered_state} \label{app:diff_modules}

Considering the setup in Fig.~\ref{fig:setup_one_direction} we are interested in the interference of two systems initially prepared in $\ket{+}_s = \frac{1}{\sqrt{2}} (\ket{0}_s + \ket{1}_s)$ with $s=A,B$.
The combined state after the input photon went through the beamsplitter is given by
\begin{align}
    \label{eq:bs1}
    \frac{1}{2} \sum_{k=0,1} ( \ket{k,u}_A \ket{+}_B + \ket{+}_A \ket{k,u}_B) ,
\end{align}
where we omit to write the ket's for the vacuum state of the optical channels for brevity.
To model the reflection we use the solution given in Eq.~\eqref{eq:final_state}, where we denote the amplitudes with $\alpha$ ($\beta$) and modes with $v$ ($w$) for system $A$ ($B$) to be able to account for differences between the systems.
Furthermore, we absorb the normalization of the total wavefunction in the prefactor such that the initial state for the nodes is respectively $\chi_u^k(0)=1/\sqrt{2}$, for $\chi=\alpha,\beta$.
The total non-normalized state after the interaction between system and pulse is given by
\begin{align}
    \label{eq:refl}
    \frac{1}{\sqrt{2}} \sum_{k,l=0,1} ( \alpha_{v_l}^k(T) \ket{k,v_l}_A \ket{+}_B + \beta_{w_l}^k(T) \ket{+}_A \ket{k,w_l}_B) ,
\end{align}
where the normalization corresponds to the probability of having a photon in one of the relevant channels.
After the pulses went through the second beamsplitter the state becomes
\begin{align}
    \label{eq:bs2}
    \frac{1}{\sqrt{8}} \sum_{k,k',l=0,1} & \ket{k}_A \ket{k'}_B
    \Big[ \alpha_{v_l}^k(T) ( \ket{v_l}_a + \ket{v_l}_b) \notag \\
    &+ \beta_{w_l}^{k'}(T) (\ket{w_l}_a - \ket{w_l}_b) \Big] .
\end{align}
As we are interested in the generation of entanglement a natural basis choice for the matter parts of the wavefunction is spanned by the Bell states
\begin{align}
    \ket{\Phi^\pm} = (\ket{0}_A\ket{0}_B \pm \ket{1}_A\ket{1}_B)/\sqrt{2} \,, \label{eq:bell_phi_app} \\
    \ket{\Psi^\pm} = (\ket{0}_A\ket{1}_B \pm \ket{1}_A\ket{0}_B)/\sqrt{2}\, \label{eq:bell_psi_app} .
\end{align}
Transforming Eq.~\eqref{eq:bs2} into this basis results in
\begin{widetext}
\begin{align}
    \ket{f} = \frac{1}{\sqrt{8}} \Big\{ 
      (\ket{\Phi^+} + \ket{\Psi^+}) & \sum_{l=0,1} \left[ \alpha_{v_l}^+ ( \ket{v_l}_a + \ket{v_l}_b ) + \beta_{w_l}^+ ( \ket{w_l}_a - \ket{w_l}_b ) \right] \notag \\
      +\ket{\Phi^-} \qquad & \sum_{l=0,1} \left[ \alpha_{v_l}^- ( \ket{v_l}_a + \ket{v_l}_b ) + \beta_{w_l}^- ( \ket{w_l}_a - \ket{w_l}_b ) \right] \notag \\
      +\ket{\Psi^-} \qquad & \sum_{l=0,1} \left[ \alpha_{v_l}^- ( \ket{v_l}_a + \ket{v_l}_b ) + \beta_{w_l}^- ( -\ket{w_l}_a + \ket{w_l}_b ) \right] 
    \Big\} ,
    \label{eq:pre_measurement_state_general}
\end{align}
\end{widetext}
where we defined $\sqrt{2} \chi_{l}^{\pm} = \chi_{l}^0(T) \pm \chi_{l}^1(T)$ for $\chi_l^k = \alpha_{v_l}^k,\beta_{w_l}^k$.
The amplitudes are determined by Eq.~\eqref{eq:MM-FT-final}, which we can rewrite to find
\begin{align}
    \label{eq:MM-FT-final-pm-app}
    \sum_{l=0,1} \alpha_{v_l}^{\pm} v_l(\omega) = r_{\pm}^A(\omega) \tilde{u}(\omega),
\end{align}
for node $A$ and analogous for node $B$.
In the above equation, we introduced
\begin{align}
    r_{\pm}^s(\omega) = \frac{r_0^s(\omega) \pm r_1^s(\omega)}{2} ,
\end{align}
where the superscript $s=A,B$ denotes that the parameters according to a certain node need to be chosen.

Eq.~\eqref{eq:pre_measurement_state_general} shows that we can in general not discriminate between $\ket{\Phi^+}$ and $\ket{\Psi^-}$ such that these states are not suited for entanglement generation within this setup.
Therefore it remains to tailor the systems $A,B$ such that the remaining part of the state can be heralded into an entangled state.
We are therefore most interested in the mode amplitude and temporal modes acting upon this state $\propto \chi_l^-$.
We remind the reader that in general the modes $v_l$ and $w_l$ are orthonormal between themselves but not each other.
However, if the systems are identical we can choose $w_l = v_l$ and $\alpha_{v_l}^{\pm} = \beta_{w_l}^{\pm}$, see Eq.~\eqref{eq:setup_state_final}.

Furthermore, we can insert the definition of the temporal modes in terms of the channel creation operators $b_p(\omega)$ ($p=a,b$), see Eq.~\eqref{eq:pulse_state}, and with this write the final state in terms of the input pulse shape and transfer functions of the nodes
\begin{widetext}
\begin{align}
    \ket{f} = \frac{1}{\sqrt{2}} \int_{-\infty}^{\infty} d\omega \Big\{ 
      (\ket{\Phi^+} + \ket{\Psi^+}) & \left[ r_+^+(\omega) \tilde{u}(\omega) b_a^{\dag}(\omega) + r_+^-(\omega) \tilde{u}(\omega) b_b^{\dag}(\omega) \right] \notag \\
      \ket{\Phi^-} \qquad & \left[ r_-^+(\omega) \tilde{u}(\omega) b_a^{\dag}(\omega) + r_-^-(\omega) \tilde{u}(\omega) b_b^{\dag}(\omega) \right] \notag \\
      \ket{\Psi^-} \qquad & \left[ r_-^-(\omega) \tilde{u}(\omega) b_a^{\dag}(\omega) + r_-^+(\omega) \tilde{u}(\omega) b_b^{\dag}(\omega) \right]
    \Big\} \ket{0}_a \ket{0}_b ,
    \label{eq:pre_measurement_state_general_ALT_NOTATION}
\end{align}
\end{widetext}
with $2 r_{\sigma}^{\pm}(\omega) = {r_{\sigma}^A(\omega) \pm r_{\sigma}^B(\omega)}$ for $\sigma=\pm$.
Note that if the systems $A,B$ are identical $r_{\pm}^-(\omega) = 0$.

\section{Photon Detection}\label{app:photon_detection}

In this section we use that within our model the final state before detection is given in terms of
\begin{align}
    \hat\rho_f = \proj{f} + \hat\rho_0 \prod_{p=a,b} \proj{0}_p ,
\end{align}
where our model enables us to calculate $\ket{f}$, Eq.~\eqref{eq:pre_measurement_state_general_ALT_NOTATION},  using a non-Hermitian Hamiltonian and the density matrix of the zero excitation subspace $\hat\rho_0$ becomes irrelevant for the state upon detection of a single photon.
The differential probability to measure a photon at time $t$ in channel $p=a,b$ is given by
\begin{align}
    \mathrm{d}P_p(t) = \braket{f|\hat{b}_p^\dagger(t) \hat b_p(t)|f} \mathrm{d}t ,
\end{align}
to evaluate this expression we use the field formulation of the photonic modes [Eq. \eqref{eq:pulse_state}]
and the bosonic commutation relation $[\hat b_p(t), \hat b^\dagger_p(t')] = \delta(t-t')$.
In the case of detection of the photon at time $t$, the qubits $A, B$ are projected into
\begin{align}
    &\ket{f_{\mathrm{det}}^p(t)} = \sqrt{\frac{\mathrm{d}t }{\mathrm{d}P_p(t)}}\,\hat{b}_p(t)\ket{f} .
\end{align}
Now, we assume that the exact time of the detection of the photon is lost and the detector integrates over the complete photon pulse duration.
Therefore, the complete detection probability is given by
\begin{align}
\label{app:eq:detection_probability}
    P_p = \int_{-\infty}^{\infty} \braket{f|\hat{b}_p^\dagger(t) \hat b_p(t)|f} \mathrm{d}t .
\end{align}
We note that extending this integral from $-\infty$ to $\infty$ does not necessitate an infinite measurement time, but is here used for convenience and can be justified by a measurement time that is longer then the relevant pulse durations. 
However, this equation can be readily restricted to a measurement time shorter than the pulse duration.
The resulting final combined state of the nodes can be expressed using the density operator
\begin{align}
\label{app:eq:detection_state}
    \hat \rho_{\mathrm{det}}^p &= \frac{1}{P_p} \int_{-\infty}^{\infty}  
    dt \, \hat{b}_p(t) \ket{f} \bra{f} \hat{b}_p^{\dag}(t) .
\end{align}
Note that if one incorporated dephasing, the dephased density matrix would still hold a single excitation which would complicate the expressions for the final state and detection probability.

\section{Detection probability and fidelity of the generated state}\label{app:diff_modules_fidelity}
We can combine Eq.~\eqref{eq:pre_measurement_state_general_ALT_NOTATION} with Eq.~\eqref{app:eq:detection_probability}
to calculate the probability of a photon detection event at one of the ports $p=a,b$
\begin{align}
    P_p
    & =
    \frac{1}{2} \int_{-\infty}^{\infty} d\omega\, R_p(\omega) |\tilde{u}(\omega)|^2 , 
\end{align}
with
\begin{align}
\label{eq:pc}
    \left.\begin{aligned}
    R_a(\omega) \\
    R_b(\omega)
    \end{aligned}\right\} = |r_-^+(\omega)|^2 + |r_-^-(\omega)|^2 + 2 |r_+^{\pm}(\omega)|^2
\end{align}
and where we used 
\begin{align}
     b_p(t) & \int_{-\infty}^{\infty} d\omega\, r(\omega)\tilde{u}(\omega) b_p^{\dag}(\omega) \ket{0}_p \notag \\
    & = \frac{1}{\sqrt{2 \pi}} \int_{-\infty}^{\infty} d\omega \, e^{i \omega t} r(\omega)\tilde{u}(\omega) \ket{0}_p ,
\end{align}
and $\delta(\omega) = \frac{1}{2 \pi} \int_{-\infty}^{\infty} dt\, e^{i \omega t}$.
Additionally, we use Eq.~\eqref{app:eq:detection_state} to calculate the two relevant fidelities
\begin{align}
    F_p= \frac{\int_{-\infty}^{\infty} d\omega \, |r_-^+(\omega)|^2 |\tilde{u}(\omega)|^2}{2P_p}, \label{eq:fidelity_ab}
\end{align}
for $p=a,b$
to prepare $\ket{\Psi^-}$ ($\ket{\Phi^-}$) upon a detection event on the lower (upper) channel.
Note that we can simplify these expressions because we assumed that the measurement time exceeds the pulse duration.

Now, we will consider a Gaussian shaped incoming photon pulse which is centered detuned from the cavity resonance by $\Delta$,
\begin{align}
    \tilde u(\omega) = (\pi \sigma_u^2)^{-1/4} \exp \left(-\frac{(\omega - \Delta)^2}{2 \sigma_u^2} \right) .
\end{align}
We assume that the linewidth $\sigma_u$ is sufficiently narrow such that $|r_{\sigma}^{\pm}(\omega)|^2$ stays approximately constant in the integrand of Eq.~\eqref{eq:fidelity_ab}.
We thus Taylor expand $|r_{\sigma}^{\pm}(\omega)|^2$ in this equation around the detuning $\Delta$, which yields
\begin{align}
    &\int_{-\infty}^{\infty} |r_{\sigma}^\pm(\omega)|^2 |\tilde{u}(\omega)|^2 \mathrm{d}\omega 
    &\approx |r_{\sigma}^\pm(\Delta)|^2 + \frac{\sigma_u^2}{4} \frac{\partial^2 |r_{\sigma}^\pm|^2}{\partial \omega^2} (\Delta)\,,
\end{align}
for $\sigma=\pm$.
Here the first order contribution vanishes because it leads to an odd function in $\omega$ that we integrate over resulting in $0$, the same holds for the third order, such that the next relevant contribution would be $\mathcal{O}\left[\sigma_u^4 \frac{\partial^4 |r_{\sigma}^\pm|^2}{\partial \omega^4} (\Delta)\right]$ also quantifying the validity of the approximation.

Inserting this approximation into the fidelity of the state after detection of a photon at channel $p=a,b$ [Eq.~\ref{eq:fidelity_ab}],
we find
\begin{align}
    \label{eq:fidelity_narrow_pulse_ab}
    F_p
    \approx \frac{|r_-^+|^2 + \frac{\sigma_u^2}{4} \frac{\partial^2 |r_-^+|^2}{\partial \omega^2}}{R_p} 
    - \frac{|r_-^+|^2}{R_p^2} \frac{\sigma_u^2}{4} \frac{\partial^2 R_p}{\partial \omega^2} ,
\end{align}
with $R_p$ according to Eq.~\eqref{eq:pc}.
For brevity of the notation we omit the explicit evaluation at $\omega=\Delta$ in this expression and until the end of this Appendix.

To gain further insight about the fidelity we now
consider two nearly identical modules (i.e., $|r_{\pm}^-| \ll |r_{\pm}^+|$) with sufficient conditional reflectivity.
For the fidelity corresponding to the measurement on port $a$
we additionally assume that the phase encoding condition is (nearly) satisfied, i.e., $|r_+^+| \ll |r_-^+|$.
With these assumptions, we Taylor expand the denominators to the first combined order in $|r_{+}^+|^2, |r_{\pm}^-|^2$ and find
Eq.~\eqref{eq:fidelity_nearly_identical_narrow_pulse} of the main text.
We stress that this corresponds to the combined second order in the variances of the parameters describing the nodes.

\section{Optical Qubit Readout\label{app:readout}}

In the following, we briefly sketch two approaches for optical readout of the qubit using the existing CQED system of the nodes.
To use our previous analysis of the system we here assume single photon pulses are used to probe the qubit states, but we note that other states, e.g., weak coherent states, can be used as well.
The first approach is compatible with the intensity encoding and does not rely on additional optics.
The other approaches allow for more flexibility in the encoding of the entanglement generation but rely on additional components for readout at the nodes.

\subsection{Reflection Intensity}

Other than for the optimization of the entanglement generation probability which favors high differences in reflection amplitudes $|r_0(\omega) - r_1(\omega)|$, for the state detection high differences in reflection intensity are favorable $||r_0(\omega)|^2 - |r_1(\omega)|^2|$ (matching with the intensity encoding).
If the photon is only reflected for one of the qubit states of the defect, a quantum non-demolition measurement of the state is possible \cite{volz_measurement_2011, nemoto_photonic_2014}. This means, that by detecting a reflected photon, the state of the defect is fully determined. Therefore a single photon measurement sequence could be applied for a few cycles while still keeping the probability of induced qubit spin flips low.
For simplicity, we consider the three-level model in the following, i.e. $C_0=C$, $\delta_0=\delta$, $\gamma_0=\gamma$, and $C_1=0$.
For the intensity encoding, we consider the two cases below, where either the reflection of the coupled ($\ket{0}$) or uncoupled ($\ket{1}$) state is suppressed.

The first case is to suppress the reflection probability if the defect is in its uncoupled $\ket{1}$ state.
The only root of the transfer function $r_1(\omega_0) = 0$ is found for a critically coupled cavity with $\kappa_1 = \kappa/2$ at $\omega_0 = 0$.
Then, the reflection transfer function has a Lorentzian minimum of the width $\kappa$, and incoming photons of a linewidth $\sigma_u \ll \kappa$ are fully transmitted.
For a photon with $\sigma_u \ll \gamma$, the reflection probability of the coupled $\ket{0}$ state is given by
\begin{align}
	|r_0(0)|^2 = \frac{C^2}{(C + 1)^2 + \frac{4\delta^2}{\gamma^2}}\,.
\end{align}
Therefore, for a high reflection probability difference between the qubit states, a cooperativity $C \gg 1$ and low detuning $\delta \ll \gamma$ is favorable.
If the defect transition is on resonance with the cavity i.e. $\delta = 0$, the reflection transfer function is a Lorentzian
\begin{align}
	|r_0(\omega)|^2 = \frac{C^2}{(C + 1)^2} \frac{1}{1 + \frac{4\omega^2}{\gamma^2(C + 1)^2}}\,,
\end{align}
with the Purcell broadened linewidth $\gamma(C + 1)$.
This makes this critically coupled cavity a favorable system for state detection if high cooperativities $C \gg 1$ are reached.
The incoming photon linewidth is limited to $\sigma_u \ll \gamma(C+1), \kappa$.

The second case is to configure the system such that for the coupled $\ket{0}$ state the reflection probability vanishes. The two possible roots $r_0(\omega_{1,2}) = 0$ of the reflection transfer function, are given by
\begin{align}
	\omega_{1,2} &= \pm \sqrt{\tilde C  - 1} \frac{\tilde \kappa}{2}\,,
\end{align}
for the respective detuning
\begin{align}
    \delta_{1,2} &= \pm \sqrt{\tilde C - 1} \frac{\tilde \kappa - \gamma}{2}\,, \label{eq:r0_0_delta}    
\end{align}
where $\tilde\kappa = 2\kappa_1 - \kappa >0$ and $\tilde C = \frac{4g^2}{\tilde \kappa \gamma} \ge 1$.
The corresponding reflection probability for the $\ket{1}$ state for photons with $\sigma_u \ll \gamma$ is given by
\begin{align}
	|r_1(\omega_{1,2})|^2 = \frac{C}{\frac{\kappa}{\tilde{\kappa}} - \frac{\tilde{\kappa}}{\kappa} + C}\,,
\end{align}
Hence, the maximum reflection contrast can be reached for an over-coupled cavity with $\kappa_1 \approx \kappa$, i.e., $\tilde \kappa \approx \kappa$ for a comparatively low $C = 1$. 
But for an over-coupled cavity, this approach relies on the spontaneous decay process to dissipate the incoming photon, thus the probe photons linewidth $\sigma_u$ must be less than $\gamma$ to assure vanishing reflection probability for $\ket{0}$ even for high $C \gg 1$.

\subsection{Reflection Phase}

\begin{figure}
    \centering
    \includegraphics[width=8cm]{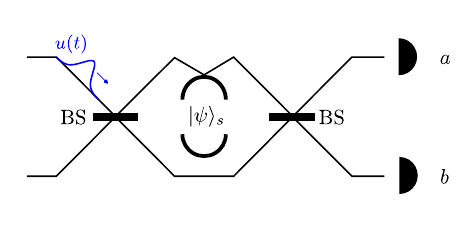}
    \caption{
    Interferometer setup to implement a readout of the qubit state $\ket{\psi}_s$ in the $z$-basis using a single photon pulse $u(t)$.
    The pulse is split and recombined using beam splitters (BS) and then detected in one of the output ports $a$ or $b$.
    If the CQED system is tuned to the phase encoding this setup enables a $z$-basis readout.
    \label{fig:phase_detection}
    }
\end{figure}

Using a Mach-Zehnder interferometer setup as displayed in Fig.~\ref{fig:phase_detection}, the qubit state $\ket{k}_s$ ($k=0,1$) of qubit $s=A,B$ can be detected using a single photon pulse $\ket{u}_a$ as input to the interferometer upper channel.
After the BS and the interaction with the qubit system the combined state of the system is given by
\begin{align}
  \ket{\Psi(T)} = \,& \frac{1}{\sqrt{2}} \left[ \sum_{k,l=0,1} \alpha_{v_l}^k(T) \ket{k}_s \ket{v_l}_a \right] \ket{0}_b \notag \\
  & + \frac{1}{\sqrt{2}} \left[ \sum_{k=0,1} \alpha_{u}^k(0) \ket{k}_s \ket{0}_a \right] \ket{u}_b , 
\end{align}
where we used Eq.~\eqref{eq:final_state} 
and note that Eq.~\eqref{eq:MM-FT-final} links the unknown initial qubit amplitudes $\alpha_u^k(0)$ to $\alpha_{v_l}^k(T)$ used in this equation.

To achieve good contrast long pulses should be used such that the transfer function can be assumed to be constant over the pulse and $v_0(t) = u(t)$, with that we find
\begin{align}
  \ket{\Psi(T)} \approx \,& \frac{1}{\sqrt{2}} \left[ \sum_{k=0,1} \alpha_{v_0}^k(T) \ket{k}_s \ket{u}_a \right] \ket{0}_b  \notag\\
  & + \frac{1}{\sqrt{2}} \left[ \sum_{k=0,1} \alpha_{u}^k(0) \ket{k}_s \ket{0}_a \right] \ket{u}_b , 
\end{align}
after passing the second BS this changes to
\begin{align}
  \ket{\Psi(T)} \approx \frac{1}{2} \sum_{k=0,1} \ket{k}_s & \Big\{ [\alpha_{v_0}^k(T) + \alpha_{u}^k(0)] \ket{u}_a \notag \\
  & + [\alpha_{v_0}^k(T) - \alpha_{u}^k(0)] \ket{u}_b \Big\}
\end{align}
where we see that if the transfer function is optimized such that 
$\alpha_{v_0}^k(T) \approx (-1)^k \alpha_{u}^k(0)$
which is compatible with phase encoding, we find
\begin{align}
  \ket{\Psi(T)} \approx 
  \alpha_{u}^0(0) \ket{0}_s \ket{u}_a - \alpha_{u}^1(0) \ket{1}_s \ket{u}_b 
  .
\end{align}
As such the single photon detection in one of the channels implements a projective measurement in the $z$-basis.

\bibliography{literatur}

\begin{thebibliography}{72}%
\makeatletter
\providecommand \@ifxundefined [1]{%
 \@ifx{#1\undefined}
}%
\providecommand \@ifnum [1]{%
 \ifnum #1\expandafter \@firstoftwo
 \else \expandafter \@secondoftwo
 \fi
}%
\providecommand \@ifx [1]{%
 \ifx #1\expandafter \@firstoftwo
 \else \expandafter \@secondoftwo
 \fi
}%
\providecommand \natexlab [1]{#1}%
\providecommand \enquote  [1]{``#1''}%
\providecommand \bibnamefont  [1]{#1}%
\providecommand \bibfnamefont [1]{#1}%
\providecommand \citenamefont [1]{#1}%
\providecommand \href@noop [0]{\@secondoftwo}%
\providecommand \href [0]{\begingroup \@sanitize@url \@href}%
\providecommand \@href[1]{\@@startlink{#1}\@@href}%
\providecommand \@@href[1]{\endgroup#1\@@endlink}%
\providecommand \@sanitize@url [0]{\catcode `\\12\catcode `\$12\catcode
  `\&12\catcode `\#12\catcode `\^12\catcode `\_12\catcode `\%12\relax}%
\providecommand \@@startlink[1]{}%
\providecommand \@@endlink[0]{}%
\providecommand \url  [0]{\begingroup\@sanitize@url \@url }%
\providecommand \@url [1]{\endgroup\@href {#1}{\urlprefix }}%
\providecommand \urlprefix  [0]{URL }%
\providecommand \Eprint [0]{\href }%
\providecommand \doibase [0]{https://doi.org/}%
\providecommand \selectlanguage [0]{\@gobble}%
\providecommand \bibinfo  [0]{\@secondoftwo}%
\providecommand \bibfield  [0]{\@secondoftwo}%
\providecommand \translation [1]{[#1]}%
\providecommand \BibitemOpen [0]{}%
\providecommand \bibitemStop [0]{}%
\providecommand \bibitemNoStop [0]{.\EOS\space}%
\providecommand \EOS [0]{\spacefactor3000\relax}%
\providecommand \BibitemShut  [1]{\csname bibitem#1\endcsname}%
\let\auto@bib@innerbib\@empty
\bibitem [{\citenamefont {Bell}(1964)}]{bell64}%
  \BibitemOpen
  \bibfield  {author} {\bibinfo {author} {\bibfnamefont {J.~S.}\ \bibnamefont
  {Bell}},\ }\href {https://doi.org/10.1103/physicsphysiquefizika.1.195}
  {\bibfield  {journal} {\bibinfo  {journal} {Physics}\ }\textbf {\bibinfo
  {volume} {1}},\ \bibinfo {pages} {195} (\bibinfo {year} {1964})}\BibitemShut
  {NoStop}%
\bibitem [{\citenamefont {Clauser}\ \emph {et~al.}(1969)\citenamefont
  {Clauser}, \citenamefont {Horne}, \citenamefont {Shimony},\ and\
  \citenamefont {Holt}}]{clauser69}%
  \BibitemOpen
  \bibfield  {author} {\bibinfo {author} {\bibfnamefont {J.~F.}\ \bibnamefont
  {Clauser}}, \bibinfo {author} {\bibfnamefont {M.~A.}\ \bibnamefont {Horne}},
  \bibinfo {author} {\bibfnamefont {A.}~\bibnamefont {Shimony}},\ and\ \bibinfo
  {author} {\bibfnamefont {R.~A.}\ \bibnamefont {Holt}},\ }\href
  {https://doi.org/10.1103/physrevlett.23.880} {\bibfield  {journal} {\bibinfo
  {journal} {Phys. Rev. Lett.}\ }\textbf {\bibinfo {volume} {23}},\ \bibinfo
  {pages} {880} (\bibinfo {year} {1969})}\BibitemShut {NoStop}%
\bibitem [{\citenamefont {Bennett}\ and\ \citenamefont
  {Brassard}(2014)}]{bennett_quantum_2014}%
  \BibitemOpen
  \bibfield  {author} {\bibinfo {author} {\bibfnamefont {C.~H.}\ \bibnamefont
  {Bennett}}\ and\ \bibinfo {author} {\bibfnamefont {G.}~\bibnamefont
  {Brassard}},\ }\href {https://doi.org/10.1016/j.tcs.2014.05.025} {\bibfield
  {journal} {\bibinfo  {journal} {Theor. Comput. Sci.}\ }\textbf {\bibinfo
  {volume} {560}},\ \bibinfo {pages} {7} (\bibinfo {year} {2014})},\ \bibinfo
  {note} {rEPRINT of a 1984 article}\BibitemShut {NoStop}%
\bibitem [{\citenamefont {Ekert}(1991)}]{ekert_quantum_1991}%
  \BibitemOpen
  \bibfield  {author} {\bibinfo {author} {\bibfnamefont {A.~K.}\ \bibnamefont
  {Ekert}},\ }\href {https://doi.org/10.1103/PhysRevLett.67.661} {\bibfield
  {journal} {\bibinfo  {journal} {Phys. Rev. Lett.}\ }\textbf {\bibinfo
  {volume} {67}},\ \bibinfo {pages} {661} (\bibinfo {year} {1991})}\BibitemShut
  {NoStop}%
\bibitem [{\citenamefont {Bennett}\ \emph {et~al.}(1992)\citenamefont
  {Bennett}, \citenamefont {Brassard},\ and\ \citenamefont
  {Mermin}}]{bennett92}%
  \BibitemOpen
  \bibfield  {author} {\bibinfo {author} {\bibfnamefont {C.~H.}\ \bibnamefont
  {Bennett}}, \bibinfo {author} {\bibfnamefont {G.}~\bibnamefont {Brassard}},\
  and\ \bibinfo {author} {\bibfnamefont {N.~D.}\ \bibnamefont {Mermin}},\
  }\href {https://doi.org/10.1103/physrevlett.68.557} {\bibfield  {journal}
  {\bibinfo  {journal} {Phys. Rev. Lett.}\ }\textbf {\bibinfo {volume} {68}},\
  \bibinfo {pages} {557} (\bibinfo {year} {1992})}\BibitemShut {NoStop}%
\bibitem [{\citenamefont {Mayers}\ and\ \citenamefont {Yao}(1998)}]{mayers98}%
  \BibitemOpen
  \bibfield  {author} {\bibinfo {author} {\bibfnamefont {D.}~\bibnamefont
  {Mayers}}\ and\ \bibinfo {author} {\bibfnamefont {A.}~\bibnamefont {Yao}},\
  }in\ \href {https://doi.org/10.1109/sfcs.1998.743501} {\emph {\bibinfo
  {booktitle} {Proceedings 39th Annual Symposium on Foundations of Computer
  Science (Cat. No.98CB36280)}}}\ (\bibinfo {year} {1998})\BibitemShut
  {NoStop}%
\bibitem [{\citenamefont {Ac{\'i}n}\ \emph {et~al.}(2007)\citenamefont
  {Ac{\'i}n}, \citenamefont {Brunner}, \citenamefont {Gisin}, \citenamefont
  {Massar}, \citenamefont {Pironio},\ and\ \citenamefont {Scarani}}]{acin07}%
  \BibitemOpen
  \bibfield  {author} {\bibinfo {author} {\bibfnamefont {A.}~\bibnamefont
  {Ac{\'i}n}}, \bibinfo {author} {\bibfnamefont {N.}~\bibnamefont {Brunner}},
  \bibinfo {author} {\bibfnamefont {N.}~\bibnamefont {Gisin}}, \bibinfo
  {author} {\bibfnamefont {S.}~\bibnamefont {Massar}}, \bibinfo {author}
  {\bibfnamefont {S.}~\bibnamefont {Pironio}},\ and\ \bibinfo {author}
  {\bibfnamefont {V.}~\bibnamefont {Scarani}},\ }\href
  {https://doi.org/10.1103/physrevlett.98.230501} {\bibfield  {journal}
  {\bibinfo  {journal} {Phys. Rev. Lett.}\ }\textbf {\bibinfo {volume} {98}},\
  \bibinfo {pages} {230501} (\bibinfo {year} {2007})}\BibitemShut {NoStop}%
\bibitem [{\citenamefont {Gisin}\ and\ \citenamefont {Thew}(2007)}]{gisin07}%
  \BibitemOpen
  \bibfield  {author} {\bibinfo {author} {\bibfnamefont {N.}~\bibnamefont
  {Gisin}}\ and\ \bibinfo {author} {\bibfnamefont {R.}~\bibnamefont {Thew}},\
  }\href {https://doi.org/10.1038/nphoton.2007.22} {\bibfield  {journal}
  {\bibinfo  {journal} {Nat. Photonics}\ }\textbf {\bibinfo {volume} {1}},\
  \bibinfo {pages} {165} (\bibinfo {year} {2007})}\BibitemShut {NoStop}%
\bibitem [{\citenamefont {Arnon-Friedman}\ \emph {et~al.}(2019)\citenamefont
  {Arnon-Friedman}, \citenamefont {Renner},\ and\ \citenamefont
  {Vidick}}]{arnon-friedman19}%
  \BibitemOpen
  \bibfield  {author} {\bibinfo {author} {\bibfnamefont {R.}~\bibnamefont
  {Arnon-Friedman}}, \bibinfo {author} {\bibfnamefont {R.}~\bibnamefont
  {Renner}},\ and\ \bibinfo {author} {\bibfnamefont {T.}~\bibnamefont
  {Vidick}},\ }\href {https://doi.org/10.1137/18m1174726} {\bibfield  {journal}
  {\bibinfo  {journal} {{SIAM} J. Comput.}\ }\textbf {\bibinfo {volume} {48}},\
  \bibinfo {pages} {181} (\bibinfo {year} {2019})}\BibitemShut {NoStop}%
\bibitem [{\citenamefont {Pirandola}\ \emph {et~al.}(2020)\citenamefont
  {Pirandola}, \citenamefont {Andersen}, \citenamefont {Banchi}, \citenamefont
  {Berta}, \citenamefont {Bunandar}, \citenamefont {Colbeck}, \citenamefont
  {Englund}, \citenamefont {Gehring}, \citenamefont {Lupo}, \citenamefont
  {Ottaviani}, \citenamefont {Pereira}, \citenamefont {Razavi}, \citenamefont
  {Shaari}, \citenamefont {Tomamichel}, \citenamefont {Usenko}, \citenamefont
  {Vallone}, \citenamefont {Villoresi},\ and\ \citenamefont
  {Wallden}}]{pirandola20}%
  \BibitemOpen
  \bibfield  {author} {\bibinfo {author} {\bibfnamefont {S.}~\bibnamefont
  {Pirandola}}, \bibinfo {author} {\bibfnamefont {U.~L.}\ \bibnamefont
  {Andersen}}, \bibinfo {author} {\bibfnamefont {L.}~\bibnamefont {Banchi}},
  \bibinfo {author} {\bibfnamefont {M.}~\bibnamefont {Berta}}, \bibinfo
  {author} {\bibfnamefont {D.}~\bibnamefont {Bunandar}}, \bibinfo {author}
  {\bibfnamefont {R.}~\bibnamefont {Colbeck}}, \bibinfo {author} {\bibfnamefont
  {D.}~\bibnamefont {Englund}}, \bibinfo {author} {\bibfnamefont
  {T.}~\bibnamefont {Gehring}}, \bibinfo {author} {\bibfnamefont
  {C.}~\bibnamefont {Lupo}}, \bibinfo {author} {\bibfnamefont {C.}~\bibnamefont
  {Ottaviani}}, \bibinfo {author} {\bibfnamefont {J.~L.}\ \bibnamefont
  {Pereira}}, \bibinfo {author} {\bibfnamefont {M.}~\bibnamefont {Razavi}},
  \bibinfo {author} {\bibfnamefont {J.~S.}\ \bibnamefont {Shaari}}, \bibinfo
  {author} {\bibfnamefont {M.}~\bibnamefont {Tomamichel}}, \bibinfo {author}
  {\bibfnamefont {V.~C.}\ \bibnamefont {Usenko}}, \bibinfo {author}
  {\bibfnamefont {G.}~\bibnamefont {Vallone}}, \bibinfo {author} {\bibfnamefont
  {P.}~\bibnamefont {Villoresi}},\ and\ \bibinfo {author} {\bibfnamefont
  {P.}~\bibnamefont {Wallden}},\ }\href {https://doi.org/10.1364/aop.361502}
  {\bibfield  {journal} {\bibinfo  {journal} {Adv. Opt. Photonics}\ }\textbf
  {\bibinfo {volume} {12}},\ \bibinfo {pages} {1012} (\bibinfo {year}
  {2020})}\BibitemShut {NoStop}%
\bibitem [{\citenamefont {Bongs}\ \emph {et~al.}(2023)\citenamefont {Bongs},
  \citenamefont {Bennett},\ and\ \citenamefont {Lohmann}}]{bongs23}%
  \BibitemOpen
  \bibfield  {author} {\bibinfo {author} {\bibfnamefont {K.}~\bibnamefont
  {Bongs}}, \bibinfo {author} {\bibfnamefont {S.}~\bibnamefont {Bennett}},\
  and\ \bibinfo {author} {\bibfnamefont {A.}~\bibnamefont {Lohmann}},\ }\href
  {https://doi.org/10.1038/d41586-023-01663-0} {\bibfield  {journal} {\bibinfo
  {journal} {Nature}\ }\textbf {\bibinfo {volume} {617}},\ \bibinfo {pages}
  {672} (\bibinfo {year} {2023})}\BibitemShut {NoStop}%
\bibitem [{\citenamefont {Xia}\ \emph {et~al.}(2023)\citenamefont {Xia},
  \citenamefont {Agrawal}, \citenamefont {Pluchar}, \citenamefont {Brady},
  \citenamefont {Liu}, \citenamefont {Zhuang}, \citenamefont {Wilson},\ and\
  \citenamefont {Zhang}}]{xia-2023-entan_enhan_optom_sensin}%
  \BibitemOpen
  \bibfield  {author} {\bibinfo {author} {\bibfnamefont {Y.}~\bibnamefont
  {Xia}}, \bibinfo {author} {\bibfnamefont {A.~R.}\ \bibnamefont {Agrawal}},
  \bibinfo {author} {\bibfnamefont {C.~M.}\ \bibnamefont {Pluchar}}, \bibinfo
  {author} {\bibfnamefont {A.~J.}\ \bibnamefont {Brady}}, \bibinfo {author}
  {\bibfnamefont {Z.}~\bibnamefont {Liu}}, \bibinfo {author} {\bibfnamefont
  {Q.}~\bibnamefont {Zhuang}}, \bibinfo {author} {\bibfnamefont {D.~J.}\
  \bibnamefont {Wilson}},\ and\ \bibinfo {author} {\bibfnamefont
  {Z.}~\bibnamefont {Zhang}},\ }\href
  {https://doi.org/10.1038/s41566-023-01178-0} {\bibfield  {journal} {\bibinfo
  {journal} {Nat. Photonics}\ }\textbf {\bibinfo {volume} {17}},\ \bibinfo
  {pages} {470} (\bibinfo {year} {2023})}\BibitemShut {NoStop}%
\bibitem [{\citenamefont {Kimble}(2008)}]{Kimble2008}%
  \BibitemOpen
  \bibfield  {author} {\bibinfo {author} {\bibfnamefont {H.~J.}\ \bibnamefont
  {Kimble}},\ }\href {https://doi.org/10.1038/nature07127} {\bibfield
  {journal} {\bibinfo  {journal} {Nature}\ }\textbf {\bibinfo {volume} {453}},\
  \bibinfo {pages} {1023} (\bibinfo {year} {2008})}\BibitemShut {NoStop}%
\bibitem [{\citenamefont {Wehner}\ \emph {et~al.}(2018)\citenamefont {Wehner},
  \citenamefont {Elkouss},\ and\ \citenamefont {Hanson}}]{Wehner2018}%
  \BibitemOpen
  \bibfield  {author} {\bibinfo {author} {\bibfnamefont {S.}~\bibnamefont
  {Wehner}}, \bibinfo {author} {\bibfnamefont {D.}~\bibnamefont {Elkouss}},\
  and\ \bibinfo {author} {\bibfnamefont {R.}~\bibnamefont {Hanson}},\ }\href
  {https://doi.org/10.1126/science.aam9288} {\bibfield  {journal} {\bibinfo
  {journal} {Science}\ }\textbf {\bibinfo {volume} {362}},\ \bibinfo {pages}
  {303} (\bibinfo {year} {2018})}\BibitemShut {NoStop}%
\bibitem [{\citenamefont {Nemoto}\ \emph {et~al.}(2014)\citenamefont {Nemoto},
  \citenamefont {Trupke}, \citenamefont {Devitt}, \citenamefont {Stephens},
  \citenamefont {Scharfenberger}, \citenamefont {Buczak}, \citenamefont
  {N{\"o}bauer}, \citenamefont {Everitt}, \citenamefont {Schmiedmayer},\ and\
  \citenamefont {Munro}}]{nemoto_photonic_2014}%
  \BibitemOpen
  \bibfield  {author} {\bibinfo {author} {\bibfnamefont {K.}~\bibnamefont
  {Nemoto}}, \bibinfo {author} {\bibfnamefont {M.}~\bibnamefont {Trupke}},
  \bibinfo {author} {\bibfnamefont {S.~J.}\ \bibnamefont {Devitt}}, \bibinfo
  {author} {\bibfnamefont {A.~M.}\ \bibnamefont {Stephens}}, \bibinfo {author}
  {\bibfnamefont {B.}~\bibnamefont {Scharfenberger}}, \bibinfo {author}
  {\bibfnamefont {K.}~\bibnamefont {Buczak}}, \bibinfo {author} {\bibfnamefont
  {T.}~\bibnamefont {N{\"o}bauer}}, \bibinfo {author} {\bibfnamefont {M.~S.}\
  \bibnamefont {Everitt}}, \bibinfo {author} {\bibfnamefont {J.}~\bibnamefont
  {Schmiedmayer}},\ and\ \bibinfo {author} {\bibfnamefont {W.~J.}\ \bibnamefont
  {Munro}},\ }\href {https://doi.org/10.1103/PhysRevX.4.031022} {\bibfield
  {journal} {\bibinfo  {journal} {Phys. Rev. X}\ }\textbf {\bibinfo {volume}
  {4}},\ \bibinfo {pages} {031022} (\bibinfo {year} {2014})}\BibitemShut
  {NoStop}%
\bibitem [{\citenamefont {Monroe}\ \emph {et~al.}(2014)\citenamefont {Monroe},
  \citenamefont {Raussendorf}, \citenamefont {Ruthven}, \citenamefont {Brown},
  \citenamefont {Maunz}, \citenamefont {Duan},\ and\ \citenamefont
  {Kim}}]{monroe_large-scale_2014}%
  \BibitemOpen
  \bibfield  {author} {\bibinfo {author} {\bibfnamefont {C.}~\bibnamefont
  {Monroe}}, \bibinfo {author} {\bibfnamefont {R.}~\bibnamefont {Raussendorf}},
  \bibinfo {author} {\bibfnamefont {A.}~\bibnamefont {Ruthven}}, \bibinfo
  {author} {\bibfnamefont {K.~R.}\ \bibnamefont {Brown}}, \bibinfo {author}
  {\bibfnamefont {P.}~\bibnamefont {Maunz}}, \bibinfo {author} {\bibfnamefont
  {L.-M.}\ \bibnamefont {Duan}},\ and\ \bibinfo {author} {\bibfnamefont
  {J.}~\bibnamefont {Kim}},\ }\href
  {https://doi.org/10.1103/PhysRevA.89.022317} {\bibfield  {journal} {\bibinfo
  {journal} {Phys. Rev. A}\ }\textbf {\bibinfo {volume} {89}},\ \bibinfo
  {pages} {022317} (\bibinfo {year} {2014})}\BibitemShut {NoStop}%
\bibitem [{\citenamefont {Gottesman}\ and\ \citenamefont
  {Chuang}(1999)}]{gottesman_demonstrating_1999}%
  \BibitemOpen
  \bibfield  {author} {\bibinfo {author} {\bibfnamefont {D.}~\bibnamefont
  {Gottesman}}\ and\ \bibinfo {author} {\bibfnamefont {I.~L.}\ \bibnamefont
  {Chuang}},\ }\href {https://doi.org/10.1038/46503} {\bibfield  {journal}
  {\bibinfo  {journal} {Nature}\ }\textbf {\bibinfo {volume} {402}},\ \bibinfo
  {pages} {390} (\bibinfo {year} {1999})}\BibitemShut {NoStop}%
\bibitem [{\citenamefont {Raussendorf}\ and\ \citenamefont
  {Briegel}(2001)}]{raussendorf_one-way_2001}%
  \BibitemOpen
  \bibfield  {author} {\bibinfo {author} {\bibfnamefont {R.}~\bibnamefont
  {Raussendorf}}\ and\ \bibinfo {author} {\bibfnamefont {H.~J.}\ \bibnamefont
  {Briegel}},\ }\href {https://doi.org/10.1103/PhysRevLett.86.5188} {\bibfield
  {journal} {\bibinfo  {journal} {Phys. Rev. Lett.}\ }\textbf {\bibinfo
  {volume} {86}},\ \bibinfo {pages} {5188} (\bibinfo {year}
  {2001})}\BibitemShut {NoStop}%
\bibitem [{\citenamefont {Eisert}\ \emph {et~al.}(2000)\citenamefont {Eisert},
  \citenamefont {Jacobs}, \citenamefont {Papadopoulos},\ and\ \citenamefont
  {Plenio}}]{eisert_optimal_2000}%
  \BibitemOpen
  \bibfield  {author} {\bibinfo {author} {\bibfnamefont {J.}~\bibnamefont
  {Eisert}}, \bibinfo {author} {\bibfnamefont {K.}~\bibnamefont {Jacobs}},
  \bibinfo {author} {\bibfnamefont {P.}~\bibnamefont {Papadopoulos}},\ and\
  \bibinfo {author} {\bibfnamefont {M.~B.}\ \bibnamefont {Plenio}},\ }\href
  {https://doi.org/10.1103/PhysRevA.62.052317} {\bibfield  {journal} {\bibinfo
  {journal} {Phys. Rev. A}\ }\textbf {\bibinfo {volume} {62}},\ \bibinfo
  {pages} {052317} (\bibinfo {year} {2000})}\BibitemShut {NoStop}%
\bibitem [{\citenamefont {Jiang}\ \emph {et~al.}(2007)\citenamefont {Jiang},
  \citenamefont {Taylor}, \citenamefont {Sørensen},\ and\ \citenamefont
  {Lukin}}]{jiang_distributed_2007}%
  \BibitemOpen
  \bibfield  {author} {\bibinfo {author} {\bibfnamefont {L.}~\bibnamefont
  {Jiang}}, \bibinfo {author} {\bibfnamefont {J.~M.}\ \bibnamefont {Taylor}},
  \bibinfo {author} {\bibfnamefont {A.~S.}\ \bibnamefont {Sørensen}},\ and\
  \bibinfo {author} {\bibfnamefont {M.~D.}\ \bibnamefont {Lukin}},\ }\href
  {https://doi.org/10.1103/PhysRevA.76.062323} {\bibfield  {journal} {\bibinfo
  {journal} {Phys. Rev. A}\ }\textbf {\bibinfo {volume} {76}},\ \bibinfo
  {pages} {062323} (\bibinfo {year} {2007})}\BibitemShut {NoStop}%
\bibitem [{\citenamefont {Doherty}\ \emph {et~al.}(2013)\citenamefont
  {Doherty}, \citenamefont {Manson}, \citenamefont {Delaney}, \citenamefont
  {Jelezko}, \citenamefont {Wrachtrup},\ and\ \citenamefont
  {Hollenberg}}]{doherty_nitrogen-vacancy_2013}%
  \BibitemOpen
  \bibfield  {author} {\bibinfo {author} {\bibfnamefont {M.~W.}\ \bibnamefont
  {Doherty}}, \bibinfo {author} {\bibfnamefont {N.~B.}\ \bibnamefont {Manson}},
  \bibinfo {author} {\bibfnamefont {P.}~\bibnamefont {Delaney}}, \bibinfo
  {author} {\bibfnamefont {F.}~\bibnamefont {Jelezko}}, \bibinfo {author}
  {\bibfnamefont {J.}~\bibnamefont {Wrachtrup}},\ and\ \bibinfo {author}
  {\bibfnamefont {L.~C.~L.}\ \bibnamefont {Hollenberg}},\ }\href
  {https://doi.org/10.1016/j.physrep.2013.02.001} {\bibfield  {journal}
  {\bibinfo  {journal} {Phys. Rep.}\ }\bibinfo {series} {The nitrogen-vacancy
  colour centre in diamond},\ \textbf {\bibinfo {volume} {528}},\ \bibinfo
  {pages} {1} (\bibinfo {year} {2013})}\BibitemShut {NoStop}%
\bibitem [{\citenamefont {Hepp}\ \emph {et~al.}(2014)\citenamefont {Hepp},
  \citenamefont {Müller}, \citenamefont {Waselowski}, \citenamefont {Becker},
  \citenamefont {Pingault}, \citenamefont {Sternschulte}, \citenamefont
  {Steinmüller-Nethl}, \citenamefont {Gali}, \citenamefont {Maze},
  \citenamefont {Atatüre},\ and\ \citenamefont
  {Becher}}]{hepp_electronic_2014}%
  \BibitemOpen
  \bibfield  {author} {\bibinfo {author} {\bibfnamefont {C.}~\bibnamefont
  {Hepp}}, \bibinfo {author} {\bibfnamefont {T.}~\bibnamefont {Müller}},
  \bibinfo {author} {\bibfnamefont {V.}~\bibnamefont {Waselowski}}, \bibinfo
  {author} {\bibfnamefont {J.~N.}\ \bibnamefont {Becker}}, \bibinfo {author}
  {\bibfnamefont {B.}~\bibnamefont {Pingault}}, \bibinfo {author}
  {\bibfnamefont {H.}~\bibnamefont {Sternschulte}}, \bibinfo {author}
  {\bibfnamefont {D.}~\bibnamefont {Steinmüller-Nethl}}, \bibinfo {author}
  {\bibfnamefont {A.}~\bibnamefont {Gali}}, \bibinfo {author} {\bibfnamefont
  {J.~R.}\ \bibnamefont {Maze}}, \bibinfo {author} {\bibfnamefont
  {M.}~\bibnamefont {Atatüre}},\ and\ \bibinfo {author} {\bibfnamefont
  {C.}~\bibnamefont {Becher}},\ }\href
  {https://doi.org/10.1103/PhysRevLett.112.036405} {\bibfield  {journal}
  {\bibinfo  {journal} {Phys. Rev. Lett.}\ }\textbf {\bibinfo {volume} {112}},\
  \bibinfo {pages} {036405} (\bibinfo {year} {2014})}\BibitemShut {NoStop}%
\bibitem [{\citenamefont {Taminiau}\ \emph {et~al.}(2012)\citenamefont
  {Taminiau}, \citenamefont {Wagenaar}, \citenamefont {van~der Sar},
  \citenamefont {Jelezko}, \citenamefont {Dobrovitski},\ and\ \citenamefont
  {Hanson}}]{taminiau_detection_2012}%
  \BibitemOpen
  \bibfield  {author} {\bibinfo {author} {\bibfnamefont {T.~H.}\ \bibnamefont
  {Taminiau}}, \bibinfo {author} {\bibfnamefont {J.~J.~T.}\ \bibnamefont
  {Wagenaar}}, \bibinfo {author} {\bibfnamefont {T.}~\bibnamefont {van~der
  Sar}}, \bibinfo {author} {\bibfnamefont {F.}~\bibnamefont {Jelezko}},
  \bibinfo {author} {\bibfnamefont {V.~V.}\ \bibnamefont {Dobrovitski}},\ and\
  \bibinfo {author} {\bibfnamefont {R.}~\bibnamefont {Hanson}},\ }\href
  {https://doi.org/10.1103/PhysRevLett.109.137602} {\bibfield  {journal}
  {\bibinfo  {journal} {Phys. Rev. Lett.}\ }\textbf {\bibinfo {volume} {109}},\
  \bibinfo {pages} {137602} (\bibinfo {year} {2012})}\BibitemShut {NoStop}%
\bibitem [{\citenamefont {Bradley}\ \emph {et~al.}(2019)\citenamefont
  {Bradley}, \citenamefont {Randall}, \citenamefont {Abobeih}, \citenamefont
  {Berrevoets}, \citenamefont {Degen}, \citenamefont {Bakker}, \citenamefont
  {Markham}, \citenamefont {Twitchen},\ and\ \citenamefont
  {Taminiau}}]{bradley_ten-qubit_2019}%
  \BibitemOpen
  \bibfield  {author} {\bibinfo {author} {\bibfnamefont {C.}~\bibnamefont
  {Bradley}}, \bibinfo {author} {\bibfnamefont {J.}~\bibnamefont {Randall}},
  \bibinfo {author} {\bibfnamefont {M.}~\bibnamefont {Abobeih}}, \bibinfo
  {author} {\bibfnamefont {R.}~\bibnamefont {Berrevoets}}, \bibinfo {author}
  {\bibfnamefont {M.}~\bibnamefont {Degen}}, \bibinfo {author} {\bibfnamefont
  {M.}~\bibnamefont {Bakker}}, \bibinfo {author} {\bibfnamefont
  {M.}~\bibnamefont {Markham}}, \bibinfo {author} {\bibfnamefont
  {D.}~\bibnamefont {Twitchen}},\ and\ \bibinfo {author} {\bibfnamefont
  {T.}~\bibnamefont {Taminiau}},\ }\href
  {https://doi.org/10.1103/PhysRevX.9.031045} {\bibfield  {journal} {\bibinfo
  {journal} {Phys. Rev. X}\ }\textbf {\bibinfo {volume} {9}},\ \bibinfo {pages}
  {031045} (\bibinfo {year} {2019})}\BibitemShut {NoStop}%
\bibitem [{\citenamefont {Childress}\ \emph {et~al.}(2005)\citenamefont
  {Childress}, \citenamefont {Taylor}, \citenamefont {Sørensen},\ and\
  \citenamefont {Lukin}}]{childress_fault-tolerant_2005}%
  \BibitemOpen
  \bibfield  {author} {\bibinfo {author} {\bibfnamefont {L.}~\bibnamefont
  {Childress}}, \bibinfo {author} {\bibfnamefont {J.~M.}\ \bibnamefont
  {Taylor}}, \bibinfo {author} {\bibfnamefont {A.~S.}\ \bibnamefont
  {Sørensen}},\ and\ \bibinfo {author} {\bibfnamefont {M.~D.}\ \bibnamefont
  {Lukin}},\ }\href {https://doi.org/10.1103/PhysRevA.72.052330} {\bibfield
  {journal} {\bibinfo  {journal} {Phys. Rev. A}\ }\textbf {\bibinfo {volume}
  {72}},\ \bibinfo {pages} {052330} (\bibinfo {year} {2005})}\BibitemShut
  {NoStop}%
\bibitem [{\citenamefont {Santori}\ \emph {et~al.}(2010)\citenamefont
  {Santori}, \citenamefont {Fattal},\ and\ \citenamefont
  {Yamamoto}}]{santori_single_2010}%
  \BibitemOpen
  \bibfield  {author} {\bibinfo {author} {\bibfnamefont {C.}~\bibnamefont
  {Santori}}, \bibinfo {author} {\bibfnamefont {D.}~\bibnamefont {Fattal}},\
  and\ \bibinfo {author} {\bibfnamefont {Y.}~\bibnamefont {Yamamoto}},\
  }\href@noop {} {\emph {\bibinfo {title} {Single-Photon Devices and
  Applications}}}\ (\bibinfo  {publisher} {Wiley-VCH},\ \bibinfo {year}
  {2010})\BibitemShut {NoStop}%
\bibitem [{\citenamefont {Kiilerich}\ and\ \citenamefont
  {M{\o}lmer}(2019)}]{kiilerich_input-output_2019}%
  \BibitemOpen
  \bibfield  {author} {\bibinfo {author} {\bibfnamefont {A.~H.}\ \bibnamefont
  {Kiilerich}}\ and\ \bibinfo {author} {\bibfnamefont {K.}~\bibnamefont
  {M{\o}lmer}},\ }\href {https://doi.org/10.1103/PhysRevLett.123.123604}
  {\bibfield  {journal} {\bibinfo  {journal} {Phys. Rev. Lett.}\ }\textbf
  {\bibinfo {volume} {123}},\ \bibinfo {pages} {123604} (\bibinfo {year}
  {2019})}\BibitemShut {NoStop}%
\bibitem [{\citenamefont {Kiilerich}\ and\ \citenamefont
  {Mølmer}(2020)}]{kiilerich_quantum_2020}%
  \BibitemOpen
  \bibfield  {author} {\bibinfo {author} {\bibfnamefont {A.~H.}\ \bibnamefont
  {Kiilerich}}\ and\ \bibinfo {author} {\bibfnamefont {K.}~\bibnamefont
  {Mølmer}},\ }\href {https://doi.org/10.1103/PhysRevA.102.023717} {\bibfield
  {journal} {\bibinfo  {journal} {Phys. Rev. A}\ }\textbf {\bibinfo {volume}
  {102}},\ \bibinfo {pages} {023717} (\bibinfo {year} {2020})}\BibitemShut
  {NoStop}%
\bibitem [{\citenamefont {Gough}\ and\ \citenamefont
  {Zhang}(2015)}]{gough_generating_2015}%
  \BibitemOpen
  \bibfield  {author} {\bibinfo {author} {\bibfnamefont {J.~E.}\ \bibnamefont
  {Gough}}\ and\ \bibinfo {author} {\bibfnamefont {G.}~\bibnamefont {Zhang}},\
  }\href {https://doi.org/10.1140/epjqt/s40507-015-0027-z} {\bibfield
  {journal} {\bibinfo  {journal} {EPJ Quantum Technol.}\ }\textbf {\bibinfo
  {volume} {2}},\ \bibinfo {pages} {1} (\bibinfo {year} {2015})}\BibitemShut
  {NoStop}%
\bibitem [{\citenamefont {Nurdin}\ \emph {et~al.}(2016)\citenamefont {Nurdin},
  \citenamefont {James},\ and\ \citenamefont {Yamamoto}}]{nurdin_perfect_2016}%
  \BibitemOpen
  \bibfield  {author} {\bibinfo {author} {\bibfnamefont {H.~I.}\ \bibnamefont
  {Nurdin}}, \bibinfo {author} {\bibfnamefont {M.~R.}\ \bibnamefont {James}},\
  and\ \bibinfo {author} {\bibfnamefont {N.}~\bibnamefont {Yamamoto}},\ }in\
  \href {https://doi.org/10.1109/CDC.2016.7798639} {\emph {\bibinfo {booktitle}
  {2016 {IEEE} 55th {Conference} on {Decision} and {Control} ({CDC})}}}\
  (\bibinfo {year} {2016})\ pp.\ \bibinfo {pages} {2513--2518}\BibitemShut
  {NoStop}%
\bibitem [{\citenamefont {Combes}\ \emph {et~al.}(2017)\citenamefont {Combes},
  \citenamefont {Kerckhoff},\ and\ \citenamefont {Sarovar}}]{combes_slh_2017}%
  \BibitemOpen
  \bibfield  {author} {\bibinfo {author} {\bibfnamefont {J.}~\bibnamefont
  {Combes}}, \bibinfo {author} {\bibfnamefont {J.}~\bibnamefont {Kerckhoff}},\
  and\ \bibinfo {author} {\bibfnamefont {M.}~\bibnamefont {Sarovar}},\ }\href
  {https://doi.org/10.1080/23746149.2017.1343097} {\bibfield  {journal}
  {\bibinfo  {journal} {Advances in Physics: X}\ }\textbf {\bibinfo {volume}
  {2}},\ \bibinfo {pages} {784} (\bibinfo {year} {2017})}\BibitemShut {NoStop}%
\bibitem [{\citenamefont {Dalibard}\ \emph {et~al.}(1992)\citenamefont
  {Dalibard}, \citenamefont {Castin},\ and\ \citenamefont
  {M{\o}lmer}}]{dalibard92}%
  \BibitemOpen
  \bibfield  {author} {\bibinfo {author} {\bibfnamefont {J.}~\bibnamefont
  {Dalibard}}, \bibinfo {author} {\bibfnamefont {Y.}~\bibnamefont {Castin}},\
  and\ \bibinfo {author} {\bibfnamefont {K.}~\bibnamefont {M{\o}lmer}},\ }\href
  {https://doi.org/10.1103/physrevlett.68.580} {\bibfield  {journal} {\bibinfo
  {journal} {Phys. Rev. Lett.}\ }\textbf {\bibinfo {volume} {68}},\ \bibinfo
  {pages} {580} (\bibinfo {year} {1992})}\BibitemShut {NoStop}%
\bibitem [{\citenamefont {M{\o}lmer}\ \emph {et~al.}(1993)\citenamefont
  {M{\o}lmer}, \citenamefont {Castin},\ and\ \citenamefont
  {Dalibard}}]{moelmer93}%
  \BibitemOpen
  \bibfield  {author} {\bibinfo {author} {\bibfnamefont {K.}~\bibnamefont
  {M{\o}lmer}}, \bibinfo {author} {\bibfnamefont {Y.}~\bibnamefont {Castin}},\
  and\ \bibinfo {author} {\bibfnamefont {J.}~\bibnamefont {Dalibard}},\ }\href
  {https://doi.org/10.1364/josab.10.000524} {\bibfield  {journal} {\bibinfo
  {journal} {J. Opt. Soc. Am. B}\ }\textbf {\bibinfo {volume} {10}},\ \bibinfo
  {pages} {524} (\bibinfo {year} {1993})}\BibitemShut {NoStop}%
\bibitem [{\citenamefont {Carmichael}(1993)}]{carmichael93}%
  \BibitemOpen
  \bibfield  {author} {\bibinfo {author} {\bibfnamefont {H.~J.}\ \bibnamefont
  {Carmichael}},\ }\href {https://doi.org/10.1103/physrevlett.70.2273}
  {\bibfield  {journal} {\bibinfo  {journal} {Phys. Rev. Lett.}\ }\textbf
  {\bibinfo {volume} {70}},\ \bibinfo {pages} {2273} (\bibinfo {year}
  {1993})}\BibitemShut {NoStop}%
\bibitem [{\citenamefont {Daley}(2014)}]{daley14}%
  \BibitemOpen
  \bibfield  {author} {\bibinfo {author} {\bibfnamefont {A.~J.}\ \bibnamefont
  {Daley}},\ }\href {https://doi.org/10.1080/00018732.2014.933502} {\bibfield
  {journal} {\bibinfo  {journal} {Adv. Phys.}\ }\textbf {\bibinfo {volume}
  {63}},\ \bibinfo {pages} {77} (\bibinfo {year} {2014})}\BibitemShut {NoStop}%
\bibitem [{\citenamefont {Tissot}\ and\ \citenamefont
  {Burkard}(2024)}]{tissot_efficient_2024}%
  \BibitemOpen
  \bibfield  {author} {\bibinfo {author} {\bibfnamefont {B.}~\bibnamefont
  {Tissot}}\ and\ \bibinfo {author} {\bibfnamefont {G.}~\bibnamefont
  {Burkard}},\ }\href {https://doi.org/10.1103/PhysRevResearch.6.013150}
  {\bibfield  {journal} {\bibinfo  {journal} {Phys. Rev. Research}\ }\textbf
  {\bibinfo {volume} {6}},\ \bibinfo {pages} {013150} (\bibinfo {year}
  {2024})}\BibitemShut {NoStop}%
\bibitem [{Note1()}]{Note1}%
  \BibitemOpen
  \bibinfo {note} {To proof this we use that we can write $r_k = 1 - \protect
  \frac {\kappa _1}{a+ib}$, where $a,b\in \protect \mathbb {R}$ and $2a >
  \kappa \ge \kappa _1$, such that $|r_k|^2 = 1 - \protect \frac {\kappa _1
  (\kappa _1 - 2a)}{a^2 + b^2} \le 1$ which follows because the second term is
  positive and $\protect \frac {\kappa _1}{a} (2 - \protect \frac {\kappa
  _1}{a}) \le 1$ for all $0 \le \protect \frac {\kappa _1}{a} \le
  2$.}\BibitemShut {Stop}%
\bibitem [{\citenamefont {Hu}\ \emph {et~al.}(2008)\citenamefont {Hu},
  \citenamefont {Young}, \citenamefont {O’Brien}, \citenamefont {Munro},\
  and\ \citenamefont {Rarity}}]{hu_giant_2008}%
  \BibitemOpen
  \bibfield  {author} {\bibinfo {author} {\bibfnamefont {C.~Y.}\ \bibnamefont
  {Hu}}, \bibinfo {author} {\bibfnamefont {A.}~\bibnamefont {Young}}, \bibinfo
  {author} {\bibfnamefont {J.~L.}\ \bibnamefont {O’Brien}}, \bibinfo {author}
  {\bibfnamefont {W.~J.}\ \bibnamefont {Munro}},\ and\ \bibinfo {author}
  {\bibfnamefont {J.~G.}\ \bibnamefont {Rarity}},\ }\href
  {https://doi.org/10.1103/PhysRevB.78.085307} {\bibfield  {journal} {\bibinfo
  {journal} {Phys. Rev. B}\ }\textbf {\bibinfo {volume} {78}},\ \bibinfo
  {pages} {085307} (\bibinfo {year} {2008})}\BibitemShut {NoStop}%
\bibitem [{\citenamefont {Hu}\ \emph {et~al.}(2009)\citenamefont {Hu},
  \citenamefont {Munro}, \citenamefont {O'Brien},\ and\ \citenamefont
  {Rarity}}]{hu-2009-propos_entan_beam_split_using}%
  \BibitemOpen
  \bibfield  {author} {\bibinfo {author} {\bibfnamefont {C.~Y.}\ \bibnamefont
  {Hu}}, \bibinfo {author} {\bibfnamefont {W.~J.}\ \bibnamefont {Munro}},
  \bibinfo {author} {\bibfnamefont {J.~L.}\ \bibnamefont {O'Brien}},\ and\
  \bibinfo {author} {\bibfnamefont {J.~G.}\ \bibnamefont {Rarity}},\ }\href
  {https://doi.org/10.1103/physrevb.80.205326} {\bibfield  {journal} {\bibinfo
  {journal} {Phys. Rev. B}\ }\textbf {\bibinfo {volume} {80}},\ \bibinfo
  {pages} {205326} (\bibinfo {year} {2009})}\BibitemShut {NoStop}%
\bibitem [{\citenamefont {Koshino}\ and\ \citenamefont
  {Matsuzaki}(2012)}]{koshino_entangling_2012}%
  \BibitemOpen
  \bibfield  {author} {\bibinfo {author} {\bibfnamefont {K.}~\bibnamefont
  {Koshino}}\ and\ \bibinfo {author} {\bibfnamefont {Y.}~\bibnamefont
  {Matsuzaki}},\ }\href {https://doi.org/10.1103/PhysRevA.86.020305} {\bibfield
   {journal} {\bibinfo  {journal} {Phys. Rev. A}\ }\textbf {\bibinfo {volume}
  {86}},\ \bibinfo {pages} {020305} (\bibinfo {year} {2012})}\BibitemShut
  {NoStop}%
\bibitem [{\citenamefont {Volz}\ \emph {et~al.}(2011)\citenamefont {Volz},
  \citenamefont {Gehr}, \citenamefont {Dubois}, \citenamefont {Estève},\ and\
  \citenamefont {Reichel}}]{volz_measurement_2011}%
  \BibitemOpen
  \bibfield  {author} {\bibinfo {author} {\bibfnamefont {J.}~\bibnamefont
  {Volz}}, \bibinfo {author} {\bibfnamefont {R.}~\bibnamefont {Gehr}}, \bibinfo
  {author} {\bibfnamefont {G.}~\bibnamefont {Dubois}}, \bibinfo {author}
  {\bibfnamefont {J.}~\bibnamefont {Estève}},\ and\ \bibinfo {author}
  {\bibfnamefont {J.}~\bibnamefont {Reichel}},\ }\href
  {https://doi.org/10.1038/nature10225} {\bibfield  {journal} {\bibinfo
  {journal} {Nature}\ }\textbf {\bibinfo {volume} {475}},\ \bibinfo {pages}
  {210} (\bibinfo {year} {2011})}\BibitemShut {NoStop}%
\bibitem [{\citenamefont {Chou}\ \emph {et~al.}(2005)\citenamefont {Chou},
  \citenamefont {de~Riedmatten}, \citenamefont {Felinto}, \citenamefont
  {Polyakov}, \citenamefont {van Enk},\ and\ \citenamefont
  {Kimble}}]{chou_measurement-induced_2005}%
  \BibitemOpen
  \bibfield  {author} {\bibinfo {author} {\bibfnamefont {C.~W.}\ \bibnamefont
  {Chou}}, \bibinfo {author} {\bibfnamefont {H.}~\bibnamefont {de~Riedmatten}},
  \bibinfo {author} {\bibfnamefont {D.}~\bibnamefont {Felinto}}, \bibinfo
  {author} {\bibfnamefont {S.~V.}\ \bibnamefont {Polyakov}}, \bibinfo {author}
  {\bibfnamefont {S.~J.}\ \bibnamefont {van Enk}},\ and\ \bibinfo {author}
  {\bibfnamefont {H.~J.}\ \bibnamefont {Kimble}},\ }\href
  {https://doi.org/10.1038/nature04353} {\bibfield  {journal} {\bibinfo
  {journal} {Nature}\ }\textbf {\bibinfo {volume} {438}},\ \bibinfo {pages}
  {828} (\bibinfo {year} {2005})}\BibitemShut {NoStop}%
\bibitem [{\citenamefont {Stockill}\ \emph {et~al.}(2017)\citenamefont
  {Stockill}, \citenamefont {Stanley}, \citenamefont {Huthmacher},
  \citenamefont {Clarke}, \citenamefont {Hugues}, \citenamefont {Miller},
  \citenamefont {Matthiesen}, \citenamefont {Le~Gall},\ and\ \citenamefont
  {Atatüre}}]{stockill_phase-tuned_2017}%
  \BibitemOpen
  \bibfield  {author} {\bibinfo {author} {\bibfnamefont {R.}~\bibnamefont
  {Stockill}}, \bibinfo {author} {\bibfnamefont {M.}~\bibnamefont {Stanley}},
  \bibinfo {author} {\bibfnamefont {L.}~\bibnamefont {Huthmacher}}, \bibinfo
  {author} {\bibfnamefont {E.}~\bibnamefont {Clarke}}, \bibinfo {author}
  {\bibfnamefont {M.}~\bibnamefont {Hugues}}, \bibinfo {author} {\bibfnamefont
  {A.}~\bibnamefont {Miller}}, \bibinfo {author} {\bibfnamefont
  {C.}~\bibnamefont {Matthiesen}}, \bibinfo {author} {\bibfnamefont
  {C.}~\bibnamefont {Le~Gall}},\ and\ \bibinfo {author} {\bibfnamefont
  {M.}~\bibnamefont {Atatüre}},\ }\href
  {https://doi.org/10.1103/PhysRevLett.119.010503} {\bibfield  {journal}
  {\bibinfo  {journal} {Phys. Rev. Lett.}\ }\textbf {\bibinfo {volume} {119}},\
  \bibinfo {pages} {010503} (\bibinfo {year} {2017})}\BibitemShut {NoStop}%
\bibitem [{\citenamefont {Humphreys}\ \emph {et~al.}(2018)\citenamefont
  {Humphreys}, \citenamefont {Kalb}, \citenamefont {Morits}, \citenamefont
  {Schouten}, \citenamefont {Vermeulen}, \citenamefont {Twitchen},
  \citenamefont {Markham},\ and\ \citenamefont
  {Hanson}}]{humphreys_deterministic_2018}%
  \BibitemOpen
  \bibfield  {author} {\bibinfo {author} {\bibfnamefont {P.~C.}\ \bibnamefont
  {Humphreys}}, \bibinfo {author} {\bibfnamefont {N.}~\bibnamefont {Kalb}},
  \bibinfo {author} {\bibfnamefont {J.~P.~J.}\ \bibnamefont {Morits}}, \bibinfo
  {author} {\bibfnamefont {R.~N.}\ \bibnamefont {Schouten}}, \bibinfo {author}
  {\bibfnamefont {R.~F.~L.}\ \bibnamefont {Vermeulen}}, \bibinfo {author}
  {\bibfnamefont {D.~J.}\ \bibnamefont {Twitchen}}, \bibinfo {author}
  {\bibfnamefont {M.}~\bibnamefont {Markham}},\ and\ \bibinfo {author}
  {\bibfnamefont {R.}~\bibnamefont {Hanson}},\ }\href
  {https://doi.org/10.1038/s41586-018-0200-5} {\bibfield  {journal} {\bibinfo
  {journal} {Nature}\ }\textbf {\bibinfo {volume} {558}},\ \bibinfo {pages}
  {268} (\bibinfo {year} {2018})}\BibitemShut {NoStop}%
\bibitem [{\citenamefont {Rozpedek}\ \emph {et~al.}(2019)\citenamefont
  {Rozpedek}, \citenamefont {Yehia}, \citenamefont {Goodenough}, \citenamefont
  {Ruf}, \citenamefont {Humphreys}, \citenamefont {Hanson}, \citenamefont
  {Wehner},\ and\ \citenamefont {Elkouss}}]{rozpedek_near-term_2019}%
  \BibitemOpen
  \bibfield  {author} {\bibinfo {author} {\bibfnamefont {F.}~\bibnamefont
  {Rozpedek}}, \bibinfo {author} {\bibfnamefont {R.}~\bibnamefont {Yehia}},
  \bibinfo {author} {\bibfnamefont {K.}~\bibnamefont {Goodenough}}, \bibinfo
  {author} {\bibfnamefont {M.}~\bibnamefont {Ruf}}, \bibinfo {author}
  {\bibfnamefont {P.~C.}\ \bibnamefont {Humphreys}}, \bibinfo {author}
  {\bibfnamefont {R.}~\bibnamefont {Hanson}}, \bibinfo {author} {\bibfnamefont
  {S.}~\bibnamefont {Wehner}},\ and\ \bibinfo {author} {\bibfnamefont
  {D.}~\bibnamefont {Elkouss}},\ }\href
  {https://doi.org/10.1103/PhysRevA.99.052330} {\bibfield  {journal} {\bibinfo
  {journal} {Phys. Rev. A}\ }\textbf {\bibinfo {volume} {99}},\ \bibinfo
  {pages} {052330} (\bibinfo {year} {2019})}\BibitemShut {NoStop}%
\bibitem [{\citenamefont {Ruf}\ \emph {et~al.}(2021)\citenamefont {Ruf},
  \citenamefont {Wan}, \citenamefont {Choi}, \citenamefont {Englund},\ and\
  \citenamefont {Hanson}}]{ruf_quantum_2021}%
  \BibitemOpen
  \bibfield  {author} {\bibinfo {author} {\bibfnamefont {M.}~\bibnamefont
  {Ruf}}, \bibinfo {author} {\bibfnamefont {N.~H.}\ \bibnamefont {Wan}},
  \bibinfo {author} {\bibfnamefont {H.}~\bibnamefont {Choi}}, \bibinfo {author}
  {\bibfnamefont {D.}~\bibnamefont {Englund}},\ and\ \bibinfo {author}
  {\bibfnamefont {R.}~\bibnamefont {Hanson}},\ }\href
  {https://doi.org/10.1063/5.0056534} {\bibfield  {journal} {\bibinfo
  {journal} {J. Appl. Phys.}\ }\textbf {\bibinfo {volume} {130}},\ \bibinfo
  {pages} {070901} (\bibinfo {year} {2021})}\BibitemShut {NoStop}%
\bibitem [{\citenamefont {Awschalom}\ \emph {et~al.}(2018)\citenamefont
  {Awschalom}, \citenamefont {Hanson}, \citenamefont {Wrachtrup},\ and\
  \citenamefont {Zhou}}]{awschalom_quantum_2018}%
  \BibitemOpen
  \bibfield  {author} {\bibinfo {author} {\bibfnamefont {D.~D.}\ \bibnamefont
  {Awschalom}}, \bibinfo {author} {\bibfnamefont {R.}~\bibnamefont {Hanson}},
  \bibinfo {author} {\bibfnamefont {J.}~\bibnamefont {Wrachtrup}},\ and\
  \bibinfo {author} {\bibfnamefont {B.~B.}\ \bibnamefont {Zhou}},\ }\href
  {https://doi.org/10.1038/s41566-018-0232-2} {\bibfield  {journal} {\bibinfo
  {journal} {Nat. Photonics}\ }\textbf {\bibinfo {volume} {12}},\ \bibinfo
  {pages} {516} (\bibinfo {year} {2018})}\BibitemShut {NoStop}%
\bibitem [{\citenamefont {Togan}\ \emph {et~al.}(2010)\citenamefont {Togan},
  \citenamefont {Chu}, \citenamefont {Trifonov}, \citenamefont {Jiang},
  \citenamefont {Maze}, \citenamefont {Childress}, \citenamefont {Dutt},
  \citenamefont {Sørensen}, \citenamefont {Hemmer}, \citenamefont {Zibrov},\
  and\ \citenamefont {Lukin}}]{togan_quantum_2010}%
  \BibitemOpen
  \bibfield  {author} {\bibinfo {author} {\bibfnamefont {E.}~\bibnamefont
  {Togan}}, \bibinfo {author} {\bibfnamefont {Y.}~\bibnamefont {Chu}}, \bibinfo
  {author} {\bibfnamefont {A.~S.}\ \bibnamefont {Trifonov}}, \bibinfo {author}
  {\bibfnamefont {L.}~\bibnamefont {Jiang}}, \bibinfo {author} {\bibfnamefont
  {J.}~\bibnamefont {Maze}}, \bibinfo {author} {\bibfnamefont {L.}~\bibnamefont
  {Childress}}, \bibinfo {author} {\bibfnamefont {M.~V.~G.}\ \bibnamefont
  {Dutt}}, \bibinfo {author} {\bibfnamefont {A.~S.}\ \bibnamefont {Sørensen}},
  \bibinfo {author} {\bibfnamefont {P.~R.}\ \bibnamefont {Hemmer}}, \bibinfo
  {author} {\bibfnamefont {A.~S.}\ \bibnamefont {Zibrov}},\ and\ \bibinfo
  {author} {\bibfnamefont {M.~D.}\ \bibnamefont {Lukin}},\ }\href
  {https://doi.org/10.1038/nature09256} {\bibfield  {journal} {\bibinfo
  {journal} {Nature}\ }\textbf {\bibinfo {volume} {466}},\ \bibinfo {pages}
  {730} (\bibinfo {year} {2010})}\BibitemShut {NoStop}%
\bibitem [{\citenamefont {Nguyen}\ \emph
  {et~al.}(2019{\natexlab{a}})\citenamefont {Nguyen}, \citenamefont {Sukachev},
  \citenamefont {Bhaskar}, \citenamefont {Machielse}, \citenamefont {Levonian},
  \citenamefont {Knall}, \citenamefont {Stroganov}, \citenamefont {Chia},
  \citenamefont {Burek}, \citenamefont {Riedinger}, \citenamefont {Park},
  \citenamefont {Lončar},\ and\ \citenamefont
  {Lukin}}]{nguyen_integrated_2019}%
  \BibitemOpen
  \bibfield  {author} {\bibinfo {author} {\bibfnamefont {C.~T.}\ \bibnamefont
  {Nguyen}}, \bibinfo {author} {\bibfnamefont {D.~D.}\ \bibnamefont
  {Sukachev}}, \bibinfo {author} {\bibfnamefont {M.~K.}\ \bibnamefont
  {Bhaskar}}, \bibinfo {author} {\bibfnamefont {B.}~\bibnamefont {Machielse}},
  \bibinfo {author} {\bibfnamefont {D.~S.}\ \bibnamefont {Levonian}}, \bibinfo
  {author} {\bibfnamefont {E.~N.}\ \bibnamefont {Knall}}, \bibinfo {author}
  {\bibfnamefont {P.}~\bibnamefont {Stroganov}}, \bibinfo {author}
  {\bibfnamefont {C.}~\bibnamefont {Chia}}, \bibinfo {author} {\bibfnamefont
  {M.~J.}\ \bibnamefont {Burek}}, \bibinfo {author} {\bibfnamefont
  {R.}~\bibnamefont {Riedinger}}, \bibinfo {author} {\bibfnamefont
  {H.}~\bibnamefont {Park}}, \bibinfo {author} {\bibfnamefont {M.}~\bibnamefont
  {Lončar}},\ and\ \bibinfo {author} {\bibfnamefont {M.~D.}\ \bibnamefont
  {Lukin}},\ }\href {https://doi.org/10.1103/PhysRevB.100.165428} {\bibfield
  {journal} {\bibinfo  {journal} {Phys. Rev. B}\ }\textbf {\bibinfo {volume}
  {100}},\ \bibinfo {pages} {165428} (\bibinfo {year}
  {2019}{\natexlab{a}})}\BibitemShut {NoStop}%
\bibitem [{\citenamefont {Pompili}\ \emph {et~al.}(2021)\citenamefont
  {Pompili}, \citenamefont {Hermans}, \citenamefont {Baier}, \citenamefont
  {Beukers}, \citenamefont {Humphreys}, \citenamefont {Schouten}, \citenamefont
  {Vermeulen}, \citenamefont {Tiggelman}, \citenamefont {dos Santos~Martins},
  \citenamefont {Dirkse}, \citenamefont {Wehner},\ and\ \citenamefont
  {Hanson}}]{pompili_realization_2021}%
  \BibitemOpen
  \bibfield  {author} {\bibinfo {author} {\bibfnamefont {M.}~\bibnamefont
  {Pompili}}, \bibinfo {author} {\bibfnamefont {S.~L.~N.}\ \bibnamefont
  {Hermans}}, \bibinfo {author} {\bibfnamefont {S.}~\bibnamefont {Baier}},
  \bibinfo {author} {\bibfnamefont {H.~K.~C.}\ \bibnamefont {Beukers}},
  \bibinfo {author} {\bibfnamefont {P.~C.}\ \bibnamefont {Humphreys}}, \bibinfo
  {author} {\bibfnamefont {R.~N.}\ \bibnamefont {Schouten}}, \bibinfo {author}
  {\bibfnamefont {R.~F.~L.}\ \bibnamefont {Vermeulen}}, \bibinfo {author}
  {\bibfnamefont {M.~J.}\ \bibnamefont {Tiggelman}}, \bibinfo {author}
  {\bibfnamefont {L.}~\bibnamefont {dos Santos~Martins}}, \bibinfo {author}
  {\bibfnamefont {B.}~\bibnamefont {Dirkse}}, \bibinfo {author} {\bibfnamefont
  {S.}~\bibnamefont {Wehner}},\ and\ \bibinfo {author} {\bibfnamefont
  {R.}~\bibnamefont {Hanson}},\ }\href
  {https://doi.org/10.1126/science.abg1919} {\bibfield  {journal} {\bibinfo
  {journal} {Science}\ }\textbf {\bibinfo {volume} {372}},\ \bibinfo {pages}
  {259} (\bibinfo {year} {2021})}\BibitemShut {NoStop}%
\bibitem [{\citenamefont {Abobeih}\ \emph {et~al.}(2018)\citenamefont
  {Abobeih}, \citenamefont {Cramer}, \citenamefont {Bakker}, \citenamefont
  {Kalb}, \citenamefont {Markham}, \citenamefont {Twitchen},\ and\
  \citenamefont {Taminiau}}]{abobeih_one-second_2018}%
  \BibitemOpen
  \bibfield  {author} {\bibinfo {author} {\bibfnamefont {M.~H.}\ \bibnamefont
  {Abobeih}}, \bibinfo {author} {\bibfnamefont {J.}~\bibnamefont {Cramer}},
  \bibinfo {author} {\bibfnamefont {M.~A.}\ \bibnamefont {Bakker}}, \bibinfo
  {author} {\bibfnamefont {N.}~\bibnamefont {Kalb}}, \bibinfo {author}
  {\bibfnamefont {M.}~\bibnamefont {Markham}}, \bibinfo {author} {\bibfnamefont
  {D.~J.}\ \bibnamefont {Twitchen}},\ and\ \bibinfo {author} {\bibfnamefont
  {T.~H.}\ \bibnamefont {Taminiau}},\ }\href
  {https://doi.org/10.1038/s41467-018-04916-z} {\bibfield  {journal} {\bibinfo
  {journal} {Nat. Commun.}\ }\textbf {\bibinfo {volume} {9}},\ \bibinfo {pages}
  {2552} (\bibinfo {year} {2018})}\BibitemShut {NoStop}%
\bibitem [{\citenamefont {Dutt}\ \emph {et~al.}(2007)\citenamefont {Dutt},
  \citenamefont {Childress}, \citenamefont {Jiang}, \citenamefont {Togan},
  \citenamefont {Maze}, \citenamefont {Jelezko}, \citenamefont {Zibrov},
  \citenamefont {Hemmer},\ and\ \citenamefont {Lukin}}]{dutt2007}%
  \BibitemOpen
  \bibfield  {author} {\bibinfo {author} {\bibfnamefont {M.~G.}\ \bibnamefont
  {Dutt}}, \bibinfo {author} {\bibfnamefont {L.}~\bibnamefont {Childress}},
  \bibinfo {author} {\bibfnamefont {L.}~\bibnamefont {Jiang}}, \bibinfo
  {author} {\bibfnamefont {E.}~\bibnamefont {Togan}}, \bibinfo {author}
  {\bibfnamefont {J.}~\bibnamefont {Maze}}, \bibinfo {author} {\bibfnamefont
  {F.}~\bibnamefont {Jelezko}}, \bibinfo {author} {\bibfnamefont
  {A.}~\bibnamefont {Zibrov}}, \bibinfo {author} {\bibfnamefont
  {P.}~\bibnamefont {Hemmer}},\ and\ \bibinfo {author} {\bibfnamefont
  {M.}~\bibnamefont {Lukin}},\ }\href@noop {} {\bibfield  {journal} {\bibinfo
  {journal} {Science}\ }\textbf {\bibinfo {volume} {316}},\ \bibinfo {pages}
  {1312} (\bibinfo {year} {2007})}\BibitemShut {NoStop}%
\bibitem [{\citenamefont {Fuchs}\ \emph {et~al.}(2011)\citenamefont {Fuchs},
  \citenamefont {Burkard}, \citenamefont {Klimov},\ and\ \citenamefont
  {Awschalom}}]{fuchs2011}%
  \BibitemOpen
  \bibfield  {author} {\bibinfo {author} {\bibfnamefont {G.}~\bibnamefont
  {Fuchs}}, \bibinfo {author} {\bibfnamefont {G.}~\bibnamefont {Burkard}},
  \bibinfo {author} {\bibfnamefont {P.}~\bibnamefont {Klimov}},\ and\ \bibinfo
  {author} {\bibfnamefont {D.}~\bibnamefont {Awschalom}},\ }\href@noop {}
  {\bibfield  {journal} {\bibinfo  {journal} {Nat. Phys.}\ }\textbf {\bibinfo
  {volume} {7}},\ \bibinfo {pages} {789} (\bibinfo {year} {2011})}\BibitemShut
  {NoStop}%
\bibitem [{\citenamefont {Kalb}\ \emph {et~al.}(2017)\citenamefont {Kalb},
  \citenamefont {Reiserer}, \citenamefont {Humphreys}, \citenamefont
  {Bakermans}, \citenamefont {Kamerling}, \citenamefont {Nickerson},
  \citenamefont {Benjamin}, \citenamefont {Twitchen}, \citenamefont {Markham},\
  and\ \citenamefont {Hanson}}]{kalb_entanglement_2017}%
  \BibitemOpen
  \bibfield  {author} {\bibinfo {author} {\bibfnamefont {N.}~\bibnamefont
  {Kalb}}, \bibinfo {author} {\bibfnamefont {A.~A.}\ \bibnamefont {Reiserer}},
  \bibinfo {author} {\bibfnamefont {P.~C.}\ \bibnamefont {Humphreys}}, \bibinfo
  {author} {\bibfnamefont {J.~J.~W.}\ \bibnamefont {Bakermans}}, \bibinfo
  {author} {\bibfnamefont {S.~J.}\ \bibnamefont {Kamerling}}, \bibinfo {author}
  {\bibfnamefont {N.~H.}\ \bibnamefont {Nickerson}}, \bibinfo {author}
  {\bibfnamefont {S.~C.}\ \bibnamefont {Benjamin}}, \bibinfo {author}
  {\bibfnamefont {D.~J.}\ \bibnamefont {Twitchen}}, \bibinfo {author}
  {\bibfnamefont {M.}~\bibnamefont {Markham}},\ and\ \bibinfo {author}
  {\bibfnamefont {R.}~\bibnamefont {Hanson}},\ }\href
  {https://doi.org/10.1126/science.aan0070} {\bibfield  {journal} {\bibinfo
  {journal} {Science}\ }\textbf {\bibinfo {volume} {356}},\ \bibinfo {pages}
  {928} (\bibinfo {year} {2017})}\BibitemShut {NoStop}%
\bibitem [{\citenamefont {Nguyen}\ \emph
  {et~al.}(2019{\natexlab{b}})\citenamefont {Nguyen}, \citenamefont {Sukachev},
  \citenamefont {Bhaskar}, \citenamefont {Machielse}, \citenamefont {Levonian},
  \citenamefont {Knall}, \citenamefont {Stroganov}, \citenamefont {Riedinger},
  \citenamefont {Park}, \citenamefont {Lončar},\ and\ \citenamefont
  {Lukin}}]{nguyen_quantum_2019}%
  \BibitemOpen
  \bibfield  {author} {\bibinfo {author} {\bibfnamefont {C.}~\bibnamefont
  {Nguyen}}, \bibinfo {author} {\bibfnamefont {D.}~\bibnamefont {Sukachev}},
  \bibinfo {author} {\bibfnamefont {M.}~\bibnamefont {Bhaskar}}, \bibinfo
  {author} {\bibfnamefont {B.}~\bibnamefont {Machielse}}, \bibinfo {author}
  {\bibfnamefont {D.}~\bibnamefont {Levonian}}, \bibinfo {author}
  {\bibfnamefont {E.}~\bibnamefont {Knall}}, \bibinfo {author} {\bibfnamefont
  {P.}~\bibnamefont {Stroganov}}, \bibinfo {author} {\bibfnamefont
  {R.}~\bibnamefont {Riedinger}}, \bibinfo {author} {\bibfnamefont
  {H.}~\bibnamefont {Park}}, \bibinfo {author} {\bibfnamefont {M.}~\bibnamefont
  {Lončar}},\ and\ \bibinfo {author} {\bibfnamefont {M.}~\bibnamefont
  {Lukin}},\ }\href {https://doi.org/10.1103/PhysRevLett.123.183602} {\bibfield
   {journal} {\bibinfo  {journal} {Phys. Rev. Lett.}\ }\textbf {\bibinfo
  {volume} {123}},\ \bibinfo {pages} {183602} (\bibinfo {year}
  {2019}{\natexlab{b}})}\BibitemShut {NoStop}%
\bibitem [{\citenamefont {Gorshkov}\ \emph {et~al.}(2007)\citenamefont
  {Gorshkov}, \citenamefont {André}, \citenamefont {Lukin},\ and\
  \citenamefont {Sørensen}}]{gorshkov_photon_2007}%
  \BibitemOpen
  \bibfield  {author} {\bibinfo {author} {\bibfnamefont {A.~V.}\ \bibnamefont
  {Gorshkov}}, \bibinfo {author} {\bibfnamefont {A.}~\bibnamefont {André}},
  \bibinfo {author} {\bibfnamefont {M.~D.}\ \bibnamefont {Lukin}},\ and\
  \bibinfo {author} {\bibfnamefont {A.~S.}\ \bibnamefont {Sørensen}},\ }\href
  {https://doi.org/10.1103/PhysRevA.76.033804} {\bibfield  {journal} {\bibinfo
  {journal} {Phys. Rev. A}\ }\textbf {\bibinfo {volume} {76}},\ \bibinfo
  {pages} {033804} (\bibinfo {year} {2007})}\BibitemShut {NoStop}%
\bibitem [{\citenamefont {Dilley}\ \emph {et~al.}(2012)\citenamefont {Dilley},
  \citenamefont {Nisbet-Jones}, \citenamefont {Shore},\ and\ \citenamefont
  {Kuhn}}]{dilley_single-photon_2012}%
  \BibitemOpen
  \bibfield  {author} {\bibinfo {author} {\bibfnamefont {J.}~\bibnamefont
  {Dilley}}, \bibinfo {author} {\bibfnamefont {P.}~\bibnamefont
  {Nisbet-Jones}}, \bibinfo {author} {\bibfnamefont {B.~W.}\ \bibnamefont
  {Shore}},\ and\ \bibinfo {author} {\bibfnamefont {A.}~\bibnamefont {Kuhn}},\
  }\href {https://doi.org/10.1103/PhysRevA.85.023834} {\bibfield  {journal}
  {\bibinfo  {journal} {Phys. Rev. A}\ }\textbf {\bibinfo {volume} {85}},\
  \bibinfo {pages} {023834} (\bibinfo {year} {2012})}\BibitemShut {NoStop}%
\bibitem [{\citenamefont {M\"ucke}\ \emph {et~al.}(2013)\citenamefont
  {M\"ucke}, \citenamefont {Bochmann}, \citenamefont {Hahn}, \citenamefont
  {Neuzner}, \citenamefont {Nölleke}, \citenamefont {Reiserer}, \citenamefont
  {Rempe},\ and\ \citenamefont {Ritter}}]{mucke_generation_2013}%
  \BibitemOpen
  \bibfield  {author} {\bibinfo {author} {\bibfnamefont {M.}~\bibnamefont
  {M\"ucke}}, \bibinfo {author} {\bibfnamefont {J.}~\bibnamefont {Bochmann}},
  \bibinfo {author} {\bibfnamefont {C.}~\bibnamefont {Hahn}}, \bibinfo {author}
  {\bibfnamefont {A.}~\bibnamefont {Neuzner}}, \bibinfo {author} {\bibfnamefont
  {C.}~\bibnamefont {Nölleke}}, \bibinfo {author} {\bibfnamefont
  {A.}~\bibnamefont {Reiserer}}, \bibinfo {author} {\bibfnamefont
  {G.}~\bibnamefont {Rempe}},\ and\ \bibinfo {author} {\bibfnamefont
  {S.}~\bibnamefont {Ritter}},\ }\href
  {https://doi.org/10.1103/PhysRevA.87.063805} {\bibfield  {journal} {\bibinfo
  {journal} {Phys. Rev. A}\ }\textbf {\bibinfo {volume} {87}},\ \bibinfo
  {pages} {063805} (\bibinfo {year} {2013})}\BibitemShut {NoStop}%
\bibitem [{\citenamefont {Morin}\ \emph {et~al.}(2019)\citenamefont {Morin},
  \citenamefont {Körber}, \citenamefont {Langenfeld},\ and\ \citenamefont
  {Rempe}}]{morin_deterministic_2019}%
  \BibitemOpen
  \bibfield  {author} {\bibinfo {author} {\bibfnamefont {O.}~\bibnamefont
  {Morin}}, \bibinfo {author} {\bibfnamefont {M.}~\bibnamefont {Körber}},
  \bibinfo {author} {\bibfnamefont {S.}~\bibnamefont {Langenfeld}},\ and\
  \bibinfo {author} {\bibfnamefont {G.}~\bibnamefont {Rempe}},\ }\href
  {https://doi.org/10.1103/PhysRevLett.123.133602} {\bibfield  {journal}
  {\bibinfo  {journal} {Phys. Rev. Lett.}\ }\textbf {\bibinfo {volume} {123}},\
  \bibinfo {pages} {133602} (\bibinfo {year} {2019})}\BibitemShut {NoStop}%
\bibitem [{\citenamefont {Barclay}\ \emph {et~al.}(2009)\citenamefont
  {Barclay}, \citenamefont {Fu}, \citenamefont {Santori},\ and\ \citenamefont
  {Beausoleil}}]{barclay_chip-based_2009}%
  \BibitemOpen
  \bibfield  {author} {\bibinfo {author} {\bibfnamefont {P.~E.}\ \bibnamefont
  {Barclay}}, \bibinfo {author} {\bibfnamefont {K.-M.~C.}\ \bibnamefont {Fu}},
  \bibinfo {author} {\bibfnamefont {C.}~\bibnamefont {Santori}},\ and\ \bibinfo
  {author} {\bibfnamefont {R.~G.}\ \bibnamefont {Beausoleil}},\ }\href
  {https://doi.org/10.1063/1.3262948} {\bibfield  {journal} {\bibinfo
  {journal} {Appl. Phys. Lett.}\ }\textbf {\bibinfo {volume} {95}},\ \bibinfo
  {pages} {191115} (\bibinfo {year} {2009})}\BibitemShut {NoStop}%
\bibitem [{\citenamefont {Faraon}\ \emph {et~al.}(2012)\citenamefont {Faraon},
  \citenamefont {Santori}, \citenamefont {Huang}, \citenamefont {Acosta},\ and\
  \citenamefont {Beausoleil}}]{faraon_coupling_2012}%
  \BibitemOpen
  \bibfield  {author} {\bibinfo {author} {\bibfnamefont {A.}~\bibnamefont
  {Faraon}}, \bibinfo {author} {\bibfnamefont {C.}~\bibnamefont {Santori}},
  \bibinfo {author} {\bibfnamefont {Z.}~\bibnamefont {Huang}}, \bibinfo
  {author} {\bibfnamefont {V.~M.}\ \bibnamefont {Acosta}},\ and\ \bibinfo
  {author} {\bibfnamefont {R.~G.}\ \bibnamefont {Beausoleil}},\ }\href
  {https://doi.org/10.1103/PhysRevLett.109.033604} {\bibfield  {journal}
  {\bibinfo  {journal} {Phys. Rev. Lett.}\ }\textbf {\bibinfo {volume} {109}},\
  \bibinfo {pages} {033604} (\bibinfo {year} {2012})}\BibitemShut {NoStop}%
\bibitem [{\citenamefont {Li}\ \emph {et~al.}(2015)\citenamefont {Li},
  \citenamefont {Schröder}, \citenamefont {Chen}, \citenamefont {Walsh},
  \citenamefont {Bayn}, \citenamefont {Goldstein}, \citenamefont {Gaathon},
  \citenamefont {Trusheim}, \citenamefont {Lu}, \citenamefont {Mower},
  \citenamefont {Cotlet}, \citenamefont {Markham}, \citenamefont {Twitchen},\
  and\ \citenamefont {Englund}}]{li_coherent_2015}%
  \BibitemOpen
  \bibfield  {author} {\bibinfo {author} {\bibfnamefont {L.}~\bibnamefont
  {Li}}, \bibinfo {author} {\bibfnamefont {T.}~\bibnamefont {Schröder}},
  \bibinfo {author} {\bibfnamefont {E.~H.}\ \bibnamefont {Chen}}, \bibinfo
  {author} {\bibfnamefont {M.}~\bibnamefont {Walsh}}, \bibinfo {author}
  {\bibfnamefont {I.}~\bibnamefont {Bayn}}, \bibinfo {author} {\bibfnamefont
  {J.}~\bibnamefont {Goldstein}}, \bibinfo {author} {\bibfnamefont
  {O.}~\bibnamefont {Gaathon}}, \bibinfo {author} {\bibfnamefont {M.~E.}\
  \bibnamefont {Trusheim}}, \bibinfo {author} {\bibfnamefont {M.}~\bibnamefont
  {Lu}}, \bibinfo {author} {\bibfnamefont {J.}~\bibnamefont {Mower}}, \bibinfo
  {author} {\bibfnamefont {M.}~\bibnamefont {Cotlet}}, \bibinfo {author}
  {\bibfnamefont {M.~L.}\ \bibnamefont {Markham}}, \bibinfo {author}
  {\bibfnamefont {D.~J.}\ \bibnamefont {Twitchen}},\ and\ \bibinfo {author}
  {\bibfnamefont {D.}~\bibnamefont {Englund}},\ }\href
  {https://doi.org/10.1038/ncomms7173} {\bibfield  {journal} {\bibinfo
  {journal} {Nat. Commun.}\ }\textbf {\bibinfo {volume} {6}},\ \bibinfo {pages}
  {6173} (\bibinfo {year} {2015})}\BibitemShut {NoStop}%
\bibitem [{\citenamefont {Riedel}\ \emph {et~al.}(2017)\citenamefont {Riedel},
  \citenamefont {Söllner}, \citenamefont {Shields}, \citenamefont
  {Starosielec}, \citenamefont {Appel}, \citenamefont {Neu}, \citenamefont
  {Maletinsky},\ and\ \citenamefont {Warburton}}]{riedel_deterministic_2017}%
  \BibitemOpen
  \bibfield  {author} {\bibinfo {author} {\bibfnamefont {D.}~\bibnamefont
  {Riedel}}, \bibinfo {author} {\bibfnamefont {I.}~\bibnamefont {Söllner}},
  \bibinfo {author} {\bibfnamefont {B.~J.}\ \bibnamefont {Shields}}, \bibinfo
  {author} {\bibfnamefont {S.}~\bibnamefont {Starosielec}}, \bibinfo {author}
  {\bibfnamefont {P.}~\bibnamefont {Appel}}, \bibinfo {author} {\bibfnamefont
  {E.}~\bibnamefont {Neu}}, \bibinfo {author} {\bibfnamefont {P.}~\bibnamefont
  {Maletinsky}},\ and\ \bibinfo {author} {\bibfnamefont {R.~J.}\ \bibnamefont
  {Warburton}},\ }\href@noop {} {\bibfield  {journal} {\bibinfo  {journal}
  {Phys. Rev. X}\ }\textbf {\bibinfo {volume} {7}},\ \bibinfo {pages} {031040}
  (\bibinfo {year} {2017})}\BibitemShut {NoStop}%
\bibitem [{\citenamefont {Bassett}\ \emph {et~al.}(2011)\citenamefont
  {Bassett}, \citenamefont {Heremans}, \citenamefont {Yale}, \citenamefont
  {Buckley},\ and\ \citenamefont {Awschalom}}]{bassett_electrical_2011}%
  \BibitemOpen
  \bibfield  {author} {\bibinfo {author} {\bibfnamefont {L.~C.}\ \bibnamefont
  {Bassett}}, \bibinfo {author} {\bibfnamefont {F.~J.}\ \bibnamefont
  {Heremans}}, \bibinfo {author} {\bibfnamefont {C.~G.}\ \bibnamefont {Yale}},
  \bibinfo {author} {\bibfnamefont {B.~B.}\ \bibnamefont {Buckley}},\ and\
  \bibinfo {author} {\bibfnamefont {D.~D.}\ \bibnamefont {Awschalom}},\ }\href
  {https://doi.org/10.1103/PhysRevLett.107.266403} {\bibfield  {journal}
  {\bibinfo  {journal} {Phys. Rev. Lett.}\ }\textbf {\bibinfo {volume} {107}},\
  \bibinfo {pages} {266403} (\bibinfo {year} {2011})}\BibitemShut {NoStop}%
\bibitem [{\citenamefont {Bernien}\ \emph {et~al.}(2013)\citenamefont
  {Bernien}, \citenamefont {Hensen}, \citenamefont {Pfaff}, \citenamefont
  {Koolstra}, \citenamefont {Blok}, \citenamefont {Robledo}, \citenamefont
  {Taminiau}, \citenamefont {Markham}, \citenamefont {Twitchen}, \citenamefont
  {Childress},\ and\ \citenamefont {Hanson}}]{bernien_heralded_2013}%
  \BibitemOpen
  \bibfield  {author} {\bibinfo {author} {\bibfnamefont {H.}~\bibnamefont
  {Bernien}}, \bibinfo {author} {\bibfnamefont {B.}~\bibnamefont {Hensen}},
  \bibinfo {author} {\bibfnamefont {W.}~\bibnamefont {Pfaff}}, \bibinfo
  {author} {\bibfnamefont {G.}~\bibnamefont {Koolstra}}, \bibinfo {author}
  {\bibfnamefont {M.~S.}\ \bibnamefont {Blok}}, \bibinfo {author}
  {\bibfnamefont {L.}~\bibnamefont {Robledo}}, \bibinfo {author} {\bibfnamefont
  {T.~H.}\ \bibnamefont {Taminiau}}, \bibinfo {author} {\bibfnamefont
  {M.}~\bibnamefont {Markham}}, \bibinfo {author} {\bibfnamefont {D.~J.}\
  \bibnamefont {Twitchen}}, \bibinfo {author} {\bibfnamefont {L.}~\bibnamefont
  {Childress}},\ and\ \bibinfo {author} {\bibfnamefont {R.}~\bibnamefont
  {Hanson}},\ }\href {https://doi.org/10.1038/nature12016} {\bibfield
  {journal} {\bibinfo  {journal} {Nature}\ }\textbf {\bibinfo {volume} {497}},\
  \bibinfo {pages} {86} (\bibinfo {year} {2013})}\BibitemShut {NoStop}%
\bibitem [{\citenamefont {Orphal-Kobin}\ \emph {et~al.}(2023)\citenamefont
  {Orphal-Kobin}, \citenamefont {Unterguggenberger}, \citenamefont
  {Pregnolato}, \citenamefont {Kemf}, \citenamefont {Matalla}, \citenamefont
  {Unger}, \citenamefont {Ostermay}, \citenamefont {Pieplow},\ and\
  \citenamefont {Schröder}}]{orphal-kobin_optically_2023}%
  \BibitemOpen
  \bibfield  {author} {\bibinfo {author} {\bibfnamefont {L.}~\bibnamefont
  {Orphal-Kobin}}, \bibinfo {author} {\bibfnamefont {K.}~\bibnamefont
  {Unterguggenberger}}, \bibinfo {author} {\bibfnamefont {T.}~\bibnamefont
  {Pregnolato}}, \bibinfo {author} {\bibfnamefont {N.}~\bibnamefont {Kemf}},
  \bibinfo {author} {\bibfnamefont {M.}~\bibnamefont {Matalla}}, \bibinfo
  {author} {\bibfnamefont {R.-S.}\ \bibnamefont {Unger}}, \bibinfo {author}
  {\bibfnamefont {I.}~\bibnamefont {Ostermay}}, \bibinfo {author}
  {\bibfnamefont {G.}~\bibnamefont {Pieplow}},\ and\ \bibinfo {author}
  {\bibfnamefont {T.}~\bibnamefont {Schröder}},\ }\href
  {https://doi.org/10.1103/PhysRevX.13.011042} {\bibfield  {journal} {\bibinfo
  {journal} {Phys. Rev. X}\ }\textbf {\bibinfo {volume} {13}},\ \bibinfo
  {pages} {011042} (\bibinfo {year} {2023})}\BibitemShut {NoStop}%
\bibitem [{\citenamefont {Meesala}\ \emph {et~al.}(2018)\citenamefont
  {Meesala}, \citenamefont {Sohn}, \citenamefont {Pingault}, \citenamefont
  {Shao}, \citenamefont {Atikian}, \citenamefont {Holzgrafe}, \citenamefont
  {Gündoğan}, \citenamefont {Stavrakas}, \citenamefont {Sipahigil},
  \citenamefont {Chia}, \citenamefont {Evans}, \citenamefont {Burek},
  \citenamefont {Zhang}, \citenamefont {Wu}, \citenamefont {Pacheco},
  \citenamefont {Abraham}, \citenamefont {Bielejec}, \citenamefont {Lukin},
  \citenamefont {Atatüre},\ and\ \citenamefont
  {Lončar}}]{meesala_strain_2018}%
  \BibitemOpen
  \bibfield  {author} {\bibinfo {author} {\bibfnamefont {S.}~\bibnamefont
  {Meesala}}, \bibinfo {author} {\bibfnamefont {Y.-I.}\ \bibnamefont {Sohn}},
  \bibinfo {author} {\bibfnamefont {B.}~\bibnamefont {Pingault}}, \bibinfo
  {author} {\bibfnamefont {L.}~\bibnamefont {Shao}}, \bibinfo {author}
  {\bibfnamefont {H.~A.}\ \bibnamefont {Atikian}}, \bibinfo {author}
  {\bibfnamefont {J.}~\bibnamefont {Holzgrafe}}, \bibinfo {author}
  {\bibfnamefont {M.}~\bibnamefont {Gündoğan}}, \bibinfo {author}
  {\bibfnamefont {C.}~\bibnamefont {Stavrakas}}, \bibinfo {author}
  {\bibfnamefont {A.}~\bibnamefont {Sipahigil}}, \bibinfo {author}
  {\bibfnamefont {C.}~\bibnamefont {Chia}}, \bibinfo {author} {\bibfnamefont
  {R.}~\bibnamefont {Evans}}, \bibinfo {author} {\bibfnamefont {M.~J.}\
  \bibnamefont {Burek}}, \bibinfo {author} {\bibfnamefont {M.}~\bibnamefont
  {Zhang}}, \bibinfo {author} {\bibfnamefont {L.}~\bibnamefont {Wu}}, \bibinfo
  {author} {\bibfnamefont {J.~L.}\ \bibnamefont {Pacheco}}, \bibinfo {author}
  {\bibfnamefont {J.}~\bibnamefont {Abraham}}, \bibinfo {author} {\bibfnamefont
  {E.}~\bibnamefont {Bielejec}}, \bibinfo {author} {\bibfnamefont {M.~D.}\
  \bibnamefont {Lukin}}, \bibinfo {author} {\bibfnamefont {M.}~\bibnamefont
  {Atatüre}},\ and\ \bibinfo {author} {\bibfnamefont {M.}~\bibnamefont
  {Lončar}},\ }\href {https://doi.org/10.1103/PhysRevB.97.205444} {\bibfield
  {journal} {\bibinfo  {journal} {Phys. Rev. B}\ }\textbf {\bibinfo {volume}
  {97}},\ \bibinfo {pages} {205444} (\bibinfo {year} {2018})}\BibitemShut
  {NoStop}%
\bibitem [{\citenamefont {Metsch}\ \emph {et~al.}(2019)\citenamefont {Metsch},
  \citenamefont {Senkalla}, \citenamefont {Tratzmiller}, \citenamefont
  {Scheuer}, \citenamefont {Kern}, \citenamefont {Achard}, \citenamefont
  {Tallaire}, \citenamefont {Plenio}, \citenamefont {Siyushev},\ and\
  \citenamefont {Jelezko}}]{metsch_initialization_2019}%
  \BibitemOpen
  \bibfield  {author} {\bibinfo {author} {\bibfnamefont {M.~H.}\ \bibnamefont
  {Metsch}}, \bibinfo {author} {\bibfnamefont {K.}~\bibnamefont {Senkalla}},
  \bibinfo {author} {\bibfnamefont {B.}~\bibnamefont {Tratzmiller}}, \bibinfo
  {author} {\bibfnamefont {J.}~\bibnamefont {Scheuer}}, \bibinfo {author}
  {\bibfnamefont {M.}~\bibnamefont {Kern}}, \bibinfo {author} {\bibfnamefont
  {J.}~\bibnamefont {Achard}}, \bibinfo {author} {\bibfnamefont
  {A.}~\bibnamefont {Tallaire}}, \bibinfo {author} {\bibfnamefont {M.~B.}\
  \bibnamefont {Plenio}}, \bibinfo {author} {\bibfnamefont {P.}~\bibnamefont
  {Siyushev}},\ and\ \bibinfo {author} {\bibfnamefont {F.}~\bibnamefont
  {Jelezko}},\ }\href {https://doi.org/10.1103/PhysRevLett.122.190503}
  {\bibfield  {journal} {\bibinfo  {journal} {Phys. Rev. Lett.}\ }\textbf
  {\bibinfo {volume} {122}},\ \bibinfo {pages} {190503} (\bibinfo {year}
  {2019})}\BibitemShut {NoStop}%
\bibitem [{\citenamefont {Bhaskar}\ \emph {et~al.}(2020)\citenamefont
  {Bhaskar}, \citenamefont {Riedinger}, \citenamefont {Machielse},
  \citenamefont {Levonian}, \citenamefont {Nguyen}, \citenamefont {Knall},
  \citenamefont {Park}, \citenamefont {Englund}, \citenamefont {Lončar},
  \citenamefont {Sukachev},\ and\ \citenamefont
  {Lukin}}]{bhaskar_experimental_2020}%
  \BibitemOpen
  \bibfield  {author} {\bibinfo {author} {\bibfnamefont {M.~K.}\ \bibnamefont
  {Bhaskar}}, \bibinfo {author} {\bibfnamefont {R.}~\bibnamefont {Riedinger}},
  \bibinfo {author} {\bibfnamefont {B.}~\bibnamefont {Machielse}}, \bibinfo
  {author} {\bibfnamefont {D.~S.}\ \bibnamefont {Levonian}}, \bibinfo {author}
  {\bibfnamefont {C.~T.}\ \bibnamefont {Nguyen}}, \bibinfo {author}
  {\bibfnamefont {E.~N.}\ \bibnamefont {Knall}}, \bibinfo {author}
  {\bibfnamefont {H.}~\bibnamefont {Park}}, \bibinfo {author} {\bibfnamefont
  {D.}~\bibnamefont {Englund}}, \bibinfo {author} {\bibfnamefont
  {M.}~\bibnamefont {Lončar}}, \bibinfo {author} {\bibfnamefont {D.~D.}\
  \bibnamefont {Sukachev}},\ and\ \bibinfo {author} {\bibfnamefont {M.~D.}\
  \bibnamefont {Lukin}},\ }\href {https://doi.org/10.1038/s41586-020-2103-5}
  {\bibfield  {journal} {\bibinfo  {journal} {Nature}\ }\textbf {\bibinfo
  {volume} {580}},\ \bibinfo {pages} {60} (\bibinfo {year} {2020})}\BibitemShut
  {NoStop}%
\bibitem [{\citenamefont {Bersin}\ \emph {et~al.}(2024)\citenamefont {Bersin},
  \citenamefont {Sutula}, \citenamefont {Huan}, \citenamefont {Suleymanzade},
  \citenamefont {Assumpcao}, \citenamefont {Wei}, \citenamefont {Stas},
  \citenamefont {Knaut}, \citenamefont {Knall}, \citenamefont {Langrock},
  \citenamefont {Sinclair}, \citenamefont {Murphy}, \citenamefont {Riedinger},
  \citenamefont {Yeh}, \citenamefont {Xin}, \citenamefont {Bandyopadhyay},
  \citenamefont {Sukachev}, \citenamefont {Machielse}, \citenamefont
  {Levonian}, \citenamefont {Bhaskar}, \citenamefont {Hamilton}, \citenamefont
  {Park}, \citenamefont {Lon{\v{c}}ar}, \citenamefont {Fejer}, \citenamefont
  {Dixon}, \citenamefont {Englund},\ and\ \citenamefont
  {Lukin}}]{bersin_telecom_2023}%
  \BibitemOpen
  \bibfield  {author} {\bibinfo {author} {\bibfnamefont {E.}~\bibnamefont
  {Bersin}}, \bibinfo {author} {\bibfnamefont {M.}~\bibnamefont {Sutula}},
  \bibinfo {author} {\bibfnamefont {Y.~Q.}\ \bibnamefont {Huan}}, \bibinfo
  {author} {\bibfnamefont {A.}~\bibnamefont {Suleymanzade}}, \bibinfo {author}
  {\bibfnamefont {D.~R.}\ \bibnamefont {Assumpcao}}, \bibinfo {author}
  {\bibfnamefont {Y.-C.}\ \bibnamefont {Wei}}, \bibinfo {author} {\bibfnamefont
  {P.-J.}\ \bibnamefont {Stas}}, \bibinfo {author} {\bibfnamefont {C.~M.}\
  \bibnamefont {Knaut}}, \bibinfo {author} {\bibfnamefont {E.~N.}\ \bibnamefont
  {Knall}}, \bibinfo {author} {\bibfnamefont {C.}~\bibnamefont {Langrock}},
  \bibinfo {author} {\bibfnamefont {N.}~\bibnamefont {Sinclair}}, \bibinfo
  {author} {\bibfnamefont {R.}~\bibnamefont {Murphy}}, \bibinfo {author}
  {\bibfnamefont {R.}~\bibnamefont {Riedinger}}, \bibinfo {author}
  {\bibfnamefont {M.}~\bibnamefont {Yeh}}, \bibinfo {author} {\bibfnamefont
  {C.}~\bibnamefont {Xin}}, \bibinfo {author} {\bibfnamefont {S.}~\bibnamefont
  {Bandyopadhyay}}, \bibinfo {author} {\bibfnamefont {D.~D.}\ \bibnamefont
  {Sukachev}}, \bibinfo {author} {\bibfnamefont {B.}~\bibnamefont {Machielse}},
  \bibinfo {author} {\bibfnamefont {D.~S.}\ \bibnamefont {Levonian}}, \bibinfo
  {author} {\bibfnamefont {M.~K.}\ \bibnamefont {Bhaskar}}, \bibinfo {author}
  {\bibfnamefont {S.}~\bibnamefont {Hamilton}}, \bibinfo {author}
  {\bibfnamefont {H.}~\bibnamefont {Park}}, \bibinfo {author} {\bibfnamefont
  {M.}~\bibnamefont {Lon{\v{c}}ar}}, \bibinfo {author} {\bibfnamefont {M.~M.}\
  \bibnamefont {Fejer}}, \bibinfo {author} {\bibfnamefont {P.~B.}\ \bibnamefont
  {Dixon}}, \bibinfo {author} {\bibfnamefont {D.~R.}\ \bibnamefont {Englund}},\
  and\ \bibinfo {author} {\bibfnamefont {M.~D.}\ \bibnamefont {Lukin}},\ }\href
  {https://doi.org/10.1103/prxquantum.5.010303} {\bibfield  {journal} {\bibinfo
   {journal} {PRX Quantum}\ }\textbf {\bibinfo {volume} {5}},\ \bibinfo {pages}
  {010303} (\bibinfo {year} {2024})}\BibitemShut {NoStop}%
\bibitem [{\citenamefont {Barrett}\ and\ \citenamefont
  {Kok}(2005)}]{barrett_efficient_2005}%
  \BibitemOpen
  \bibfield  {author} {\bibinfo {author} {\bibfnamefont {S.~D.}\ \bibnamefont
  {Barrett}}\ and\ \bibinfo {author} {\bibfnamefont {P.}~\bibnamefont {Kok}},\
  }\href {https://doi.org/10.1103/PhysRevA.71.060310} {\bibfield  {journal}
  {\bibinfo  {journal} {Phys. Rev. A}\ }\textbf {\bibinfo {volume} {71}},\
  \bibinfo {pages} {060310} (\bibinfo {year} {2005})}\BibitemShut {NoStop}%
\bibitem [{\citenamefont {Hermans}\ \emph {et~al.}(2022)\citenamefont
  {Hermans}, \citenamefont {Pompili}, \citenamefont {Beukers}, \citenamefont
  {Baier}, \citenamefont {Borregaard},\ and\ \citenamefont
  {Hanson}}]{hermans_qubit_2022}%
  \BibitemOpen
  \bibfield  {author} {\bibinfo {author} {\bibfnamefont {S.~L.~N.}\
  \bibnamefont {Hermans}}, \bibinfo {author} {\bibfnamefont {M.}~\bibnamefont
  {Pompili}}, \bibinfo {author} {\bibfnamefont {H.~K.~C.}\ \bibnamefont
  {Beukers}}, \bibinfo {author} {\bibfnamefont {S.}~\bibnamefont {Baier}},
  \bibinfo {author} {\bibfnamefont {J.}~\bibnamefont {Borregaard}},\ and\
  \bibinfo {author} {\bibfnamefont {R.}~\bibnamefont {Hanson}},\ }\href
  {https://doi.org/10.1038/s41586-022-04697-y} {\bibfield  {journal} {\bibinfo
  {journal} {Nature}\ }\textbf {\bibinfo {volume} {605}},\ \bibinfo {pages}
  {663} (\bibinfo {year} {2022})}\BibitemShut {NoStop}%
\end{thebibliography}%

\end{document}